# Constrained minimization problems and FFT-based solvers: application to local Dirichlet boundary conditions and contact mechanics.

Lionel GÉLÉBART[a,*], Yaovi Armand AMOUZOU-ADOUN[a]

[a]*Université Paris-Saclay, CEA,SRMA, Gif-sur-Yvette, 91191, FRANCE*

Abstract

The present paper focuses on a recent and active research axis to overcome the limitations of FFT-based solvers in order to apply various types of boundary conditions (BCs) and not only periodic BCs. A complete framework based on discrete trigonometric transforms (DTTs) of various types, possibly combined with discrete Fourier transform, has been proposed recently to account for any type of BCs defined per face of the unit-cell and per component of the displacement or traction vector. A parallel implementation by Amouzou-Adoun et al. [1] recently proved its robustness and versatility.

The main novelty consists of few modifications introduced in the displacement-based FFT/DTT iterations in order to account for Dirichlet BC on a selected part of a boundary face (and Neumann BC on the other part), or on interior points, and also to apply kinematic relations between displacements on different nodes. Introduced from the viewpoint of local equations, these modifications are then discussed from the perspective of constrained minimization and saddle point systems. The current proposition is identified as a 'null-space' method in the category of 'segregated' methods [2]: the reaction forces (acting locally to enforce the Dirichlet BC) are evaluated a posteriori once the displacement field is found. This approach is original in the context of FFT-based solver's extension to Dirichlet BCs.

Starting from this modified FFT/DTT-based algorithm to account for equality constraints, the second important novelty consists in embedding this algorithm in a conventional active-set strategy to account for inequality constraints, with an application in the context of contact mechanics. Two implementations have been proposed, in the context of small strain and finite strains.

It is believed that these propositions significantly expand the field of applications of FFT-based solvers.

Keywords : FFT-based solver; Lagrange multipliers; Constrained minimization; Contact Mechanics

1 : Introduction

Solving mechanical problems defined on heterogeneous unit-cells, subjected to periodic boundary conditions (BCs), with fast Fourier transform (FFT)-based algorithms was proposed initially by Moulinec and Suquet [3]. Due to their simplicity, efficiency and ability for parallelism, they have become very popular and various improvements have been introduced since the original proposition. The reader is invited to consult the review papers of Schneider [4], Lucarini et al. [5] or Gierden et al. [6], for an overview of the different advances. The next three paragraphs position our choices in terms of algorithm, spatial discretization of equations and extension to non-periodic BCs. The last paragraph introduces the central question of local Dirichlet BCs and unilateral contact, which can be regarded as constrained minimization problems. The main goal of the present paper is to introduce simple modifications in an FFT-based algorithm in order to solve these types of problems.

Regarding algorithms, the original proposition consists of a simple fixed-point algorithm [3] whose convergence strongly depends on the mechanical contrast and the choice of the reference material. To improve the efficiency of the fixed-point algorithm, various propositions have been made by Eyre and Milton for the accelerated scheme [7], by Michel et al. for the Augmented Lagrangian scheme [8], and Monchiet and Bonnet for the polarization-based scheme [9]. Note that the aforementioned two methods have been recognized as special cases of the last scheme [10]. Another approach consists in switching from the fixed-point algorithm to the standard conjugate gradient algorithm, as proposed simultaneously by Zeman et al. [11] and by Brisard and Dormieux [12]. A combination of this conjugate gradient method for linear equations with a standard Newton-Raphson algorithm was then proposed, by Gélébart and Mondon-Cancel [13] and Kabel et al. [14], to solve non-linear problems. Note that the conjugate gradient method belongs to the class of Krylov-type methods that can be used as well (the most popular being MINRES and BiCGStab, see review in [4]). The reinterpretation of the Moulinec-Suquet fixed-point algorithm as a gradient descent method opened the door to the use of fast gradient methods [15] [16] and to the Barzilaï-Borwein method that proved to be very efficient, with low memory footprint and adapted to non-linear behaviors [17]. In practice, since 2014 in the AMITEX_FFTP code [18] and as depicted in [19], the fixed-point algorithm is accelerated with the same convergence acceleration technique as used in the finite element (FE) code CAST3M [20] to accelerate a modified Newton-Raphson algorithm (i.e. using a constant stiffness matrix). This implementation was recognized as an Anderson technique [21] with a depth of 4 and a periodicity of 2 or 3. In addition to an efficient convergence acceleration, with a reduced sensitivity to the mechanical contrast and to the reference material choice, it has the advantage of the quasi-Newton methods (and other solvers), not requiring the material tangent (whose implementation may be tedious). A drawback of the method is that it needs the storage of various (4 in our implementation) couples of solutions and residual fields that increases the memory footprint of the method (especially when compared to the original fixed-point algorithm). It must be emphasized that in the present paper, taking into account local Dirichlet BCs, the convergence acceleration of the fixed-point algorithm is essential (convergence fails when deactivating the acceleration).

Regarding the spatial discretization of the mechanical equations, leading to a linear system solved using discrete Fourier transform (DFT, abusively called FFT in the following, the fast Fourier transform being an efficient method used to evaluate the DFT), various approaches can be followed. The initial proposition of Moulinec and Suquet [3] can be regarded as a collocation method, starting from the continuous equation and using a truncation of Fourier series to solve the so-called Lippman-Schwinger equation for an auxiliary problem defined on a homogeneous reference material. This allows us to define a discrete Green operator (DGO) that can be applied in Fourier space using forward and backward FFT. A second approach relies on the use of variational formulations to build the DGO [22][23]. Finally, the third approach, which is adopted in the present paper, consists in using a finite difference (FD) scheme to build discrete differential operators (divergence, gradient) and then to build the DGO directly from the local equations. In practice the FD-based DGO is very similar to the collocation-based DGO, replacing the Fourier frequencies by "modified frequencies", functions of the Fourier frequencies, whose definitions derive from the FD scheme. An example of using a FD scheme to build the DGO in mechanical problems can be found in the work of Wiegmann in 1999 [24]. Since then, various FD schemes have been proposed and the hexahedral FD scheme (so-called "rotated scheme" by Willot [25]) proves to be very useful in the context of mechanics of heterogeneous materials: it improves both the convergence of the solver and the quality of the solution with an efficient reduction of spurious oscillations. In addition, this hexahedral FD scheme is strictly equivalent to a hexahedral linear finite element with reduced integration [26], which allows for rigorous cross-validation as done recently in [27] for conduction problems. Quite recently, Finel proposed an alternative FD scheme, referred to in this work as TETRA2, based on the two symmetric tetrahedra present within a 3D voxel [28]. This novel scheme demonstrated an improvement of the stress and strain fields, solutions of elastic problems, with respect to the use of the hexahedral scheme (referred to as HEX1). These two schemes are used in the present paper.

FFT-based solvers are classically developed in the framework of periodicity. The question of extending the field of application of FFT-based methods from full periodic BCs to various types of BCs, has attracted increasing interest in recent years, with different proposed strategies. The first strategy relies on a buffer zone around the unit-cell so that the periodicity is applied at the boundary of the buffer [29]. This approach has been validated for both Neumann and Dirichlet BCs [30]. However, it increases the problem size and the convergence depends on the stiffness and on the thickness of the buffer (especially for Dirichlet BCs). A buffer zone was also considered in the work of Rutka et al. using immersed interface methods on homogeneous materials [31]. Additional jumps defined at the interfaces were evaluated from the Schur complement to find the jumps using an FFT-based solver. However, for heterogeneous materials, the FFT-based solver is prohibited [32]. The buffer zone strategy has also been recently revisited by Zecevic and Lebensohn [33], using an elaborated two-level augmented Lagrangian algorithm to enforce Dirichlet BCs. Alternatively, To et al. [34] proposed to introduce source terms at the interfaces between the buffer and the domain, and to determine them taking advantage of exact Green operator and form factors to describe boundaries. The extension of this method to non-linear behaviors and/or finite-strain problems does not seem straightforward. A similar approach proposed by Sancho et al. in [35] consists in introducing artificial-body-forces to impose Dirichlet BCs. We believe that this approach is very close, if not the same, to the introduction of Lagrange multipliers, that can be regarded as the reaction forces introduced in the present paper. However, in our approach, these reaction forces are not in the vector of unknowns, they are evaluated a posteriori once the convergence is achieved on the unknown displacement field. Another approach consists in using symmetrized microstructures and loadings, so that Neumann or Dirichlet conditions can be applied to the unit-cell. Yet, this approach multiplies the problem size by 4 (in 2D) or 8 (in 3D) [36]. A penalty-like approach has also been proposed to deal with null Dirichlet BCs through the introduction of an additional thin stiff layer as proposed by Bodecker et al. [37]. However, penalty methods come with an important drawback: increasing the penalty factor improves the precision on the Dirichlet condition but degrades the convergence of iterative solvers. In the present work, the Dirichlet BC is strictly satisfied and no additional parameter is added. The last approach consists in taking advantage of discrete trigonometric (sine or cosine) transforms (DTTs) to apply whether Neumann or Dirichlet BCs. This approach was first introduced (to the authors' knowledge) in 1999 by Wiegmann [24] for homogeneous materials. Shortly, the idea was to establish the connection between Neumann or Dirichlet BCs, and field symmetries, and to deduce from these symmetries the use of DTTs (cosine or sine transforms) instead of the DFT used for periodic BCs. It was then extended to heterogeneous materials, in mechanics [38], but for a limited number of "mixed" loading cases, or in conductivity for a rather exhaustive combinations of loadings, including Neumann, Dirichlet, and Periodic on each face (or couple of face for periodicity)[27]. In [39], based on DTTs, Risthaus and Schneider proposed a unified framework to account for full-Dirichlet, Neumann or periodicity in the context of thermal homogenization. They also proposed an implementation of the full-Dirichlet BC for mechanical problems [40] using the rotated scheme of Willot [41] and a conjugate gradient solver [42]. Almost in the same time, similar works have been proposed using a Fourier-Galerkin based discretizations instead of FDs [43][39][40][44][45]. Finally, the authors recently proposed a robust and efficient parallel implementation for any type of boundary condition, per face and per component, for linear and non-linear behavior, using the small strain and finite strain frameworks, based on the hexahedral and double tetrahedra FD schemes [1]. As the present work builds on this implementation, a brief description of this starting point, including algorithm, spatial discretization and DTTs based non-periodic BCs, is given in section 3.

As mentioned above, using DTTs allows us to account for Dirichlet or Neumann or periodic BCs, per component of the displacement (or traction) vector, and per face of the unit-cell. However, the type of BC (i.e. Dirichlet, Neumann, periodic) is imposed for the whole face so that it is not possible to mix for example Neumann and Dirichlet BCs on two parts of the same face. In addition, it is neither possible to apply Dirichlet conditions to inner points of the unit-cell, nor linear relations between displacement vectors on different points. All these questions can be treated in FE codes through the introduction of Lagrange multipliers in order to solve a constrained minimization problem (equality constraint). Hence, the main goal of section 4 is to solve such problems using an efficient FFT-based solver. The proposed method has the advantage of being very simple in terms of concept and implementation. It relies on a displacement-based accelerated fixed-point algorithm, for which, at each iteration, both the kinematic conditions and the corresponding equilibrium condition are enforced. It results from a simple idea: the stress divergence is balanced by a reaction force where the displacement is prescribed. The method is first explained in sections 4.1 and 4.2, starting from the set of local discrete equations, taking into account reaction forces. In section 4.3, the method is analyzed from the viewpoint of constrained minimization problems and the reaction forces are directly associated with Lagrange multipliers. According to Benzi et al. [2], the numerical solution of saddle point problems can be divided into two broad categories: segregated and coupled methods. Coupled methods solve at the same time both the main unknown (in our case the displacement field) and the Lagrange multipliers. The main originality of the present paper is that the proposed FFT-based solver belongs to the category of segregated methods [2] as only the displacement field is solved, reactions forces being evaluated as a post-processing. The connection with the so-called null-space methods [2] is established in section 4.4. Alternatively, most of the FFT-based methods introduced above are based on the identification of additional unknown fields (whether reaction forces or displacements in a surrounding domain) to fulfill Dirichlet condition.

The application proposed in section 4.1 for testing and validating the method concerns the simulation of a half-"CT-like" specimen. Besides its prime interest to the mechanics of materials community, it requires imposing displacement at points located inside the unit-cell (to apply the loading), and, null-Neumann BC on the cracked part together with mixed Neumann/Dirichlet BCs on the other part of the same face. In section 4.2, especially with the aim of reducing oscillations arising with HEX1 and TETRA2 schemes, alternative ways to apply (local) Dirichlet BCs are proposed. One of them consists in dispatching the reaction force on a set of neighboring nodes. The other one prescribes the displacement at a center of a voxel through a linear combination of the displacement at nodes. This second approach demonstrates the ability of the method to take into account linear relations between displacement at different nodes. The simulations performed in section 4, completely out of reach of "standard" FFT-based solvers, demonstrate the significant potential of the proposed method to widely expand their field of application. Here, 'standard' FFT-based solvers stand for generic iterative solvers built on a full periodic discrete Green operator.

After considering equality constraints, the purpose of section 5 is to tackle the more complex case of minimization problems subject to inequality constraints. A typical example, once again of prime interest to the mechanics of materials community, is the simulation of contact mechanics and especially of indentation. The question has already been investigated by Zecevic and Lebensohn [33] using a general approach based on a buffer layer and a rather complex algorithm (two-level augmented Lagrangian). Their approach is also general but very different from ours which does not introduce any buffer. An original treatment of contact mechanics between different solids, compatible with FFT-based solvers, has also been proposed by Leichner et al. [46] taking into account the position of the solid interfaces through the transport of level-set functions and the contact conditions through the introduction of an anisotropic 'contact' material filling the gap between two solids. However, this approximation of the contact condition introduces a gap w.r.t. analytical solution. On the other hand, many specific solvers using FFT (in 2D or 3D), devoted to contact have been investigated by the contact mechanics community. The reader is invited to consult the review of Wang et al. [47] for a deeper investigation. Briefly, for elastic contacts, the displacement of a surface subjected to distributed local pressure is obtained through a 2D convolution product using the Boussinesq solution. This convolution can be evaluated whether in real space [48] or in Fourier space as proposed by Ju and Farris [49], Stanley and Kato [50] followed by Liu et al. to reduce the effect of periodic BC [51]. To account for an elasto-plastic behavior, Frérot et al. [52] utilize the Mindlin solution that provides the displacement field in a half-space specimen for a given distribution of eigenstresses. In that case the problem is solved in 3D and 2D-FFT are applied to avoid the issue with periodicity in the third direction. In the present approach, the problem does not take into account half-space solutions, it is directly solved in 3D with chosen BCs on the five faces that are not concerned by contact mechanics. A drawback of the method is its sensitivity w.r.t. the choice of BCs and the size of the domain. However, this drawback could be reduced using non-regular grids (as in [53]) or multiscale methods (as in [30]) to perform the simulation on a larger domain keeping a high resolution below the contact. The advantage of the method is its versatility as it should extend straightforwardly to heterogeneous materials and non-linear behaviors. Finally, it must be kept in mind that our goal is to propose a generic FFT-based solver able to account for unilateral constraints, contact mechanics being one possible application but not the single target. For the sake of simplicity, a spherical and rigid indenter is considered together with frictionless contact. The algorithm proposed in section 4 is adapted to account for a so-called active-set method (the active contact zone is subject to Dirichlet constraints, i.e. sliding on the sphere surface, and the obtained solution is analyzed to detect the points where contact conditions are violated, then the active contact zone is updated and the procedure repeated until convergence). The update of the active contact zone is based on a naïve primal-dual active-set strategy: the active-set is updated if the gap (primal quantity) is not strictly positive somewhere out of the contact zone or if the reaction forces (dual quantity) are positive somewhere in the contact zone. Modifying this update condition to account for the primal-dual strategy proposed by Hintermüller et al. [54] could be done easily (for the present application, the naïve condition allows for convergence so we have not used Hintermüller's condition, whose drawback is the introduction of an additional algorithm parameter $c$ that combines the two conditons in a single one). Note that, in an original way, the proposed algorithm updates the contact zone within the iterations of the iterative solver as soon as the stability of the domain violating the two conditions is detected. Similarly, Polonsky and Keer [48] proposed to modify the active-set within the iterations of a conjugate gradient. The algorithm is presented using the small strain (SS) approximation or the finite strain (FS) framework, with minor differences to switch from one to another. To validate the method, results are compared to the Hertz theory on the reaction force as a function of the indenter depth (additional validations are reported in appendix A1). For small ($\delta$ (depth)/$R$(indenter radius)) ratios, both the SS and FS results are very close to the Hertz theory. The stability of the results w.r.t. spatial resolution and loading discretization is then demonstrated from small to large $\delta/R$ ratios. For large $\delta/R$ ratios, the SS and FS curves depart from each other.

To summarize, section 3 briefly recalls the foundations on which the present paper relies, section 4 introduces the ability of applying equality constraints using reaction forces (i.e. Lagrange multipliers), section 5 extends section 4 to inequalities with an application to indentation. In section 4 and 5, the algorithm is first introduced from the viewpoint of local equations and then discussed from the viewpoint of constrained minimization problems. Conclusion and future prospects are proposed in section 6.

## 2 : Abbreviations and notations

The following abbreviations are often used in the paper: small strain (SS), finite strain (FS), boundary condition (BC), integration point (IP), volume element (VE), finite difference (FD), discrete Fourier transform (DFT), discrete trigonometric transform (DTT), discrete transform (DT) (DTs include the DTTs and the DFT), iDFT, iDTT and iDT their inverse.

For the notations: $\boldsymbol{a}$ denotes a vector, $\underline{A}$ a second-order tensor, and $\mathbb{C}$ or $\mathbb{R}$ a fourth-order tensor. The divergence of a second-order tensor is $\mathbf{div}(\underline{A})$, the gradient of a vector is $\underline{\nabla}(\boldsymbol{a})$ and its symmetric part is $\underline{\nabla}^{\mathrm{S}}(\boldsymbol{a})$. An index $\mathrm{D}$ denotes discrete divergence $\mathbf{div_D}$ or gradient $\underline{\nabla_{\mathrm{D}}}$ operators (evaluated for example with a given FD scheme $\mathrm{D}$).

Finally, to locate the position of multiple points in grid, the *fortran*-like (or *matlab*, or *numpy*…) notation is used. Hence the notation (0:10,10,37) refers to line of 11 nodes whose coordinates are $x_1 = 0,1,\dots,9,10$, $x_2 = 10$ and $x_3 = 37$. This notation extends to a surface of nodes for example (0:10, 10:20,18) for 11x11 nodes. In the application below, the nodes (corners of voxels) are located at integer positions and voxel centers at half position, for example (0.5:9.5, 10.5,37.5) for the 10 voxel centers equally distributed on a line located with $x_2 = 10.5$ and $x_3 = 37.5$.

## 3 : FFT-based solver for periodic and non-periodic BCs

This section summarizes the choices made for the FFT-based solver used as a starting point for the extensions proposed in section 4 and 5. Its massively parallel implementation in the code AMITEX* has demonstrated its robustness through various applications in linear, non-linear, small strain and finite strain frameworks for non-trivial BCs [1]. For the sake of simplicity, the presentation builds upon the SS approximation and linear elasticity, but the extension to non-linear behaviors and FS is straightforward (see in [1]). Note that FS are also considered in section 5 to deal with contact mechanics.

### 3.1 : FFT-based solver with periodic BCs

#### 3.1.1 : Problem under periodic BCs

The set of equations corresponding to a linear elastic mechanical problem, defined on a heterogeneous volume element $\Omega$ subjected to periodic BCs is given below in equation (1).

$$
\begin{cases}
\mathbf{div}(\underline{\sigma}) = \mathbf{0} & \text{in } \Omega \,/\, \partial\Omega \\
\underline{\tilde{\varepsilon}} = \underline{\nabla}^{s}(\widetilde{\boldsymbol{u}}) & \text{in } \Omega \,/\, \partial\Omega \\
\underline{\sigma} = \mathbb{C} : (\underline{\varepsilon}^{*} + \underline{\tilde{\varepsilon}}) & \text{in } \Omega \,/\, \partial\Omega \\
\widetilde{\boldsymbol{u}}(\boldsymbol{x} + \boldsymbol{h_d}) = \widetilde{\boldsymbol{u}}(\boldsymbol{x}) & \text{on opposite faces of } \partial\Omega \\
(\underline{\sigma}.\boldsymbol{n})(\boldsymbol{x} + \boldsymbol{h_d}) = -(\underline{\sigma}.\boldsymbol{n})(\boldsymbol{x}) & \text{on opposite faces of } \partial\Omega
\end{cases}
\tag{1}
$$

$\underline{\sigma}$ is the second-order stress tensor, $\mathbb{C}$ the fourth-order heterogeneous elastic stiffness tensor and $\widetilde{\boldsymbol{u}}$ the fluctuation displacement whose symmetric gradient is the fluctuation strain $\underline{\tilde{\varepsilon}}$ around an arbitrary prescribed strain field $\underline{\tilde{\varepsilon}}^{*}$(compatible with a field $\boldsymbol{u}^{*}$, $\underline{\varepsilon}^{*} = \underline{\nabla}^{s}(\boldsymbol{u}^{*})$). In the context of first order homogenization $\underline{\varepsilon}^{*}$ is generally chosen homogeneous, but it is not mandatory (see for example [55] in the context of beam or plates homogenization). The boundary conditions are explicitly defined on $\partial\Omega$ (with $\boldsymbol{h_d}$ the periodicity vector between two opposite faces in direction $d$). This description is well suited for an FE-based implementation. For an FFT-based implementation, this problem is equivalently modified into problem (*2*), with $\widetilde{\boldsymbol{u}}$ and $\underline{\sigma}$ extended by periodicity.

$$
\begin{cases}
\mathbf{div}(\underline{\sigma}) = \mathbf{0} & \text{in } \Omega^{\infty} \\
\underline{\tilde{\varepsilon}} = \underline{\nabla}^{s}(\widetilde{\boldsymbol{u}}) & \text{in } \Omega^{\infty} \\
\underline{\sigma} = \mathbb{C} : \left(\underline{\varepsilon}^{*} + \underline{\tilde{\varepsilon}}\right) & \text{in } \Omega \\
\widetilde{\boldsymbol{u}} \quad \Omega - \text{periodic} & \text{in } \Omega^{\infty} \\
\underline{\sigma} \quad \Omega - \text{periodic} & \text{in } \Omega^{\infty}
\end{cases}
\tag{2}
$$

Using extended fields, the derivation (of $\widetilde{\boldsymbol{u}}$ or $\underline{\sigma}$) around a point located at the boundary $\partial\Omega$ is the same as around any other point in the volume $\Omega$. Considering a regular grid discretization, discrete derivation is then invariant by translation and can be evaluated using discrete Fourier transform.

In the following, the spatial discretization of $\Omega$ is considered with a regular grid of cubic elements, or voxels (for simplicity, but a grid of parallelepiped elements can be considered as well), with displacement $\widetilde{\boldsymbol{u}}^{\boldsymbol{i}}$ and $\mathbf{div}(\underline{\sigma})^{i}$ defined at nodes $i$, and stress $\underline{\sigma}^{i}$ and strain $\underline{\tilde{\varepsilon}}^{i}$ defined at integration points of element $i$. For more than a single integration point, this simplified notation $\underline{\sigma}^{i}$ gathers the stress evaluated at all the integration points of the element $i$. In the present paper, focus is put on the discrete derivation schemes (see Figure 1): the HEX1 corresponding to linear FE with a single IP [25][26], and the TETRA2 scheme recently proposed by Finel [28] (and described in a simpler and more practical way in [56]), with 2 IPs defined at the center of each element (i.e. voxel). For each IP, the strain is deduced from the displacement of nodes located on one of the two tetrahedra defined from the 8 nodes of the cube.

For the sake of simplicity, exponent $i$ is omitted unless necessary (as in section 4 where displacement is prescribed on specified nodes).

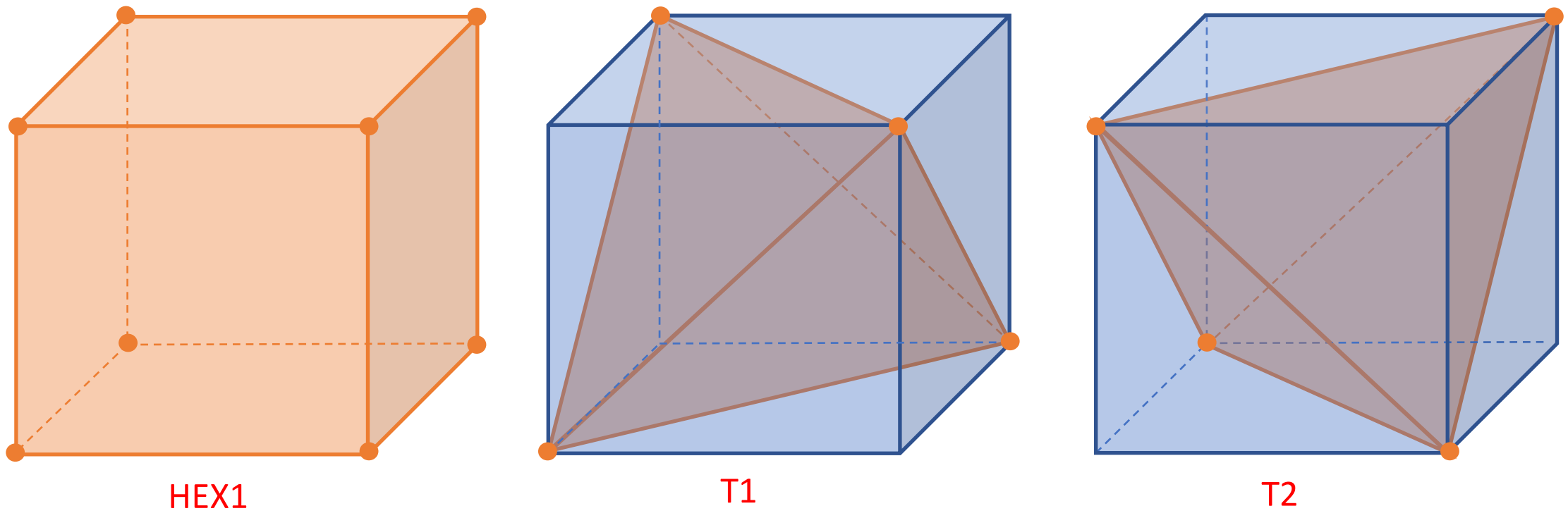

*Figure 1 : Description of the HEX1 and TETRA2 FD schemes. For the HEX1 scheme, displacement gradient is evaluated at a single integration point (center of the voxel), and for the TETRA2 scheme, two gradients are evaluated at two IPs located at voxel center, each value associated with one of the two tetrahedra T1 and T2.*

3.1.2 : Discrete Green operator and fixed-point algorithm.

Considering the discretized equations, the periodic extension used in problem (*2*), and introducing the term $\mathbb{R}_0 : \underline{\nabla}_{\mathrm{D}}(\widetilde{\boldsymbol{u}})$, problem (*2*) can be re-written as problem (*3*) - left part.

$$\begin{cases} \underline{\sigma} = \mathbb{C} : \left( \underline{\varepsilon}^* + \underline{\nabla}^{\mathrm{S}}_{\mathrm{D}}(\widetilde{\boldsymbol{u}}) \right) \\ \boldsymbol{p} = \mathbf{div_D} \left( \underline{\sigma} - \mathbb{R}_0 : \underline{\nabla}_{\mathrm{D}}(\widetilde{\boldsymbol{u}}) \right) \\ \mathbf{div_D} \left( \mathbb{C}_0 : \underline{\nabla}^{\mathrm{S}}_{\mathrm{D}}(\widetilde{\boldsymbol{u}}) \right) = -\boldsymbol{p} \end{cases} \Leftrightarrow \begin{cases} \underline{\sigma} = \mathbb{C} : (\underline{\varepsilon}^* + \underline{\nabla}^{\mathrm{S}}_{\mathrm{D}}(\widetilde{\boldsymbol{u}})) \\ \boldsymbol{p} = \mathbf{div_D} \left( \underline{\sigma} - \mathbb{R}_0 : \underline{\nabla}_{\mathrm{D}}(\widetilde{\boldsymbol{u}}) \right) \\ \widehat{\widetilde{\boldsymbol{u}}} = \widehat{\mathbb{G}}_{\mathrm{D}}(\widehat{\boldsymbol{p}}) \end{cases} \quad (3)$$

Here, $\mathbb{R}_0$ is a fourth-order tensor, generally associated with an isotropic behavior (this is no more the case for general non-periodic BCs, see section 3.2.3). In addition, a subscript $\mathrm{D}$ has been added to derivation operators, $\mathbf{div}$ and $\underline{\nabla}$, in order to emphasise the discrete derivation and, consequently the discrete nature of these equations. In the present paper, $\mathrm{D}$ corresponds to the HEX1 or TETRA2 scheme. For these two schemes, the discrete displacement and stress divergence fields are defined at voxel corners (nodal fields) and the stress and displacement gradients at voxel centers (element fields).

Solving problem *(*3) - left part, assuming periodic extensions of $\widetilde{\boldsymbol{u}}$ and $\underline{\sigma}$, can be done using discrete Fourier transform (with the notation $DFT(f) = \hat{f}^{DFT}$ or simply $\hat{f}$) as depicted in problem *(*3) - right part, that defines the following fixed-point algorithm: first step, for a given $\widetilde{\boldsymbol{u}}$ evaluate $\underline{\sigma}$ and then $\boldsymbol{p}$ in real space, second step, evaluate a new $\widetilde{\boldsymbol{u}}$ from the application of the discrete Green operator $\mathbb{G}_{\mathrm{D}}$ (using a back and forth in Fourier space), and iterate until convergence. This displacement-based algorithm requires the evaluation of gradient and divergence in real space (using HEX1 or TETRA2 scheme), which is quite different from the strain-based algorithm originally proposed in [3] and still widely used. Note that working with the displacement field as unknown, instead of the strain field, simplifies the application of local Dirichlet conditions in section 4.

For the HEX1 scheme, using an isotropic elastic behavior for $\mathbb{R}_0$, with Lamé coefficients $(\lambda_0, \mu_0)$, the expression of the discrete Green operator in Fourier space reads:

$$\widehat{\mathbf{u}} = \widehat{\mathbb{G}}_{\mathrm{D}}(\widehat{\boldsymbol{p}}) = -\frac{\widehat{\boldsymbol{p}}}{\mu_0\|\boldsymbol{\xi}_{\boldsymbol{D}}\|^2} + \frac{(\lambda_0+\mu_0)\widehat{\boldsymbol{p}}.\boldsymbol{\xi}_{\boldsymbol{D}}}{(\lambda_0+2\mu_0)\|\boldsymbol{\xi}_{\boldsymbol{D}}\|^4}\boldsymbol{\xi}_{\boldsymbol{D}} \tag{4}$$

In this equation, $\boldsymbol{\xi}_{\boldsymbol{D}}$ represents the modified wave vectors associated with the HEX1 scheme (D=HEX1). As well as $\widehat{\mathbf{u}}$ and $\widehat{\boldsymbol{p}}$, $\boldsymbol{\xi}_{\boldsymbol{D}}$ is defined on a 3D grid discretizing the Fourier space, with the same grid size as the grid in physical space (in practice, as the physical fields are real-valued, the 3D-DFT only stores a half of the grid in Fourier space, generally the first direction is reduced). This equation is evaluated locally at each point of Fourier space. The expression of $\boldsymbol{\xi}_{\boldsymbol{D}}$ can be found in [25] for the HEX1 scheme. The expression of the discrete Green operator is a bit more complex for the TETRA2 scheme and can be found in [28] and [1].

To improve the convergence of the fixed-point algorithm, the convergence acceleration technique proposed by the developers of the CAST3M FE-code to accelerate the modified-Newton Raphson algorithm for solving non-linear problems [20] is directly applied to the fixed-point algorithm. This acceleration convergence technique was identified as an Anderson convergence acceleration method [57].

Finally, the convergence criteria used to stop the iterative algorithm is evaluated on the equilibrium condition. It reads for cubic voxels of size $d$ (an expression is proposed in [30] for non-cubic voxels):

$$\frac{d\left\|\mathbf{div}_{\mathbf{D}}(\underline{\sigma})\right\|_2}{\left\|\underline{\sigma}\right\|_2} < tol_{eq} \tag{5}$$

The norm of a discrete vector field $||\boldsymbol{p}||_2$ defined at nodes is $\sqrt{\sum_{i\in G_N}\sum_{j=1}^{3}\left(p_j^i\right)^2}$, and the norm of a second order tensor field $\left\|\underline{\sigma}\right\|_2$defined at integration points is $\sqrt{\sum_{i\in G_C}\sum_{j=1}^{3}\sum_{k=1}^{3}\left(\sigma_{jk}^i\right)^2}$ ($G_N$ and $G_C$ gather all the nodes, or centres, of the grid). For the TETRA2 scheme, the stress tensor is the average stress over the two integration points.

To sum-up, this section provides a quick overview of the FFT-based solver used in the present paper. If the use of a displacement-based algorithm, with evaluation of derivatives in real space, can be regarded as an originality also described in a recent companion paper [1], the other aspects are rather standard and can be found in review articles focused on FFT-based solvers [4][5].

### 3.2 : Extension to non-periodic BC using DTTs

Recently, the question of the extension of FFT-based solvers to non-periodic BCs has been addressed by several teams, using appropriate discrete trigonometric transforms instead of a discrete Fourier transform in each direction for each component of the displacement or traction vector [27][58][44]. This approach allows us to apply whether Neumann or Dirichlet or periodic BCs on each face and for each component of the displacement/traction vector. The purpose of the present section is to give an overview of this approach, the interested reader is invited to read [1] to go deeper into details.

#### 3.2.1 : Correspondence between boundary conditions and symmetry extensions

First of all, the problem to solve is the same as problem (1) except the boundary conditions. In the general case, for each face, each component of the displacement or traction vector is subjected to whether Neumann, Dirichlet or periodic BCs (considering couple of opposite faces in case of periodic BC). As in the previous section, the first step consists in replacing the problem with boundary conditions (similar to problem (1)) by a problem with appropriate field extensions for the fluctuation displacement $\tilde{\boldsymbol{u}}$ and stress $\underline{\sigma}$ (similar to problem (*2*)). In the case of null-Neumann or Dirichlet conditions, the extensions are symmetries: even-symmetry (S for Symmetry) or odd-symmetry (A for Antisymmetry). Hence, the problem can be written in the most general case as follows:

$$\begin{cases} \mathbf{div}(\underline{\sigma}) = \mathbf{0} & \text{in } \Omega^{\infty} \\ \underline{\tilde{\varepsilon}} = \underline{\nabla}^{s}(\tilde{\boldsymbol{u}}) & \text{in } \Omega^{\infty} \\ \underline{\sigma} = \mathbb{C}:\left(\underline{\varepsilon}^{*} + \underline{\tilde{\varepsilon}}\right) & \text{in } \Omega \\ \tilde{\boldsymbol{u}} \ \ \text{appropriate symmetries} & \text{in } \Omega^{\infty} \\ \underline{\sigma} \ \ \text{appropriate symmetries} & \text{in } \Omega^{\infty} \end{cases} \tag{6}$$

The question is now to define the "appropriate symmetries". For that purpose, the 6 faces bounding the volume element are denoted $S_J^0$ and $S_J^1$ $(J = 1{:}3)$: $S_J^0$ with an outer normal vector $\boldsymbol{N_J^0} = -\boldsymbol{e_J}$, positioned at $x_J = 0$, and $S_J^1$ positioned at $x_J = L_J$ with an outer normal $\boldsymbol{N_J^1} = \boldsymbol{e_J}$. The appropriate symmetries of $\tilde{\boldsymbol{u}}$ and $\underline{\sigma}$ can be completely determined according to Table 1, considering each component $i$ and each face $S_J^{\alpha}$.

| | $\tilde{u}_i$ | $\sigma_{iJ}$ | $\sigma_{ij(j\neq J)}$ |
|---|---|---|---|
| Dirichlet | A | S | A |
| Neumann | S | A | S |
| Periodicity | P | P | P |

*Table 1: Extensions used to satisfy Dirichlet, Neumann or Periodic condition for component $i$ on a face $S_J^{\alpha}$ (A for Antisymmetry (odd-symmetry), S for even-symmetry, and P for periodicity)*

To clarify the idea behind this table, it must be noted that an odd-symmetry (A) enforces a field to vanish at boundary and, conversely, an even-symmetry (S) leaves the quantity free (unconstrained) at boundary. Hence, for Dirichlet condition on component $i$ on a face $S_J^{\alpha}$, the extension of $\tilde{u}_i$ is A (imposed null) and $\sigma_{iJ}$ is S (free). Using the symmetries proposed for the other components of $\underline{\sigma}$ in the last column of Table 1, it can be verified that the component $i$ of $\mathbf{div_D}(\underline{\sigma})$ systematically vanishes at the boundary (extension of $\left(\mathbf{div_D}(\underline{\sigma})\right)_i$ is A). In other words, as the component $i$ of the unknown fluctuation displacement is imposed at the boundary, by construction (using symmetry extensions), the corresponding equilibrium equation is eliminated, by construction. On the other hand, for null-Neumann condition on component $i$ on a face $S_J^{\alpha}$, the extension of $\sigma_{iJ}$ is A (imposed null) and $\tilde{u}_i$ is S (free). Using the symmetries proposed in Table 1, the component $i$ of $\mathbf{div_D}(\underline{\sigma})$ is left completely free (extension of $\left(\mathbf{div_D}(\underline{\sigma})\right)_i$ is S). In other words, the component $i$ of the unknown fluctuation displacement is left free at the boundary, by construction (using symmetry extensions), and the corresponding equilibrium equation is fully active (i.e. not imposed to zero by construction).

### 3.2.2 : Correspondence between symmetry extensions and DTTs

In section 3.1, a discrete transform, the discrete Fourier transform, is associated with a periodicity extension. Here, a discrete trigonometric transform is associated with a symmetry extension (whether S or A on opposite faces). The corresponding association is given in Table 2 below, using the same designation as in the *fftw* library [59] for discrete sine and cosine transforms).

| Symmetries | Nodal field ($\tilde{\boldsymbol{u}}$) | Element field ($\underline{\sigma}$) |
|---|---|---|
| SS | DCT1 | DCT2 |
| SA | DCT3 | DCT4 |
| AA | DST1 | DST2 |
| AS | DST3 | DST4 |

*Table 2 : Correspondence between symmetries on opposite faces and the appropriate Discrete Trigonometric Transform*

Note that this correspondence table assumes that boundary nodes coincide with the faces of the VE, and, consequently, that voxel centers do not. As $\tilde{\boldsymbol{u}}$ is a nodal field and $\underline{\sigma}$ is an element field defined at voxel centers, their corresponding DTTs differ. In practice, DTTs are only applied to nodal fields for the algorithm proposed in the present paper.

3.2.2 : Correspondence between DTTs and DFT

The last important point for this extension to non-periodic BCs lies in the relationship between DTTs applied to the current cell and DFT applied on extended cells. This relation can be written when considering 4-times extended cells following equation (*7*) together with Table 3 below.

$$S_{4F}(\widehat{q_4}) = \alpha DTT(S_R(q)) \tag{7}$$

| DTT | $S_R$: selection in $[0{:}n]$ (Real space) | $S_{4F}$: selection in $[0{:}4n-1]$ (Fourier space) | $N$: number of selected points | $\alpha$ |
|---|---|---|---|---|
| DCT1 | $[0{:}n]$ | $2\times[0{:}n]$ $(0,2,4,\dots,2n)$ | $n+1$ | $2$ |
| DCT3 | $[0{:}n-1]$ | $2\times[0{:}n-1]+1$ $(1,3,5,\dots,2n-1)$ | $n$ | $2$ |
| DST1 | $[1{:}n-1]$ | $2\times[1{:}n-1]$ $(1,3,5,\dots,2n-1)$ | $n-1$ | $-2i$ |
| DST3 | $[1{:}n]$ | $2\times[0{:}n-1]+1$ $(0,2,4,\dots,2n)$ | $n$ | $-2i$ |

*Table 3 : Selection of points in real space (to be used as input of DTTs) and selection of points in Fourier space for the 4 times extended signals.*

Equation (*7*) deals with a nodal signal $q$ (i.e. a component of $\tilde{\boldsymbol{u}}$ or $\boldsymbol{p}$ (see definition in equation (*3*)), which can be extended using the appropriate symmetries (see correspondence Table 2) to define a 4-times larger signal $q_4$ that is periodic on which DFT can be applied to obtain $\widehat{q_4} = DFT(q_4)$. The signal $q$ being defined on $[0{:}n]$ (i.e. $n+1$ points), DTT is applied to a selected subset of $q$ defined by $S_R$, a selection operator. In practice, the points concerned by an antisymmetry condition are not selected (for example with DST3, corresponding to symmetry AS, the first point is not selected and $S_R$ selects the subset $[1{:}n]$). Finally, equation (*7*) exhibits that, up to a factor $\alpha$, each point of the DTT signal corresponds to a specific selected point in the DFT of the 4-times extended signal. The selection operator $S_{4F}$ acting on the DFT of the extended signal is given in Table 3.

3.2.3 : Application of the discrete Green operator

An important point when dealing with non-periodic BCs through DTTs is that the usual reference material $\mathbb{R}_0$(with $\mathbb{R}_0{:}\underline{\nabla u} = \lambda_0 Tr(\underline{\nabla u})\underline{I} + 2\mu_0 \underline{\nabla^S u}$ ) associated with an isotropic elastic stiffness tensor can only be used for a limited set of BCs corresponding to periodic and normal mixed BCs (see for instance [44], [58] or [1]). In the most general case, one can use $\mathbb{R}_0{:}\underline{\nabla u} = 2\mu_0\underline{\nabla u}$ or, as proposed by Paux et al. [44], $\mathbb{R}_0{:}\underline{\nabla u} = \lambda_0 \underline{\nabla u}\odot\underline{I} + 2\mu_0\underline{\nabla u}$ (with $\odot$ the Hadamard product, so that $(\underline{A}\odot\underline{B})_{ij} = A_{ij}B_{ij}$, no summation on indices). In the different applications proposed in the present paper, this last proposition is employed.

Finally, the discrete Green operator defined for periodic boundary conditions (see equation (*4*), to be adapted according to the choice on $\mathbb{R}_0$ just above) can be used as well on the 4-times extended signal (periodic). Importantly, the application of the discrete Green operator is never evaluated on the complete DFT of the 4-times extended signal but only to the points selected by $S_{4F}$. Fields are 'virtually' extended but the extensions are never stored in memory.

### 3.2.4 : Consequences on the algorithm and its implementation

As a consequence, from an algorithm point of view, the fixed-point defined by the 3 successive steps corresponding to each line of problem (*3*)-right part, is exactly the same for non-periodic BCs and the convergence acceleration can be applied identically. The most important differences concern the application of different types of DT (or iDT) (possibly different for each component (of $\widetilde{\boldsymbol{u}}$) and for each direction) on a carefully selected subset ($S_R$ selection), and inversely, the application of inverse DTTs on $S_{4F}(\widehat{\boldsymbol{p}_4})$ taking care of the $S_R$ selection to obtain $\boldsymbol{p}$. In addition, it should be noted that combining periodic and non-periodic boundary conditions, the components (of $\widetilde{\boldsymbol{u}}$ or $\boldsymbol{p}$) in Fourier space can be purely real (if non-periodic in the 3 directions) or complex (if at least one direction is subjected to periodicity). In that second case, as the input field is real, the size of the transformed field is half-sized in the direction of the first periodicity condition. Hence, the implementation must handle fields in Fourier space having, for each component, possibly different types (real and complex), and for complex components, possibly different shapes. Implementation details can be found in [1], especially for a massively parallel (distributed memory) implementation using the *2decomp&fft* library [60] enhanced for that purpose.

To summarize, the extension of FFT-based solvers to non-periodic BCs through the use of DTTs (instead of DFT for non-periodic BCs) is not conceptually complicated but the implementation in the most general case requires a careful attention. Such an implementation was performed as a partial re-writing of the massively parallel code AMITEX [18] into a new version AMITEX*, implemented for linear and non-linear behaviors, in the small and finite strain frameworks [1].

However, if it greatly expands the field of application of FFT-based solvers, the limitation of this approach is that the type of condition (Neumann, Dirichlet or periodic) is homogeneous per face: it is not possible to mix Dirichlet condition on a part of a face and Neumann on the other part of the same face. The second limitation is that the conditions are applied at the faces of the VE but cannot be applied inside the VE. As FFT-based solvers are defined on grids, it can be useful to apply Dirichlet conditions in the VE: a typical example proposed in the present paper concerns the simulation of a Compact Tension (CT)-like specimen (a simplified version of a Compact Tension specimen).

## 4 : Local Dirichlet conditions – Minimization under equality constraints

Using DTTs in addition to FFT, as summarized in the previous section, allows us to significantly extend the field of applications of FFT-based solvers (once again, DTTs can be regarded as an FFT on a larger cell and the designation FFT-based solver is kept). However, this extension still suffers a limitation: if the type of BC, whether Neumann, Dirichlet or periodicity depends on the component of the displacement or traction vector, it must be the same for all the points belonging to a given face of the VE. Hence, it is not possible to have a surface that is free almost everywhere but with prescribed displacement on some few points, lines or areas (as classically done in standard FE solvers), nor to apply displacements at interior points within the VE. Such capabilities will open the door of structural analysis to FFT-based solver, with the advantage for the user to benefit from massively parallel implementations. The approach for prescribing local Dirichlet BC either at the free surface or in the volume is introduced below. A first proposition is made to apply strongly the local BCs, however it appears to be not well-suited to TETRA2 or HEX1 schemes, giving rise to important oscillations. To circumvent this issue, two options are proposed to relax the first proposition and vanish oscillations. Finally, the method introduced from the viewpoint of local equations is regarded from the perspective of minimization problem under equality constraints associated with the introduction of Lagrange multipliers.

### 4.1 : Strongly applied conditions

#### 4.1.1: Method and algorithm

The first proposition consists in strongly applying the Dirichlet conditions on a given set of nodes $I_0$ in directions gathered in the set $J_0$ (i.e. each element $I_0(k) = i_0$ is a node number, and each corresponding element of $J_0(k) = j_0$ is an integer between 1 and 3, representing the prescribed direction for node $i_0$). The prescribed displacements are denoted $(\boldsymbol{U}^{I_0})_{J_0}$. Assuming the existence of reaction forces acting, as consequence of the applied displacement, on the same nodes and in the same directions, denoted $(\boldsymbol{f}^{I_0})_{J_0}$, the equations of the problem are given below.

$$\begin{cases} \mathbf{div}(\underline{\sigma}) + (\boldsymbol{f}^{I_0})_{J_0} = 0 & \text{in } \Omega^\infty \\ \underline{\tilde{\varepsilon}} = \underline{\nabla^s}(\tilde{\boldsymbol{u}}) & \text{in } \Omega^\infty \\ \underline{\sigma} = \mathbb{C} : \left(\underline{\tilde{\varepsilon}}\right) & \text{in } \Omega \\ (\tilde{\boldsymbol{u}}^{I_0})_{J_0} = (\boldsymbol{U}^{I_0})_{J_0} & \\ \tilde{\boldsymbol{u}} \ \text{ appropriate symmetries} & \text{in } \Omega^\infty \\ \underline{\sigma} \ \text{ appropriate symmetries} & \text{in } \Omega^\infty \end{cases} \tag{8}$$

For the sake of simplicity, it is assumed that $\boldsymbol{u}^*$, and consequently $\underline{\varepsilon}^*$, is set to zero so that the displacement condition is directly applied to $\tilde{\boldsymbol{u}}$. Note that if $\boldsymbol{u}^*$ is not set to zero, then the stress is $\underline{\sigma} = \mathbb{C} : \left(\underline{\varepsilon}^* + \underline{\tilde{\varepsilon}}\right)$ as in problem (*6*), and the local boundary condition on the fluctuation displacement is $(\tilde{\boldsymbol{u}}^{I_0})_{J_0} = (\boldsymbol{U}^{I_0})_{J_0} - (\boldsymbol{u}^{*I_0})_{J_0}$. Definition of “appropriate symmetries” refers to the previous section and corresponds to Neumann or Dirichlet or periodic conditions applied on the 6 faces. Note that the local Dirichlet conditions are applied whether on interior nodes or on nodes of the faces on which null-Neumann conditions are applied. In addition, we assume that “enough” Dirichlet BCs are applied through symmetry extensions to remove any free rigid body motion. To solve this problem, the fixed-point algorithm defined by the set of successive equations (*9*) is proposed.

$$\begin{cases} proj_U : \ (\tilde{\boldsymbol{u}}^{I_0})_{J_0} = (\boldsymbol{U}^{I_0})_{J_0} \\ \underline{\sigma} = \mathbb{C} : \left(\underline{\nabla^s_{\mathrm{D}}}(\tilde{\boldsymbol{u}})\right) \\ \boldsymbol{p} = \mathbf{div_D}(\underline{\sigma}) \\ proj_P : \ (\boldsymbol{p}^{I_0})_{J_0} = 0 \\ \boldsymbol{p} = \boldsymbol{p} - \mathbf{div_D}\left(\mathbb{R}_0 : \underline{\nabla_{\mathrm{D}}}(\tilde{\boldsymbol{u}})\right) \\ \hat{\tilde{\boldsymbol{u}}} = \hat{\mathbb{G}}_{\mathrm{D}}(\hat{\boldsymbol{p}}) \end{cases} \tag{9}$$

The differences with the algorithm described in problem (*3*)-right part are minor and correspond to the addition of the lines referenced $proj_U$ and $proj_P$. Line $proj_U$ is the projection of the current fluctuation field $\tilde{\boldsymbol{u}}$ on the space of kinematically admissible fields (i.e. simply set the prescribed displacements) and line $proj_P$ is the projection of the current residual field $\boldsymbol{p}$ on the set of statically admissible fields (i.e. simply set $\left(\mathbf{div}(\underline{\sigma})^{I_0}\right)_{J_0} + (\boldsymbol{f}^{I_0})_{J_0} = 0$). In other words, at locations where the displacement is imposed, the stress divergence is systematically balanced by a reaction force to satisfy equilibrium. Note that the reaction forces $(\boldsymbol{f}^{I_0})_{J_0}$ can be evaluated as a post-processing, once convergence achieved.

Remark 1: The reaction forces $(\boldsymbol{f}^{I_0})_{J_0}$ correspond to the definition of the Lagrange multipliers commonly introduced to apply Dirichlet BC in FE direct solvers (see section 4.3). In these solvers, the Lagrange multipliers are added to the vector of nodal unknowns (in addition to the vector of nodal displacements), and the system to be solved is enlarged accordingly. In the present proposition, the same iterative solver as introduced in section 3 (i.e. an accelerated fixed-point algorithm), is used with minor modifications that simply consists of a projection of the displacement field ($\widetilde{\boldsymbol{u}}$, forcing displacements to fulfil Dirichlet conditions) and a projection of the residual field ($\boldsymbol{p}$).

Remark 2: The reaction forces $(\boldsymbol{f}^{I_0})_{J_0}$ should be called reaction force 'densities' as their physical dimension is [N/mm$^3$], the same as $\mathbf{div_D}(\underline{\sigma})$. Consequently, the reaction force (densities) $(\boldsymbol{f}^{I_0})_{J_0}$ must be rescaled to obtain the reaction forces $(\boldsymbol{r}^{I_0})_{J_0}$(dimension [N]). For a point located in the volume, it simply consists in multiplying the value by the volume of the voxel $dV$. For a point located at the surface (subjected to null-Neuman BCs through appropriate symmetries), the value is multiplied by $dV/2$. If the point is located at an edge the value is multiplied by $dV/4$ and by $dV/8$ if it is a corner.

The algorithm is detailed below with emphasis on the parts added to the classical fixed-point algorithm written in black (see Algorithm 1). In blue, the convergence accelerations parts are depicted. Note that the recent proposition [1] allows to accelerate whether the displacement field $\widetilde{\boldsymbol{u}}$, if $ACV_U$ is true, or the quantity $\boldsymbol{p} - \mathbf{div_D}\left(\mathbb{R}_0 : \underline{\nabla_D}(\widetilde{\boldsymbol{u}})\right)$, if $ACV_P$ is true. In practice, the function $ACV_{storeR}$ stores the residual fields (difference between two successive solutions, i.e between the end and the beginning of the iteration) while $ACV_{storeS}$ stores the corresponding solution fields (at the beginning of the iteration). The two corresponding tables have four indices and the fields are stored at each iteration using cycling indices. After the four first iterations and every $mod_{ACV}$ iterations (generally 2 or 3), an accelerated solution can be proposed from the four solution and residual fields stored in the tables (following an Anderson acceleration technique described in [19] in the context of FFT-based solvers).

1. Initial $\widetilde{\boldsymbol{u}} = \mathbf{0}$
2. $proj_U : \ (\widetilde{\boldsymbol{u}}^{I_0})_{J_0} = (\boldsymbol{U}^{I_0})_{J_0}$
3. Loop $i \geq 0$
4. $\quad$ if $(ACV_U)$
5. $\quad\quad$ if $(i > 0)$ $ACV_{storeR}(i, \widetilde{\boldsymbol{u}})$
6. $\quad\quad$ if $(i > 0 \ \& \ \text{modulo}(i, mod_{ACV}) = 0)$ $ACV_{accelerate} \rightarrow \widetilde{\boldsymbol{u}}$
7. $\quad\quad$ $ACV_{storeS}(i, \widetilde{\boldsymbol{u}})$
8. $\quad$ $\underline{\sigma} = \mathbb{C} : \underline{\nabla^{s}_{\mathrm{D}}}(\widetilde{\boldsymbol{u}})$
9. $\quad$ $\boldsymbol{p} = \mathbf{div}_{\mathbf{D}}(\underline{\sigma})$
10. $\quad$ $proj_P : \ (\boldsymbol{p}^{I_0})_{J_0} = 0$
11. $\quad$ if equilibrium$(\boldsymbol{p})$: EXIT Loop *(see equation (5))*
12. $\quad$ $\boldsymbol{p} = \boldsymbol{p} - \mathbf{div}_{\mathbf{D}}\left(\mathbb{R}_0 : \underline{\nabla_{\mathrm{D}}}(\widetilde{\boldsymbol{u}})\right)$
13. $\quad$ if $(ACV_P)$
14. $\quad\quad$ if $(i > 0)$ $ACV_{storeR}(i, \widetilde{\boldsymbol{p}})$
15. $\quad\quad$ if $(i > 0 \ \& \ \text{modulo}(i, mod_{ACV}) = 0)$ $ACV_{accelerate} \rightarrow \widetilde{\boldsymbol{p}}$
16. $\quad\quad$ $ACV_{storeS}(i, \widetilde{\boldsymbol{p}})$

1. $\quad$ $\widehat{\boldsymbol{p}} = DT(\boldsymbol{p})$
2. $\quad$ $\widehat{\widetilde{\boldsymbol{u}}} = \widehat{\mathbb{G}}_{\mathrm{D}}(\widehat{\boldsymbol{p}})$
3. $\quad$ $\widetilde{\boldsymbol{u}} = iDT(\widehat{\widetilde{\boldsymbol{u}}})$

17. $\quad$ $proj_U : \ (\widetilde{\boldsymbol{u}}^{I_0})_{J_0} = (\boldsymbol{U}^{I_0})_{J_0}$

*Algorithm 1 : Description of the FFT-based algorithm proposed to account for local kinematic constraints (Dirichlet constraints) defined at nodes. In black, the lines corresponding to the standard fixed-point algorithm, in blue, the convergence acceleration (whether on $\widetilde{\boldsymbol{u}}$ if $ACV_U$ is true, or on $\boldsymbol{p} - \mathbf{div}_{\mathbf{D}}\left(\mathbb{R}_0 : \underline{\nabla_D}(\widetilde{\boldsymbol{u}})\right)$ if $ACV_P$ is true) and, in red, the lines to be added to apply the constraints.*

To sum up, compared to the displacement-based algorithm recently proposed in [1], accounting for non-periodic BCs through DTTs, the introduction of local displacement constraints is very simple (see red lines in Algorithm 1). This algorithm is tested and validated below on a CT-like specimen associated with a loading case out of reach for standard FFT-based solvers (periodic) and even to non-periodic solvers based solely on DTTs.

Note that the criterion used for the convergence (see “equilibrium” function in line 11) is the same as the standard FFT-based solver introduced in equation (5). A typical tolerance of $10^{-4}$ is often used n practice.

4.1.2: Results on CT-like specimen

The CT-like specimen is both interesting for the mechanics of materials community and challenging for the FFT-based solvers community. Actually, in addition to non-periodic boundary conditions with Neumann and Dirichlet on a same face, together with Dirichlet on interior points, the proposed simulation has the originality of simulating an accurate crack (i.e. a zero thickness crack instead of a crack with a one voxel thickness). The description of the simulated unit-cell is proposed on Figure 2. This setup comes after different unfruitful attempts when trying to apply null-Neumann BCs on the 6 faces and constraints on local vertical displacement (on the red line and surface). Actually, the algorithm did not converge, probably due to an issue in dealing with rigid body motion that should remain free but that, in practice, is imposed through an average displacement applied in Fourier space (for null frequency vector). That condition comes in conflict with the Dirichlet BC imposed locally. We believe that this issue can be overcome but we have not yet found the solution and this question remains open. A first way to circumvent it is to apply mixed Dirichlet-Neumann BC using DTTs to account for CT-symmetry and represent the crack by a half line of void voxel. However, considering the symmetry, the crack has a thickness of two voxels. The chosen approach introduces an additional void layer but takes into account a zero-thickness crack that is quite a novelty for FFT-based solvers. Hence, it is proposed to use a void buffer layer (green on Figure 2) that allows us to apply Dirichlet condition (vertical) on face $S_3^0$ (yellow face on Figure 2), null-Neumann on the other faces, and local constraints on the vertical displacement on red line and surface. Boundary conditions on faces are applied using symmetry extensions (through DTTs) and local constraints are applied using the proposed strategy in Algorithm 1.

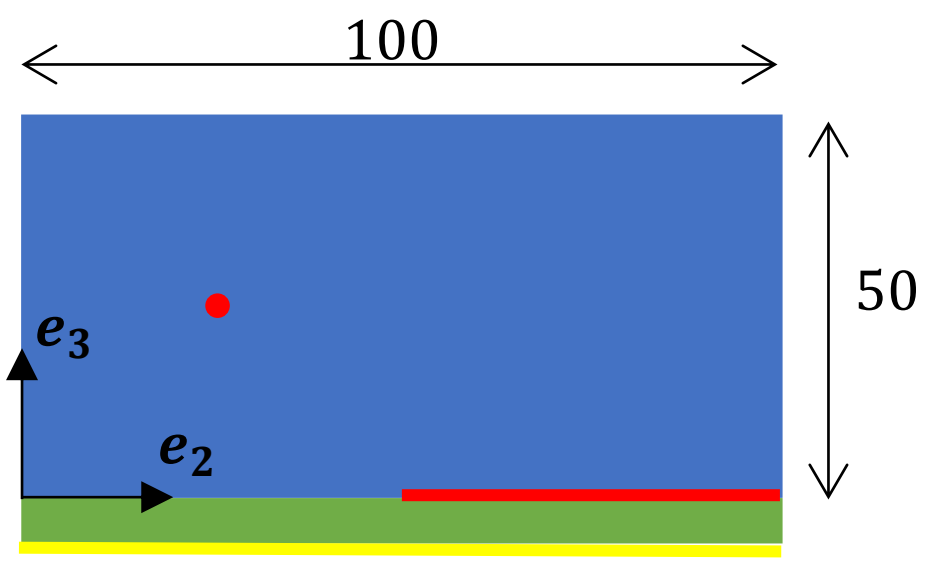

*Figure 2: Half CT-like specimen description: the blue part represents the half CT specimen, the green part a void buffer layer (1 voxel thickness), the red dot represents the projection along the direction $\boldsymbol{e}_1$ of the line (position (0:10,25,25)), which is subjected to a displacement of 10 in direction $\boldsymbol{e}_3$ (vertical), the red line the projection along the direction $\boldsymbol{e}_1$of the surface (position (0:10,50:100,0)), which is subjected to a null displacement in direction $\boldsymbol{e}_3$ (vertical) and the yellow surface (0:10,0:100,-1) is subjected to a null displacement in direction $\boldsymbol{e}_3$ (vertical).*

The used discretization is 10x100x(50+1) voxels with a single voxel of void layers (green). The void layer is considered as an elastic material with null properties and the CT-like specimen material is elastic isotropic with $(4.04 \times 10^4, 2.70 \times 10^4)$MPa for the Lamé coefficients $(\lambda, \mu)$. The discrete Green operator is defined with the $\mathbb{R}_0$ tensor proposed by Paux et al. [44] for general non-periodic BCs (see section 3.2.3) using $(\lambda_0, \mu_0)$=$(\lambda/2, \mu/2)$. All the results presented below have been obtained using the TETRA2 finite derivation scheme. Actually, the simulations with HEX1 scheme did not converge with strongly applied BCs. As mentioned in section 3.1.1, the use of the HEX1 finite difference scheme is equivalent to the use of linear finite element with reduced integration (with a single integration point). The drawback of this finite element is the possible development of hourglass deformation modes that produce zero strain at the integration point. The interpretation of the non-convergence with strongly applied BCs is that the loading applied to the line of nodes strongly excites an hourglass mode and the algorithm fails to converge. Alternatively, applying weakly BCs (see section 4.2) does not excite the hourglass mode and convergence is achieved, without spurious oscillations (see Figure 3). To circumvent the problem with this reduced integration FE, various hourglass control methods have been proposed, see for instance the work of Belytschko and co-workers 40 years ago [61], or recently, the method proposed by Schneider [62] in the context of FFT-based solvers that requires a displacement-based implementation compatible with the present solver.

Despite the expected complexity of the simulation for FFT-based solvers, it should be noted that convergence is achieved without any particular issue using convergence acceleration (it fails without). As observed in Table 4 the new proposition $ACV_P$ exhibits a better convergence and lower iteration numbers are obtained with $mod_{ACV} = 2$.

| | $mod_{ACV} = 2$ | $mod_{ACV} = 3$ |
|---|---|---|
| $ACV_U$ | 399 | 539 |
| $ACV_P$ | 198 | 321 |

*Table 4 : Number of iterations at convergence with different choices of convergence acceleration (convergence tolerance $10^{-4}$, using TETRA2 scheme). With $ACV_U$, acceleration is applied to fluctuation displacement fields, with $ACV_P$ acceleration is applied to the field $\boldsymbol{p} - \boldsymbol{div_D}\left(\mathbb{R}_0 : \underline{\nabla_D}(\tilde{\boldsymbol{u}})\right)$ (see Algorithm 1).*

A representation of the deformed configuration together with the stress field $\sigma_{33}$ in Figure 3 (– top), shows the ability of the FFT-based solver to reproduce this CT-like simulation. It must be emphasized that no spurious oscillations can be observed on the two representations (deformed shape and stress field). This observation, confirmed with higher resolutions (i.e voxel size divided by 2 and 4), tends to demonstrate the ability of the TETRA2 scheme to produce smooth results (note that the represented stress field is evaluated on one of the two integration points, it is not an average over the two IPs that could have reduce oscillations). The other results in Figure 3 (– middle and bottom) correspond to the loading applied "weakly" with the TETRA2 scheme (– middle) and the HEX1 scheme (– bottom), as proposed in the following section 4.2.

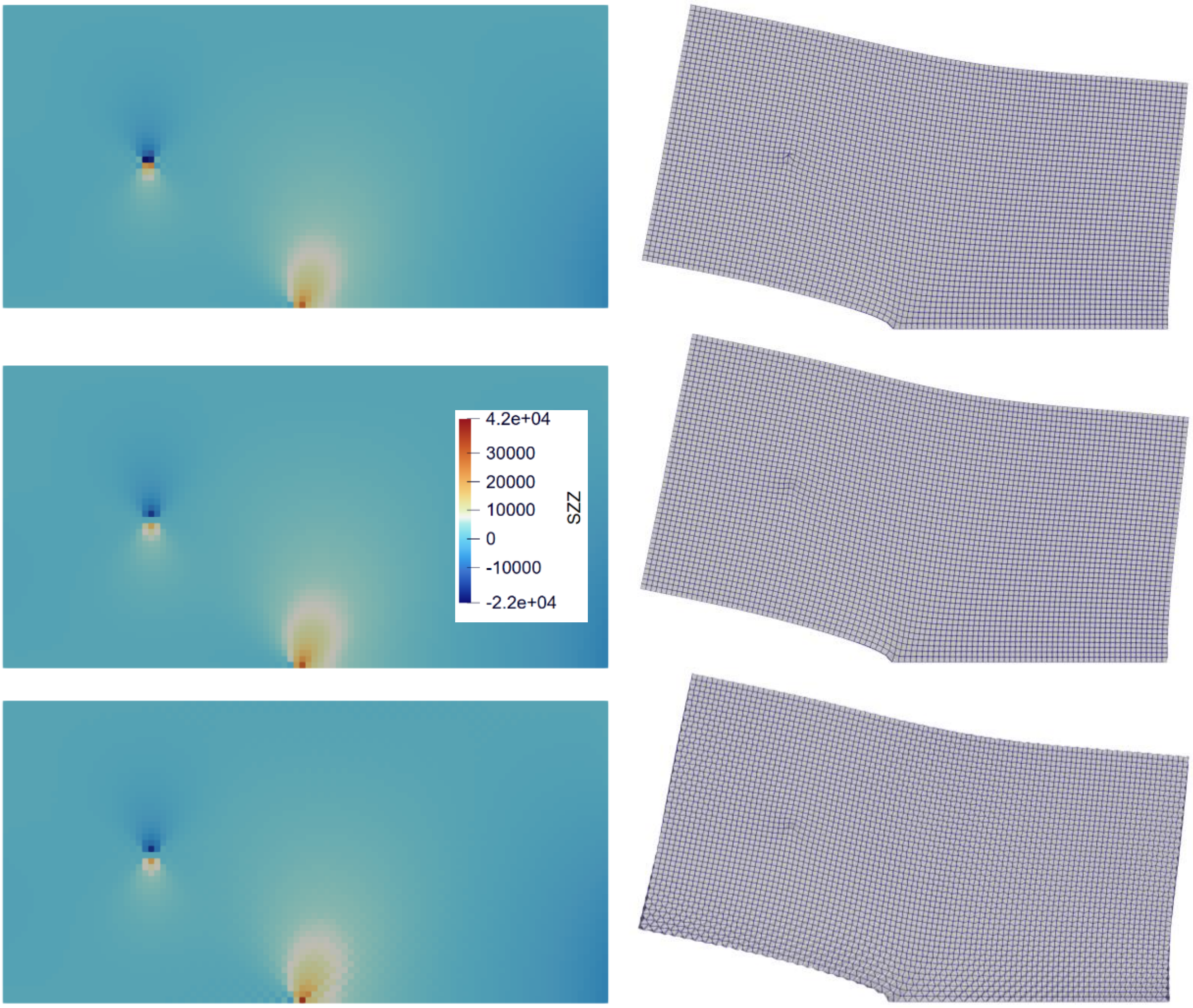

*Figure 3: Representation (buffer layer removed) of the deformed configuration and $\sigma_{33}$ stress field (on one of the two integration points) for three simulations: (top) TETRA2 elements with displacement strongly applied on the line of nodes (0:10,24,24), (middle) TETRA2 elements with displacement applied weakly to the line of elements (0.5:9.5,24.5,24.5), (bottom) HEX1 elements with displacement applied weakly to the line of elements (0.5:9.5,24.5,24.5). See section 4.2.1 for the meaning of "displacement applied weakly".*

4.1.3 : Validation and convergence sensitivity w.r.t. mesh refinement

The procedure for applying local Dirichlet BC is validated from a cross-comparison with the same simulation performed with the FE code CAST3M [20] that uses a direct solver. The two simulations have the same regular mesh, the same material parameters and the same loading as described in the previous section. The only difference is the type of element: linear hexahedral elements with full integration (8 integration points) for the FE simulation and TETRA2 elements for the present FFT-based simulation. For this comparison between a direct and an iterative solver, the tolerance of the iterative FFT-based solver (see equation (5)) is lowered to $10^{-12}$. The equilibrium criterion is reached after 576 iterations. The comparison is made between vertical stress field ($\sigma_{33}$), using the value interpolated at the center of the element for the FE simulation and the average of the two values for the TETRA2 element (i.e. one value for each tetrahedra of a same voxel). In addition, the resultant reaction forces $R_S$ and $R_L$ are respectively evaluated on the surface of nodes located at the crack tip subjected to null vertical displacement $\tilde{u}_3$, and on the line of nodes on which the opening displacement $\tilde{u}_3$ is prescribed. Interestingly, it is observed that the macroscopic equilibrium condition $R_S + R_L = 0$ is not met at the beginning of the iterative algorithm but is almost reached at convergence with a relative error of $5.10^{-12}$ (i.e. $(|R_S| - |R_L|)/|R_L|$ with $|R_S| = 8418.91935121\ N$ and $|R_L| = 8418.91935126\ N$). When comparing the norm of the reactions for the FE simulation ($|R_L^{FE}| = 8505.04\ N$) and the present FFT-based simulation ($|R_L^{FFT}| = 8418.92\ N$), the relative error, $(|R_L^{FFT}| - |R_L^{FE}|)/|R_L^{FE}|$, is $1.02\ 10^{-2}$ (1.02%). Note that dividing by 2 the dimension of elements (i.e. multiplying by 8 their total number), for both FE and FFT-based simulations, the relative error decreases down to 0.78% (with $|R_L^{FE}| = 8220.07\ N$ and $|R_L^{FFT}| = 8155.14\ N$). This quantitative comparison is extremely satisfying if we consider that the two simulations have been run with different types of elements and a moderate discretization. To further explore this comparison, stress fields on two different planes are displayed in Figure 4. This qualitative comparison between the FE simulation (upper image in Figure 4) and the FFT-based simulation (lower image in Figure 4) exhibits a very nice agreement. The most important differences are observed at the crack tip and around the loading line, but they remain limited. It can be noticed that the particular shape of the stress field at the crack type in the $(\mathrm{e}_1, \mathrm{e}_2)$ plane is also reproduced (the stress field is not invariant in the $e_1$ direction). As a conclusion, this cross-comparison validates the proposed method and its implementation.

As the convergence of iterative methods under grid refinement is an important question for solvers of saddle point systems (see section 4.3 for the link with saddle point problems), the number of iterations is given below for three increasingly refined meshes (the number of voxels is multiplied by 8 from one mesh to another, the definition of the first mesh being defined above). Note that the layer of voxels with null elastic properties remains a layer with a single voxel thickness for the three refinements. The numbers of iterations are given in Table 5 for a low ($10^{-6}$) and two very low ($3.2 \times 10^{-12}$ and $10^{-12}$) tolerances. Note that for the highest resolution, the criterion reaches $3.2 \times 10^{-12}$ after 1300 iterations but its evolution becomes very flat and oscillating after, so that the number of iterations to reach $10^{-12}$ is very high. Omitting the tolerance $10^{-12}$ and keeping the results of $10^{-6}$ and $3.2 \times 10^{-12}$, the number of iterations at convergence is approximately multiplied by 1.5 when the mesh size is divided by 2, corresponding to a moderate dependence in $d^{-0.6}$ (with $d$ the voxel size).

However, it must be kept in mind that such analysis is heavily case dependent and the convergence analysis reported for the beam bending test in section 4.2.4 exhibits a rather mesh-independent results (and even improved convergence when increasing refinement). Note also that the convergence analysis reported in Appendix A-2 for an indentation problem using a fixed active-set and equality constraints also exhibits a very low mesh sensitivity (a factor of 1.3 when increasing the resolution from 33 to 129, i.e. multiplying the number of voxels by ~60). These results on mesh sensitivity should be considered in the context of the extensive literature on numerical methods for saddle point systems, and in particular the review by Benzi et al. [2]. The convergence of iterative methods for saddle point problems may be compromised as the mesh is refined when the spectral properties of the discretized system or of its preconditioned form deteriorate with decreasing mesh size, resulting in an increasing number of iterations. Appropriate preconditioning techniques can mitigate this mesh dependence and, in particular, constraint preconditioners have been proposed that can lead to mesh-independent convergence rates under suitable assumptions. However, as discussed in Section 4.4, the proposed method should be interpreted as a null-space method rather than as a constraint preconditioning approach in the sense of Benzi et al. [2].

Finally, it can be noticed that the buffer zone method proposed by the first author to apply Dirichlet BC on the whole surface of the domain, with a full-periodic FFT-based solver, was also exhibiting a degradation of convergence with mesh refinement as soon as the size of the buffer zone was kept constant [29].

| $h$ \tolerance | $10^{-6}$ | $3.2\ 10^{-12}$ | $10^{-12}$ |
|---|---|---|---|
| 1 | 283 | 558 | 576 |
| 0.5 | 465 | 832 | 838 |
| 0.25 | 665 | 1300 | 4436 |

*Table 5 : Number of iterations at convergence, for two different tolerances and 3 increasing mesh refinement ($h$ represents the voxel size, so that from 1 to 0.5, or 0.5 to 0.25, the number of voxels is multiplied by 8)*

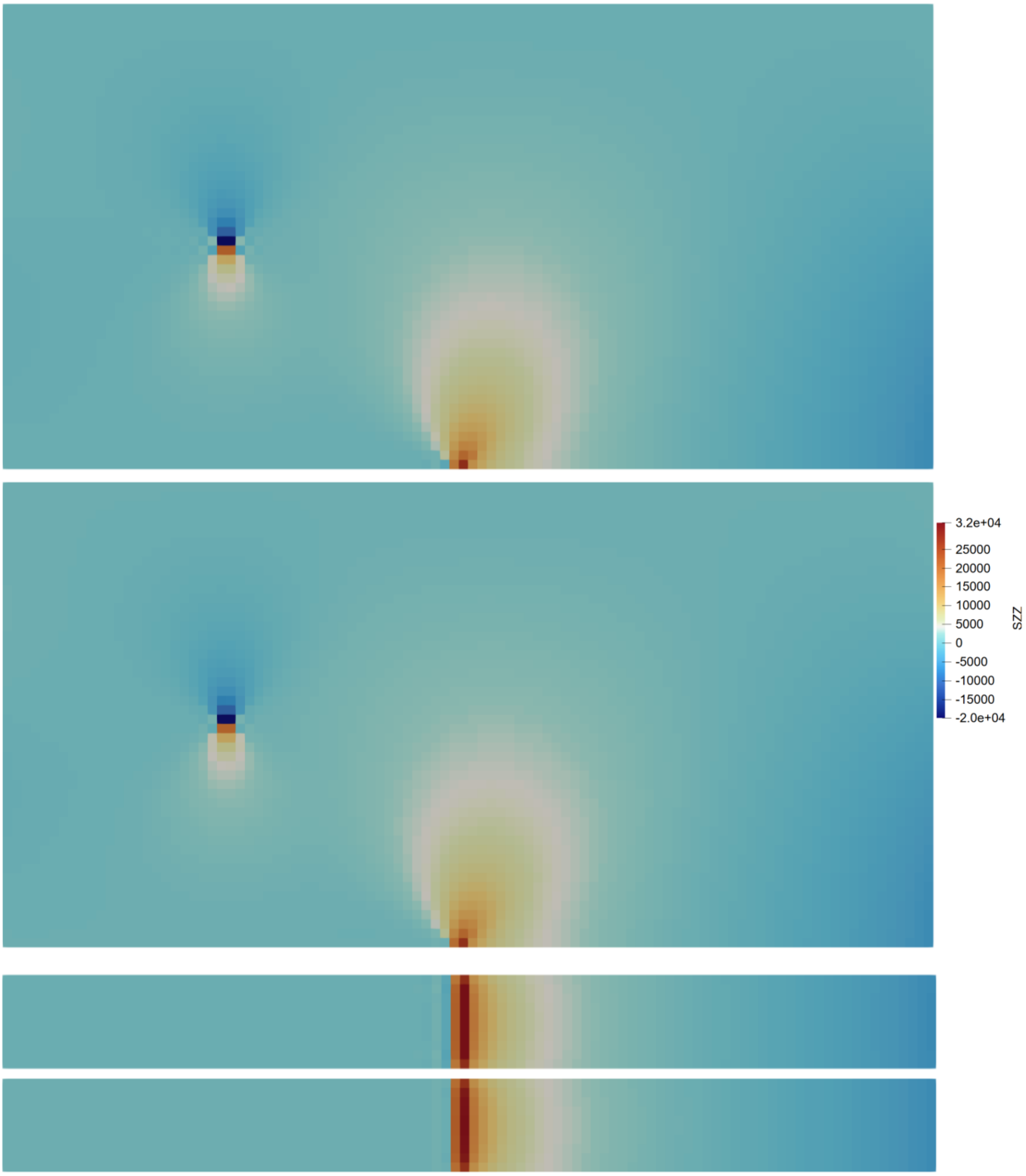

*Figure 4: Comparison of vertical stress fields ($\sigma_{33}$) in planes ($e_2, e_3$) and ($e_1, e_2$). For each plane the upper field is the FFT-based simulation (using TETRA2 elements) and the lower is the FE simulation (using CAST3M code and linear hexahedral element with full integration).*

### 4.2 : Alternatives : Weakly applied conditions or distributed reaction forces

When strongly applying kinematic constraints to isolated nodes, or lines of nodes, non-convergence issue for HEX1 scheme (see previous section), or very important spurious oscillations for TETRA2 scheme (see Figure 5), are reported. Compared to HEX1, TETRA2 scheme exhibit very smooth stress fields and less convergence issue. This section focuses on TETRA2 scheme. To overcome the oscillation issue with TETRA2 scheme, two alternative methods are proposed:

1. Weakly applied conditions: the displacement is applied to the center of a cube or a square (whether the condition is applied in the volume or at the surface) using an average displacement and reaction forces defined on the same support (8 nodes of the cube or 4 nodes of the square),
2. Distributed reaction forces: the displacement is applied to a given node, but reaction forces are smoothly distributed on a larger support (27 nodes or 9 nodes whether the condition is applied in the volume or at the surface).

In both cases and consistently with the previous subsection, the modification in Algorithm 1 is limited to the two functions $proj_U$ and $proj_P$.

Below, the descriptions of the methods, as well as the application, assume a constraint on a single point and a single component. Note that the extension to multiple components and multiple points is straightforward as soon as the larger supports used in these propositions do not overlap. The question of overlapping supports remains a future prospect.

4.2.1 : Weakly applied conditions

Considering a condition in the volume, instead of applying a displacement to a single node, the displacement is now applied to a voxel center. As the grid of voxel centers are not the support of the discrete displacement field, an interpolation is used from nodes to center, so that instead of the strong condition $\tilde{u}_{j_0}^{i_0} = U_{j_0}^{i_0}$ applied to node $i_0$ for component $j_0$, the condition applied to the center of element $i_0$ reads:

$$\frac{1}{8} \sum_{i \in V(i_0)} \tilde{u}_{j_0}^{i} = U_{j_0}^{i_0} \qquad \text{with } V(i_0) = 8 \text{ nodes of element } i_0 \tag{10}$$

This condition is a linear relationship between displacement of different nodes. To introduce this condition in Algorithm 1, the current displacement field is corrected in $proj_U$ with:

$$proj_U\colon \tilde{u}_{j_0}^{i} = \tilde{u}_{j_0}^{i} + U_{j_0}^{i_0} - \frac{1}{8} \sum_{k \in V(i_0)} \tilde{u}_{j_0}^{k} \quad \text{for } i \in V(i_0) \tag{11}$$

When considering the static condition applied strongly, the associated equation consists in forcing the equilibrium with $\operatorname{div_D}(\underline{\sigma})_{j_0}^{i_0} + f_{j_0}^{i_0} = 0$, so that in the algorithm is simply $p_{j_0}^{i_0} = 0$ in $proj_P$ (see section 4.1.1). For the weak condition here, the reaction force $f_{j_0}^{i_0}$ is now distributed on the 8 nodes of the element so that the equilibrium condition on these nodes reads:

$$\operatorname{div_D}(\underline{\sigma})_{j_0}^{i} + \frac{1}{8} f_{j_0}^{i_0} = 0 \quad \text{for } i \in V(i_0) \tag{12}$$

And, consequently:

$$\mathrm{div}_{\mathrm{D}}\left(\underline{\sigma}\right)_{j_0}^{i} - \frac{1}{8}\sum_{i\in V(i_0)} \mathrm{div}_{\mathrm{D}}\left(\underline{\sigma}\right)_{j_0}^{i} = 0 \quad \text{for } i \in V(i_0) \tag{13}$$

To introduce this static condition in Algorithm 1, the current field $\boldsymbol{p}$ must be corrected in $proj_P$ with:

$$proj_P: p_{j_0}^{i} = p_{j_0}^{i} - \frac{1}{8}\sum_{k\in V(i_0)} p_{j_0}^{k} \quad \text{for } i \in V(i_0) \tag{14}$$

The present condition applied in volume to voxel centers, can be straightforwardly transposed to a condition applied at the surface to the center of a voxel face located at the boundary. The only difference is that $V(i_0)$ consists of four nodes (i.e. the corners of the voxel face) instead of eight nodes (i.e. the corners of the voxel).

4.2.2 : Distributed reaction forces

Another proposition to avoid the oscillations using TETRA2 scheme with kinematic conditions on isolated nodes, consists in smoothly redistributing the reaction forces on neighboring nodes.

Considering a kinematic condition applied to a node $i_0$, located in the volume, for component $j_0$, Algorithm 1 is applied with the same correction as in section 4.1 for a strongly applied condition:

$$proj_U: \tilde{u}_{j_0}^{i_0} = U_{j_0}^{i_0} \tag{15}$$

The difference with a strongly applied condition (section 4.1) is that the reaction force located at node $i_0$ is now smoothed over the 26 neighboring nodes, i.e. the corners of the 2x2x2 elements cell centered on node $i_0$, according to different weights $w_{ME}$, $w_{MF}$ and $w_{Co}$, where the subscript $ME$ stands for nodes located at the middle of edges (12 points), $MF$ for nodes located at the middle of faces (6 points), and $Co$ for corners (8 points). To account for reaction redistribution, the reaction $f_{j_0}^{i_0}$ evaluated at point $i_0$ (first line below) is complemented by additional forces located on the 26 neighboring nodes. Then, equilibrium equations around node $i_0$ are modified as follows:

$$\begin{aligned} &\mathrm{div}_{\mathrm{D}}\left(\underline{\sigma}\right)_{j_0}^{i_0} + f_{j_0}^{i_0} = 0 \\ &\mathrm{div}_{\mathrm{D}}\left(\underline{\sigma}\right)_{j_0}^{i} + w_i f_{j_0}^{i_0} = 0 \qquad \text{for } i \in V'(i_0) = 26 \text{ nodes neighbors of } i_0 \end{aligned} \tag{16}$$

with $w_i \in \{w_{ME}, w_{MF}$ and $w_{Co}\}$.

It should be noted that, in spite of considering 26 additional nodes, no additional unknown has been added to $f_{j_0}^{i_0}$ which remains the only unknown associated to the prescribed condition on $\tilde{u}_{j_0}^{i_0}$. The introduction of the above equilibrium equations is done in Algorithm 1 through a correction of the current field $\boldsymbol{p}$ applied in $proj_P$ with:

$$proj_P: f_{j_0}^{i_0} = -p_{j_0}^{i_0}$$
$$p_{j_0}^{i_0} = 0 \qquad (= p_{j_0}^{i_0} + f_{j_0}^{i_0}) \tag{17}$$
$$p_{j_0}^{i} = p_{j_0}^{i} + w_i f_{j_0}^{i_0} \qquad \text{for } i \in V'(i_0) = 26 \text{ nodes neighbors of } i_0$$

The method has been introduced for a condition applied to a node located in the volume, its modification towards a node located at the surface is straightforward: instead of considering a neighborhood of 26 nodes, only the 8 neighbors located at the surface are considered. Finally, the definition of the weights is given in table below for the two cases. The total reaction force associated with the kinematic condition is then equal to $\left(1+\frac{12}{4}+\frac{6}{2}+\frac{8}{8}\right) f_{j_0}^{i_0} = 8 f_{j_0}^{i_0}$ and $\left(1+\frac{4}{2}+\frac{4}{4}\right) f_{j_0}^{i_0} = 4 f_{j_0}^{i_0}$, respectively for volume and surface condition.

| | Volume | Surface |
|---|---|---|
| $w_{ME}$ | $1/4$ | $1/2$ |
| $w_{MF}$ | $1/2$ | × |
| $w_{Co}$ | $1/8$ | $1/4$ |

*Table 6 : Definition of the weights used to smooth the reaction force for the cases of a node located in the volume (redistribution over 27 nodes) or at the surface (redistribution over 9 nodes). Co, ME, MF, stand for nodes located at Corner, Middle Edge and Middle Face of a 2x2x2 volume elements cell (or a 2x2 surface elements cell).*

The loading points located at edges and corners have not been considered in the paper. A possible treatment of an edge point is to consider it as a surface point, with 8 points located in its neighborhood and the same weights as in Table 6. The only difference would be that for a surface point, the 8 points belong to the same surface, whereas for an edge point, 2 points belong to the edge and 6 points to the two connected faces (3 per face). For a corner, the same strategy could be used but considering the six closest neighboring points of the corner: 3 on the connected edges and 3 on the connected faces. A similar definition of the weights would give $w_E = 2/3$ and $w_F = 1/3$.

Finally, note that the present proposition assumes a redistribution on a local neighborhood (26 or 8 nodes depending on whether the loading point is located in the volume or at the surface). As a future prospect, the redistribution could also be defined on a larger neighborhood with a given physical size (so that the number of nodes in the neighborhood would increase with mesh resolution) and a given redistribution profile (for example the Hertz profile).

#### 4.2.3 : Results on a CT-like specimen

First, weakly applied conditions have been used (see Figure 3 - bottom) to overcome the convergence issue observed using HEX1 scheme together with strong BCs applied to the complete line of nodes located at (0:10,25,25). The condition is weakly applied to the line of elements with centers located at (0.5:9.5,24.5,24.5) using the same equations (*10*) to (*14*), just replacing $V(i_0)$ the set of nodes of element $i_0$ by $V^*$ the set of 44 nodes belonging to the concerned elements. Convergence is achieved with HEX1 scheme and, despite small oscillations, stress field compares nicely to the results obtained with the TETRA2 scheme. However, these spurious oscillations, associated with hourglass phenomenon, are more visible on the displacement field as observed on the deformed configuration.

Secondly, to demonstrate the issue arising with TETRA2 scheme when a displacement is applied to a single node, the same configuration as shown in Figure 2 is considered. The only difference is that the load is applied to a single node located at (2,24,24) rather than to the entire line of nodes (0:10,24,24). As observed on Figure 5-right, despite the convergence of the algorithm, the solution is massively affected by artifacts observed on the displacement field. At first sight, this result is quite surprising.

However, the explanation can be found in the analysis made by Finel [28] when introducing TETRA2 scheme. Actually, if the simplified description proposed in [56] is of practical interest, the original description [28] demonstrates that the problem solved on the complete grid consists of two decoupled problems, defined on two different geometric supports (i.e. the two subgrids of tetrahedra reported on Figure 6), solved simultaneously. As a consequence, for this type of FD scheme, applying the loading to a single node will let one of the two problems unloaded. Indeed, when looking at Figure 5-right part, it can be noticed that half of the nodes remains on the undeformed configuration and the other part has a displacement consistent with Figure 5-left part obtained with weakly applied condition to an element centered at (2.5,24.5,24.5).

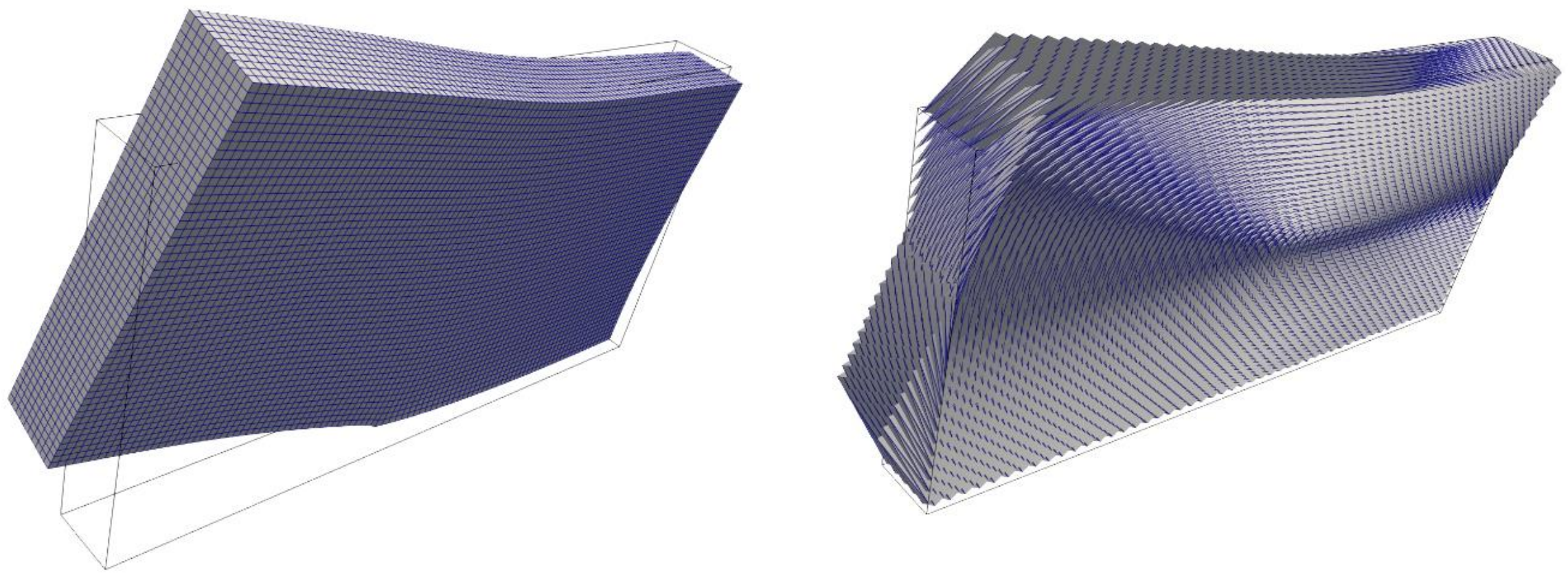

*Figure 5: Representation of the deformed configuration of a CT-like specimen subjected to a vertical load on a single point that do not belong to the plane of symmetry so that the specimen opens and twists as expected. On the right, the displacement is applied strongly to a single node (position (2,24,24)) exhibiting important artifacts of the displacement field associated with the TETRA2 elements. On the left, the displacement is applied weakly to an element center (position (2.5,24.5,24.5)) with a complete disappearance of artifacts.*

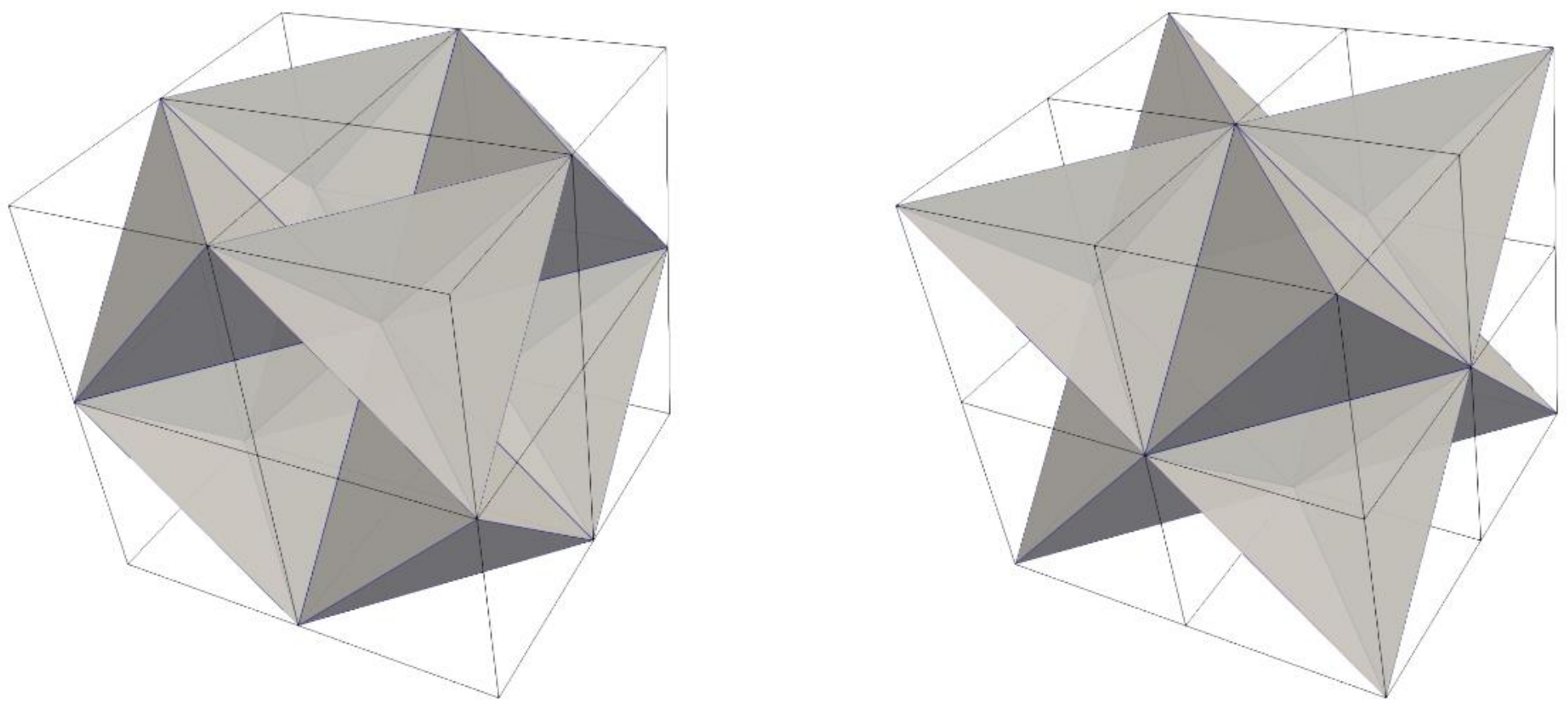

*Figure 6 : Representation of the two subgrids of tetrahedra, supports of the two decoupled problems simultaneously solved when considering TETRA2 elements (see [28] for a deeper explanation).*

Hence, a simulation with distributed reaction forces and the prescribed displacement applied at node (2,24,24) has been performed. The result is very similar w.r.t. the simulation using weakly applied BCs at the center of the voxel at (2.5,24.5,24.5). A slight discrepancy is observed due to the minor difference of the location where the displacement is applied (a half-voxel difference). Because of the strong similarity with Figure 5-left part, this additional result is not reported in a figure. More interestingly, it should be noted that the number of iterations at convergence is the same (332) for both simulations.

Finally, for the TETRA2 scheme, we have not been able to find convenient constraints on the support of strongly applied Dirichlet BC that would guarantee the absence of oscillations on the displacement field. Unfortunately, after many attempts and practical tests, this question remains open. However, we have proposed two different ways to overcome such an issue arising when the loading support is a single node and we believe that they can be adapted to other situations if necessary.

#### 4.2.4 : Comparison between the strong Dirichlet BC (on a single node) and the two alternatives

We have proposed two different ways to apply local Dirichlet BCs compatible with the TETRA2 scheme for which applying 'strongly' a displacement to a single node is not possible. Applying a displacement to a single node raises a question regarding mesh sensitivity, however it is quite common in FE simulations and we would like to address this question with the present solver. Actually, applying a displacement to a single point generates a singularity around that point so that the numerical solution becomes mesh-dependent when refining the mesh around that point. However, in practical structural analysis simulations, Saint-Venant principle is advocated, the detailed solution around the loading point is not required and the mesh remains coarse around that point. Conversely, if the mesh refinement is uniform, as in the present FFT-based method, a mesh dependence may appear. The purpose is then to quantify this dependence and compare, in terms of reaction forces, strongly applied Dirichlet BC on a single node (using a FE code as it is not possible with the TETRA2 FD scheme) with the two alternatives. For the FFT-based simulations, the convergence rate is also explored as a function of the refinement.

In order to keep exactly the same geometry and the same position of the loading point for various mesh refinements, a simple bending test is proposed. The dimensions of the beam are 20x1x1, null-Dirichlet BC are applied to the left-hand side and the loading point is located in the volume at position (19,0.5,0.5). For the strong Dirichlet BC, as mentioned before, the simulations have been performed with the FE code CAST3M using hexahedral elements with quadratic interpolation (i.e. with 20 nodes and 27 Gauss points) and also with linear interpolation (i.e. 8 nodes and 8 Gauss points). The regular mesh is defined by $20n_y \times n_y \times n_y$ cubic elements, $n_y$ being an even number so that the loading point corresponds to a node. For the distributed reaction force approach, the mesh definition is the same. For weakly applied BC, the dimensions of the mesh ($n_x \times n_y \times n_y$ and $d_x \times d_y \times d_y$ respectively for the number of voxels and voxel size) have to be adjusted so that the loading point corresponds exactly to an element center. For a given number of voxels in the beam thickness $n_y$, the voxel size is $d_y = 1/n_y$, the number of voxels in the axial direction is $n_x = n_y + 1 + (19(2n_y + 1) - 1)/2$, and $d_x = 1/n_x$. Hence, the mesh consists of slightly non-cubic voxels, taken into account in the present implementation of the FFT-based solver. For FE simulations, the code CAST3M is not parallel in distributed memory and the simulations were limited to resolutions $n_y$ = 8, 16, 32, 64 and 96 for linear elements and $n_y$ = 4, 8, 16, 32 and 48 for quadratic elements. For FFT-based simulations, taking advantage of a massively parallel implementation in the code AMITEX* [1] (currently in progress), a wider range of resolution has been investigated with $n_y$ = 8, 16, 32, 64, 128, 256 and 512. The highest resolution was $n_y$=512 leading to a problem with $20 \times 512^3 \times 3 = 8.10^9$ dof, solved in 2 hours on 1024 cores (distributed on 9 nodes of 128 cores). Finally, considering a 1D beam approximation (Euler-Bernoulli), with a length of 19 and a section of 1x1, the reaction force $R_{1D}$ evaluated analytically is used as a common reference value to compare the different numerical variants. The reaction forces for the three investigated approaches are denoted $R_S$, $R_D$ and $R_W$, with subscripts corresponding respectively to "strongly", "distributed", and "weakly". The relative difference w.r.t. the 1D common reference $(R_X - R_{1D})/R_{1D}$, $X \in \{S, D, W\}$, is plotted in Figure 7 as function of the number of d.o.f. for the various simulations. Note that this relative difference is not a relative error as the 1D solution is an approximate solution. However, regarding the beam aspect ratio this reference can be considered as the target value and this relative difference will allow to quantify the detrimental effect of excessively refining the mesh.

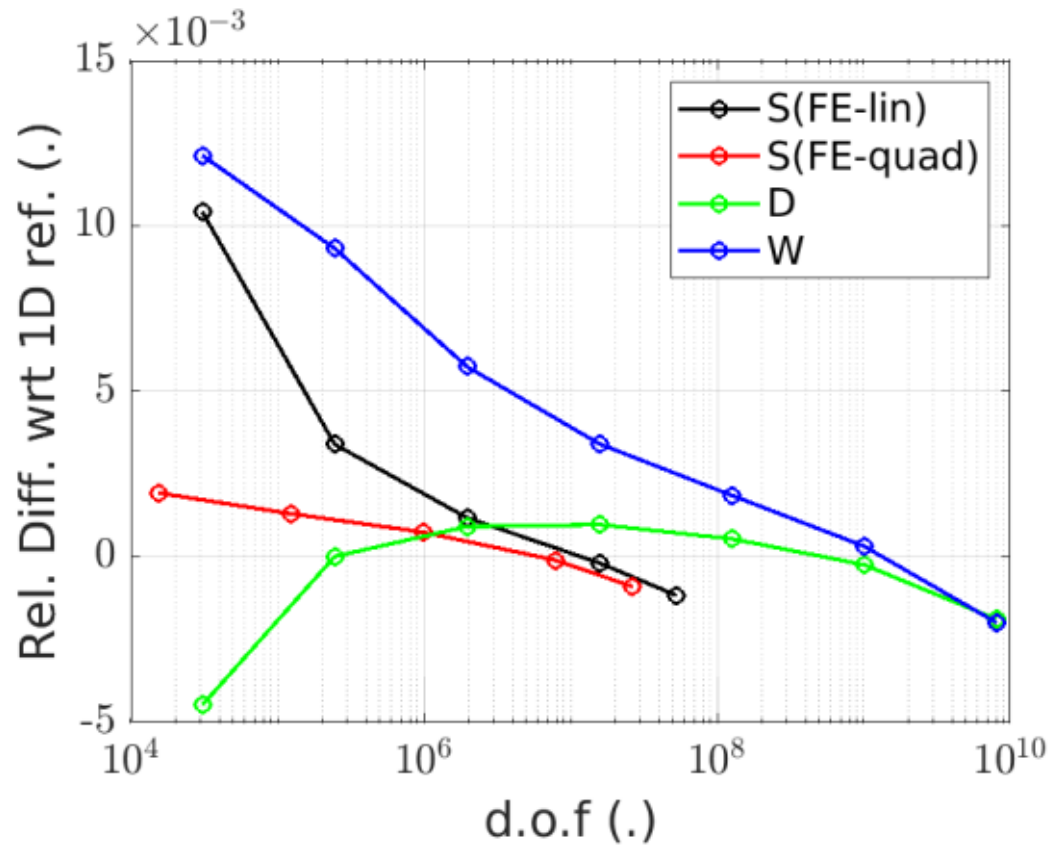

*Figure 7: Relative difference of the reaction force w.r.t. the 1D approximation for the different variants (S: strongly applied Dirichlet BC, with linear or quadratic elements, D: distributed reaction forces, W: weakly applied Dirichlet BC, D and W simulations are performed with the present FFT-based solver and TETRA2 FD scheme).*

Considering FE simulations with quadratic elements (red curve), using strongly applied BC, the relative difference remains in the range [-0.1%, 0.2%] for all the resolutions. As expected from the singularity around the loading point, it can be observed that the curve does not converge when increasing the refinement (and a similar observation is made with linear elements, black curve). However, in the present case of beam bending with a high aspect ratio, the effect of this singularity remains acceptable for the investigated resolutions. Considering the FFT-based solver with the "distributed reaction forces" variant (green curve), omitting the lowest resolution, all the values remain in the range [-0.2%,0.1%] even for extremely high resolutions. The difference between the "weakly applied BC" variant and the "distributed reaction forces" variant is the most pronounced for the lowest resolution (1.2% vs -0.4%) and decreases when increasing the resolution. From this numerical experiment, it seems that the "weakly applied" variant provides a stiffer behavior than the "distributed reaction forces" variant (both being simulated with the same FD scheme), which should probably be preferred when considering the quite flat behavior of the curve with variation between -0.02% to 0.1% in the wide range of resolutions $n_y$ =16 to 256.

As a future prospect, the "distributed reaction forces" variant could also be considered to take into account a domain of physical size and redistribute the forces on an increasing number of neighboring nodes when increasing the spatial resolution ($n_y$). In that case, the singularity is removed and the reaction force is expected to converge with the spatial resolution. For a condition applied at the surface, a Hertz profile could be considered for the redistribution.

Finally, the number of iterations at convergence for a tolerance of $10^{-6}$ is reported in Table 7 for the two approaches and the different mesh refinements. It is quite interesting to observe that convergence is improved for higher resolutions. For the weakly applied BC approach, a gap occurs between 32 and

64. It must be emphasized that the convergence of iterative solvers for saddle point systems (see section 4.3) is generally not expected to improve when increasing the resolution [2]. Hence, the interpretation of these observations remains an open question.

| variant \ $n_y$ | 8 | 16 | 32 | 64 | 128 | 256 | 512 |
|---|---|---|---|---|---|---|---|
| W | 1164 | 1048 | 1017 | 564 | 480 | 459 | 430 |
| D | 634 | 598 | 524 | 518 | 527 | 446 | 385 |

*Table 7 : Number of iterations at convergence for a tolerance of $10^{-6}$. The variant W stands for weakly applied BC and D for distributed reaction forces approach.*

4.3 : Discussion : minimization under equality constraint

In the previous section, the resolution of the discrete problem was proposed from the perspective of local equations. The purpose of the section is to look at the same problems from the perspective of constrained minimization problems and associated linear systems. This approach is well known in the FE community and can be found in different handbooks (see for example section 4.6 devoted to Lagrange multipliers in the book of Yastrebov [63], also used in the next section to deal with contact mechanics).

For strongly applied conditions (see section 4.1), the set of local equations (*8*) for the discrete problem can be re-formulated as a minimization problem, subject to equality constraints as follows (note that here, and throughout this section, the unknown vectors belong to finite-dimensional real spaces equipped with the standard Euclidean inner product):

$$\inf_{[\tilde{u}]\in\mathbb{R}^n}(J([\tilde{u}])) = \inf_{[\tilde{u}]\in\mathbb{R}^n}\left(\frac{1}{2}[\tilde{u}]^T[\![K]\!][\tilde{u}]\right) \quad \text{with } [\tilde{u}] \text{ subject to } \left[\tilde{u}_{J_0}^{I_0}\right] = \left[U_{J_0}^{I_0}\right] \tag{18}$$

Vector $[\tilde{u}] \in \mathbb{R}^n$ gathers the unknown displacements vector (3 components per node) at all the nodes that are not imposed to zero at boundary through Anti-Symmetry extension (i.e. Dirichlet condition imposed through the use of DTT). Vector $[U_{J_0}^{I_0}] \in \mathbb{R}^m$ gathers the values of the imposed displacements to $[\tilde{u}_{J_0}^{I_0}] \in \mathbb{R}^m$, a subset of $[\tilde{u}]$ (see section 4.1 for the definition of $I_0$ and $J_0$, gathering respectively the node location and the components submitted to local Dirichlet conditions). $m$ is the number of constraints. Finally, matrix $[\![K]\!] \in \mathbb{R}^{n\times n}$ is built from the local discrete operator $\mathbf{div_D}\left(\mathbb{C}:\left(\underline{\nabla_\mathrm{D}^\mathrm{S}}(\tilde{\boldsymbol{u}})\right)\right)$ so that, for example, the equilibrium condition $\mathbf{div_D}\left(\mathbb{C}:\left(\underline{\nabla_\mathrm{D}^\mathrm{S}}(\tilde{\boldsymbol{u}})\right)\right) = \mathbf{0}$ for any grid point is equivalent to $[\![K]\!][\tilde{u}] = [0]$. Then, the constrained problem (*18*) can be rewritten using its Lagrangian formulation leading to problem (*19*), corresponding to the definition of a saddle point.

$$\sup_{\left[f_{J_0}^{I_0}\right]\in\mathbb{R}^m}\ \inf_{[\tilde{u}]\in\mathbb{R}^n}\left(J([\tilde{u}]) + \left[f_{J_0}^{I_0}\right]^T\left(\left[\tilde{u}_{J_0}^{I_0}\right] - \left[U_{J_0}^{I_0}\right]\right)\right) \tag{19}$$

In this formulation, $[\tilde{u}]$ is no longer constrained and a vector of Lagrange multipliers $[f_{J_0}^{I_0}] \in \mathbb{R}^m$ (interpreted in section 4.1 as reaction forces) is added to $[\tilde{u}]$ as unknown vector. Note that there is one Lagrange multiplier per relation so that the size of $\left[f_{J_0}^{I_0}\right]$ is equal to the size of $\left[\tilde{u}_{J_0}^{I_0}\right]$. Finally, the first-order optimality conditions (derivation w.r.t. to $[\tilde{u}]$ and $[f_{J_0}^{I_0}]$) allows us to build the following linear system:

$$\begin{bmatrix} [\![K]\!] & [\![A_{J_0}^{I_0}]\!]^T \\ [\![A_{J_0}^{I_0}]\!] & [\![0]\!] \end{bmatrix} \begin{bmatrix} [\tilde{u}] \\ [f_{J_0}^{I_0}] \end{bmatrix} = \begin{bmatrix} [0] \\ [U_{J_0}^{I_0}] \end{bmatrix} \tag{20}$$

The matrix $[\![A_{J_0}^{I_0}]\!] \in \mathbb{R}^{m\times n}$ has the dimension (size of $[f_{J_0}^{I_0}]$, size of $[\tilde{u}]$ ). It ensures the correspondence between the indices of $[\tilde{u}]$ subject to constraints and the corresponding values imposed in vector $[U_{J_0}^{I_0}]$. On the other hand, $[\![A_{J_0}^{I_0}]\!]^T$ ensure the correspondence between the reaction forces $[f_{J_0}^{I_0}]$ and the indices of $[\tilde{u}]$ subject to constraints. In that case, the matrix $[\![A_{J_0}^{I_0}]\!]$ has a rather simple form with zeros almost everywhere and a few ones to ensure correspondence. For example, for a single constraint on node $i_0$ and component $j_0$, matrix $[\![A_{j_0}^{i_0}]\!] \in \mathbb{R}^{1\times n}$ has a single line and a single "one" value at the position of the unknown $\tilde{u}_{j_0}^{i_0}$ in $[\tilde{u}]$ as shown below.

$$[\![A_{j_0}^{i_0}]\!] = [0 \dots 010 \dots 0] \tag{21}$$

Adding another constraint of the same type is simply done by adding a new line, of the same type, to build matrix $[\![A]\!]$.

To summarize, the linear system (equation (*20*)), which is equivalent to the set of local equations (*8*), provides the solution of a constrained minimization problem, by solving a saddle point problem while introducing Lagrange multiplier. The method is very common and standard in FE codes. The present paper demonstrates the ability of the proposed FFT-based solver to solve this type of problems with simple equality constraints such as the strongly applied conditions (i.e. $(\tilde{\boldsymbol{u}}^{I_0})_{J_0} = (\boldsymbol{U}^{I_0})_{J_0}$).

For weakly applied conditions, compared to the previous case, equation (*10*) now imposes linear relations between kinematic unknowns. The same approach as before can be used, associating a Lagrange multiplier to any relation. Matrix $[\![A_{J_0}^{I_0}]\!]$ now stores the linear relations corresponding to equation (*10*) and, for a single constraint on element $i_0$ and component $j_0$, it has the form:

$$[\![A_{j_0}^{i_0}]\!] = [0..\frac{1}{8}0..0\frac{1}{8}0..0\frac{1}{8}0..0\frac{1}{8}0..0\frac{1}{8}0..0\frac{1}{8}0..0\frac{1}{8}0..0\frac{1}{8}..0] \tag{22}$$

The eight values $\frac{1}{8}$ are located at positions of $\tilde{u}_{j_0}^{i}$ in $[\tilde{u}]$ (for the eight nodes $i$, corners of element $i_0$). Regarding equation (*20*), matrix $[\![A_{j_0}^{i_0}]\!]^T$ will dispatch the reaction force $f_{j_0}^{i_0}$ on the eight nodes $i$, corners of element $i_0$, as done in the local equations (*12*). Therefore, the present FFT-based solver demonstrates its ability to solve minimization problems with linear equality constraints. The application in section 4.2.1 is an example of this possibility, but the same approach can be followed to apply any other kinematic constraints.

For distributed reaction forces (see section 4.2.2), deriving the linear system associated with the problem is less straightforward. A single node $i_0$ and a single component $j_0$ are considered for simplicity. The set of nodes surrounding $i_0$ for the redistribution is noted $V_0$ (i.e. 26 nodes if the node $i_0$ is located in the volume). First, let us consider a problem with a strong Dirichlet condition on node $i_0$, component $j_0$, and known forces on $V_0$ nodes, denoted $\left[h_{j_0}^{V_0}\right] \in \mathbb{R}^{26}$. This problem reads:

$$\inf_{[\tilde{u}]\in\mathbb{R}^n} \left(\frac{1}{2}[\tilde{u}]^T[\![K]\!][\tilde{u}] - \left[h_{j_0}^{V_0}\right]^T \left[\tilde{u}_{j_0}^{V_0}\right]\right) \quad \text{with } [\tilde{u}] \text{ subject to } \left[\tilde{u}_{j_0}^{i_0}\right] = \left[U_{j_0}^{i_0}\right] \tag{23}$$

$\left[\tilde{u}_{j_0}^{V_0}\right] \in \mathbb{R}^{26}$, a subset of $[\tilde{u}]$, gathers the 26 displacements (component $j_0$) in the set of nodes $V_0$.

After introducing the Lagrange multiplier $\left[f_{j_0}^{i_0}\right] \in \mathbb{R}^1$, for the single constraint on node $i_0$, component $j_0$, the problem reads:

$$\sup_{\left[f_{j_0}^{i_0}\right]\in\mathbb{R}^1} \inf_{[\tilde{u}]\in\mathbb{R}^n} \left(J([\tilde{u}]) - \left[h_{j_0}^{V_0}\right]^T \left[\tilde{u}_{j_0}^{V_0}\right] + \left[f_{j_0}^{i_0}\right]^T \left(\left[\tilde{u}_{j}^{i_0}\right] - \left[U_{j_0}^{i_0}\right]\right)\right) \tag{24}$$

The optimality conditions (derivation w.r.t. $[\tilde{u}]$ and $[f_{j_0}^{i_0}]$, the unknowns) allow to build the following linear system, where the definition of $[\![A]\!]$-matrices corresponds to the description given above for strong Dirichlet BC.

$$\begin{bmatrix} [\![K]\!] & [\![A_{j_0}^{i_0}]\!]^T & -[\![A_{j_0}^{V_0}]\!]^T \\ [\![A_{j_0}^{i_0}]\!] & [\![0]\!] & [\![0]\!] \end{bmatrix} \begin{bmatrix} [\tilde{u}] \\ \left[f_{j_0}^{i_0}\right] \\ \left[h_{j_0}^{V_0}\right] \end{bmatrix} = \begin{bmatrix} [0] \\ \left[U_{j_0}^{i_0}\right] \end{bmatrix} \tag{25}$$

Finally, to obtain the equivalence with the problem defined by the local equations in section 4.2.2, the constraint that enforces $\left[h_{j_0}^{V_0}\right]$ to be equal to the distributed reaction forces is introduced using the relation $\left[h_{j_0}^{V_0}\right] = [\![W_{j_0}^{V_0}]\!]\left[f_{j_0}^{i_0}\right]$, where $[\![W]\!] \in \mathbb{R}^{26\times 1}$ is a 26x1 matrix that stores the 26 weights defined in Table 6. Introducing this relation in place of $\left[h_{j_0}^{V_0}\right]$ in the linear system leads to:

$$\begin{bmatrix} [\![K]\!] & [\![A_{j_0}^{i_0}]\!]^T - [\![A_{j_0}^{V_0}]\!]^T [\![W_{j_0}^{V_0}]\!] \\ [\![A_{j_0}^{i_0}]\!] & [\![0]\!] \end{bmatrix} \begin{bmatrix} [\tilde{u}] \\ \left[f_{j_0}^{i_0}\right] \end{bmatrix} = \begin{bmatrix} [0] \\ \left[U_{j_0}^{i_0}\right] \end{bmatrix} \tag{26}$$

To sum up, the problem described by local equations in section 4.2.2 is equivalent to problem (*26*), or alternatively, to problem (*23*) itself subjected to the additional constraint on distributed reaction forces $\left[h_{j_0}^{V_0}\right] = [\![W_{j_0}^{V_0}]\!]\left[f_{j_0}^{i_0}\right]$. It should be noted that the global matrix in problem (*26*) is no longer symmetric. Such problems are referred to as *generalized saddle point problems* in the review paper of Benzi et al. [2].

4.4 : Discussion : connection with null-space methods [2]

The present FFT-based method for solving Dirichlet conditions, belongs to the category of null-space methods referenced in the review of Benzi et al. for the numerical solution of saddle point problems [2]. For a saddle point system defined by:

$$\begin{bmatrix} [\![K]\!] & [\![A]\!]^T \\ [\![A]\!] & [\![0]\!] \end{bmatrix} \begin{bmatrix} [u] \\ [f] \end{bmatrix} = \begin{bmatrix} [0] \\ [u_d] \end{bmatrix} \tag{27}$$

with $[u] \in \mathbb{R}^n$, $[f] \in \mathbb{R}^m$, $[u_d] \in \mathbb{R}^m$, $[\![K]\!] \in \mathbb{R}^{n\times n}$ and $[\![A]\!] \in \mathbb{R}^{m\times n}$. Null-space methods rely on a given vector $[u']$ kinematically admissible (so that $[\![A]\!][u']$=$[u_d]$) and a matrix $[\![Z]\!] \in \mathbb{R}^{n\times(n-m)}$ so that $[\![A]\!][\![Z]\!] = [\![O]\!]$. For any vector $[v] \in \mathbb{R}^{n-m}$, the vector $[u] = [\![Z]\!][v] + [u']$ satisfies the kinematic condition. Pre-multiplying the first block of the aforementioned equation (*27*) by $[\![Z]\!]^T$, the following system of reduced dimension $(n-m)^2$ is obtained:

$$[\![Z]\!]^T[\![K]\!][\![Z]\!][v] = -[\![Z]\!]^T[\![K]\!][u'] \tag{28}$$

Once obtained the solution $[v]$ of this system, $[u]$ and $[f]$ can be evaluated as a post-processing from $[u] = [\![Z]\!][v] + [u']$ and $[\![A]\!]^T[f] = -[\![K]\!][u]$.

For the sake of simplicity, the link with our work is established for a single strongly applied BC (equation (*20*)), with $[\![A]\!] = [\![A_{j_0}^{i_0}]\!] \in \mathbb{R}^{1\times n}$ a matrix that selects node $i_0$, component $j_0$ in the vector $[u]$ (see equation (*21*)) and $[f] = [f_{j_0}^{i_0}] \in \mathbb{R}^1$ a reaction force vector. Let us define the diagonal matrix $[\![S_{j_0}^{i_0}]\!] \in \mathbb{R}^{n\times n}$, equal to the identity matrix but with a null coefficient at the position corresponding to node $i_0$, component $j_0$. Hence, this matrix acts on a vector $[p]$ as the projection $proj_P$ in algorithm (*9*) that systematically vanishes $(\boldsymbol{p}^{i_0})_{j_0}$. Noting that $[\![S_{j_0}^{i_0}]\!][\![A_{j_0}^{i_0}]\!]^T[f_{j_0}^{i_0}] = [0]$, multiplying the first block equation in (*27*) by matrix $[\![S_{j_0}^{i_0}]\!]$ gives:

$$[\![S_{j_0}^{i_0}]\!][\![K]\!][u] = [0] \tag{29}$$

In this system, the line corresponding to node $i_0$, component $j_0$, is simply $0 = 0$. On the other hand, $[\![S_{j_0}^{i_0}]\!]$ also acts on $[u]$ as $proj_U$ in algorithm (*9*) to impose the displacement $(\boldsymbol{u}^{i_0})_{j_0}$ as follows:

$$[u] = [\![S_{j_0}^{i_0}]\!][u] + [u'^{i_0}_{j_0}] \tag{30}$$

Combining these two equations, one gets:

$$[\![S_{j_0}^{i_0}]\!][\![K]\!][\![S_{j_0}^{i_0}]\!][u] = -[\![S_{j_0}^{i_0}]\!][\![K]\!][u'^{i_0}_{j_0}] \tag{31}$$

This equation is then a reformulation of equation (*28*), considering $[v] \in \mathbb{R}^{n-1}$ as a copy of vector $[u]$ removing the line corresponding to node $i_0$, component $j_0$ and $[\![Z_{j_0}^{i_0}]\!] \in \mathbb{R}^{n\times(n-1)}$ a matrix chosen as the identity matrix with a missing column corresponding to node $i_0$, component $j_0$, so that $[v] = [\![Z_{j_0}^{i_0}]\!]^T[u]$. Matrix $[\![S_{j_0}^{i_0}]\!] \in \mathbb{R}^{n\times n}$ corresponds to $[\![Z_{j_0}^{i_0}]\!]$ with an additional column of zeros at position corresponding to node $i_0$, component $j_0$, so that $[\![S_{j_0}^{i_0}]\!] = [\![Z_{j_0}^{i_0}]\!][\![Z_{j_0}^{i_0}]\!]^T$. Hence, equation (*31*) is obtained from equation (*28*) by replacing $[\![Z]\!]$ by $\left[\!\left[Z_{j_0}^{i_0}\right]\!\right]$ and then pre-multiplying the equation by $\left[\!\left[Z_{j_0}^{i_0}\right]\!\right]$ and using $[v] = \left[\!\left[Z_{j_0}^{i_0}\right]\!\right]^T [u]$.

As a common practice for building fixed-point FFT-based iterative solvers, the quantity $-[\![K^0]\!][u]$ is added on both sides of equation (*29*), with matrix $[\![K^0]\!]$ built as matrix $[\![K]\!]$ using a reference homogeneous 'behavior' $\mathbb{R}_0$. Then, multiplying by $[\![K^0]\!]^{-1}$ gives:

$$[u] = -[\![K^0]\!]^{-1}(\left[\!\left[S_{j_0}^{i_0}\right]\!\right][\![K]\!][u] - [\![K^0]\!][u]) \tag{32}$$

Finally, introducing this equation in equation (*30*) leads to the equation below that defines a fixed-point algorithm:

$$[u] = -\left[\!\left[S_{j_0}^{i_0}\right]\!\right][\![K^0]\!]^{-1}\left(\left[\!\left[S_{j_0}^{i_0}\right]\!\right][\![K]\!][u] - [\![K]\!][u]\right) + \left[u'^{i_0}_{j_0}\right] \tag{33}$$

The fixed-point algorithm defined by equation (*33*) is equivalent to the fixed-point algorithm (*9*) considering that:

- $[\![K^0]\!]^{-1}$ corresponds directly to the Green operator $-\mathbb{G}_{\mathrm{D}}$,
- $[p] = [\![K]\!][u]$ stores the stress divergence $\mathbf{div}_{\mathbf{D}}(\mathbb{C}:\left(\underline{\nabla_{\mathrm{D}}^{\mathrm{S}}}(\boldsymbol{u})\right))$ for all the grid points,
- $\left[\!\left[S_{j_0}^{i_0}\right]\!\right][p]$ corresponds to the projection $proj_P$,
- $\left[\!\left[S_{j_0}^{i_0}\right]\!\right][u] + \left[u'^{i_0}_{j_0}\right]$ corresponds to the projection $proj_U$.

Hence, the method proposed in the present paper belongs to the category of the so-called null-space methods [2] (to eliminate the Lagrange multipliers in the vector of unknowns) associated with the standard construction of an FFT-based fixed-point algorithm. Demonstration has been made for strongly applied BCs. The two other propositions (i.e. for weakly applied BC and 'distributed reaction force'), following a similar strategy, also correspond to a null-space method.

As the proposed FFT-based solver has the ability to solve minimization problems subject to equality constraints, the next section focuses on a further extension towards inequality constraints as observed in unilateral contact mechanics.

## 5 : Contact mechanics – application to spherical indentation

In this section, the FFT-based solver proposed in Algorithm 1 to account for equality constraints is modified to account for unilateral constraints associated with rigid and frictionless contact mechanics. A homogeneous cubic volume of size $L^3$ is considered with the face $S_1^0$ subjected to null-Dirichlet BCs, null-Neumann BCs on the four faces $S_J^\alpha$ ($\alpha = 0{,}1$ and $J = 2{,}3$), and contact conditions on face $S_1^1$ (see definition of $S_J^\alpha$ in section 3.2.1). The rigid indenter is a sphere $\mathcal{S}$ of radius $R_s$ and center $O_s$ initially located at position $[L + R, 0.5L, 0.5L]$. The loading is given by the displacement of the sphere $\boldsymbol{U_S} = U_S\boldsymbol{e_1}$. The contact conditions on surface $S_1^1$, reported in problem (*34*), are associated with the existence of a contact zone $\mathcal{C}$, defined on the deformed configuration, through the Karush-Kuhn-Tucker (KKT) conditions under the frictionless assumption. In the discrete problem, the contact zone $\mathcal{C}$ gathers the active nodes constrained by the last three equations of problem (*34*), and the other part $S_1^1/\mathcal{C}$ gathers inactive nodes constrained by the two first equations.

$$\begin{cases} \underline{\sigma}.\boldsymbol{n} = \boldsymbol{0} & \text{on } S_1^1/\mathcal{C} \\ g > 0 & \text{on } S_1^1/\mathcal{C} \\ \boldsymbol{n_c}.\underline{\sigma}.\boldsymbol{n_c} < 0 & \text{on } \mathcal{C} \\ g = 0 & \text{on } \mathcal{C} \\ \underline{\sigma}.\boldsymbol{n_c} - (\boldsymbol{n_c}.\underline{\sigma}.\boldsymbol{n_c})\boldsymbol{n_c} = \boldsymbol{0} & \text{on } \mathcal{C} \end{cases} \tag{34}$$

In this set of contact conditions, the outer normal is $\boldsymbol{n}$. For any point $M'$ of the surface $S_1^1$ in the deformed configuration, $M_s'$ is the radial projection of $M'$ on the sphere located at $O_s'$ (position of the sphere center), and the potential contact normal is defined by $\boldsymbol{n_c} = -\boldsymbol{O_s'M'}/\|\boldsymbol{O_s'M'}\|$. The gap is then defined as $g = (\boldsymbol{O_s'M_s'} - \boldsymbol{O_s'M'}).\boldsymbol{n_c} = R - R_s$ (with $R = \|\boldsymbol{O_s'M'}\|$). Hence, out of the contact zone, the surface is free (1st equation in problem (*34*)) and the gap is positive (2nd equation in problem (*34*)). In the contact zone, the gap is null (4th equation in problem (*34*)), the normal traction vector is negative (3rd equation in problem (*34*)) and the tangential traction vector is null (5th equation in problem (*34*)) because of the frictionless assumption. It is worth noting that the condition on the gap does not impose the position at the surface of the sphere that remains free.

Following the previous section, the introduction of the FFT-based solver is proposed from local equations, in the small and finite strain setting, and, after an analysis of the application results, the viewpoint of minimization problem under inequality constraint is adopted.

5.1 : Small strain implementation

In the present small strain implementation, the contact normal $\boldsymbol{n_c}$ is evaluated in the deformed configuration. This normal is used to define the gap and the orientation of the reaction force. Note that in the small strain framework, the difference between the initial and deformed configurations are not considered in the field equations (equilibrium, behavior) leading to a small ambiguity regarding the configuration choice. This ambiguity disappears when using the finite strain setting as proposed below (section 5.2).

After regular discretization of the VE, the discrete problem is solved using the active-set strategy (see for instance the handbook of Yastrebov about numerical methods in contact mechanics [63]) to determine the contact zone. Starting with an initial active contact zone $\mathcal{C}$ (i.e. active-set) defined by the set of discrete points violating the gap condition after a displacement increment of the sphere, the algorithm consists in applying iteratively the following two steps:

1. Solve the problem with the BCs:

$$\begin{cases} \underline{\sigma}.\boldsymbol{e_1} = \boldsymbol{0} & \text{on } S_1^1/\mathcal{C} \\ g = 0 & \text{on } \mathcal{C} \\ \underline{\sigma}.\boldsymbol{n_c} - (\boldsymbol{n_c}.\underline{\sigma}.\boldsymbol{n_c})\boldsymbol{n_c} = \boldsymbol{0} & \text{on } \mathcal{C} \end{cases} \tag{35}$$

2. Update the active contact zone $\mathcal{C}$ as follows:

$$\begin{cases} \text{for } M \in S_1^1/\mathcal{C} \quad : \quad \text{if } (g(M) < 0) \quad M \in \mathcal{C} \\ \text{for } M \in \mathcal{C} \quad : \quad \text{if } \big((\boldsymbol{n_c}.\underline{\sigma}.\boldsymbol{n_c})(M) > 0\big) \quad M \notin \mathcal{C} \end{cases} \tag{36}$$

The two steps are detailed below and then incorporated in the accelerated fixed-point FFT-based solver proposed in the previous section.

5.1.1: Step 1

After spatial discretization, the resolution of step 1 is done using reaction forces (i.e. Lagrange multipliers) as done in problem (*8*). The scalar $f_{n_c}^{\mathcal{C}}$ are the values of the reaction forces oriented by the vector $\boldsymbol{n_c}$ for all the points of $\mathcal{C}$. The kinematic condition (fourth equation in problem below) imposes that any point of $\mathcal{C}$ must be located at the surface of the sphere in the deformed configuration. Importantly, as the vector normal is defined by $\boldsymbol{n_c} = -\boldsymbol{O'_s M'}/\|\boldsymbol{O'_s M'}\|$, this fourth equation does not impose any constraint on the position of $M'$ (with $\tilde{\boldsymbol{u}}(M) = \boldsymbol{MM'}$ ) on the sphere, it only enforces the gap condition.

$$\begin{cases} \mathbf{div_D}\left(\underline{\sigma}\right) + f_{n_c}^{\mathcal{C}} \boldsymbol{n_c} = 0 & \text{in } \Omega^\infty \\ \underline{\tilde{\varepsilon}} = \underline{\nabla_D^s}(\tilde{\boldsymbol{u}}) & \text{in } \Omega^\infty \\ \underline{\sigma} = \mathbb{C} : \underline{\tilde{\varepsilon}} & \text{in } \Omega \\ \tilde{\boldsymbol{u}}(M) = \boldsymbol{MO'_s} + R_s \boldsymbol{n_c} & \text{in } \mathcal{C} \\ \tilde{\boldsymbol{u}} \text{ appropriate symmetries} & \text{in } \Omega^\infty \\ \underline{\sigma} \text{ appropriate symmetries} & \text{in } \Omega^\infty \end{cases} \tag{37}$$

In the present case, the appropriate symmetries are associated with null-Dirichlet on face $S_1^0$ and null-Neumann on the other faces, hence all the components of $\tilde{\boldsymbol{u}}$ have the symmetries A on face $S_1^0$ and S on the five other faces. The symmetries of $\underline{\sigma}$, not reported here, are consistent with Table 1. Note that the condition $\underline{\sigma}.\boldsymbol{e_1} = \boldsymbol{0}$ on $S_1^1/\mathcal{C}$ is then satisfied by this choice of symmetries.

The resolution can be done as in section 4, with a fixed-point algorithm corresponding to the succession of equations given in problem (*38*), omitting convergence acceleration for the sake of clarity.

$$\begin{cases} proj_U : \tilde{\boldsymbol{u}}(M) = \boldsymbol{MO'_s} + R_s \boldsymbol{n_c} \quad \forall M \in \mathcal{C} \\ \qquad \underline{\sigma} = \mathbb{C} : \underline{\nabla_D^s}(\tilde{\boldsymbol{u}}) \\ \qquad \boldsymbol{p} = \mathbf{div_D}\left(\underline{\sigma}\right) \\ proj_P : (\boldsymbol{p}.\boldsymbol{n_c})(M) = 0 \quad \forall M \in \mathcal{C} \\ \qquad \boldsymbol{p} = \boldsymbol{p} - \mathbf{div_D}\left(\mathbb{R}_0 : \underline{\nabla_D}(\tilde{\boldsymbol{u}})\right) \\ \qquad \hat{\tilde{\boldsymbol{u}}} = \hat{\mathbb{G}}_D(\hat{\boldsymbol{p}}) \end{cases} \tag{38}$$

As in section 4, the local Dirichlet conditions are imposed through projections $proj_U$ and $proj_P$. On the one hand, $proj_U$ modifies the current field $\tilde{\boldsymbol{u}}$, so that the points of $\mathcal{C}$ in the deformed configuration are located on the sphere. As $\boldsymbol{n_c} = -\boldsymbol{O'_s M'}/\|\boldsymbol{O'_s M'}\|$ is recomputed at each iteration, this equation just projects the current point $M'$ (of the given iteration) on the sphere but does not constrain the position of $M'$ on the sphere. On the other hand, $proj_P$ vanishes $\boldsymbol{p}.\boldsymbol{n_c}$ on $\mathcal{C}$ so that systematically $\mathbf{div_D}\left(\underline{\sigma}\right).\boldsymbol{n_c} + f_{n_c}^{\mathcal{C}} = 0$, with $f_{n_c}^{\mathcal{C}}$ the normal reaction force to be evaluated as a post-processing once achieved the convergence of the algorithm. At convergence, the points of $\mathcal{C}$ are located at the surface of the sphere without any tangential stress.

5.1.2: Step 2

The first equation in problem (*36*) consists in evaluating the gap $g(M) = R - R_S$ (with $R = \|\boldsymbol{O'_S M'}\|$) for any point of the surface $S_1^1/\mathcal{C}$. If the gap is negative, the point $M$ is added to $\mathcal{C}$. Instead of a strict comparison to zero, a small tolerance was applied using $R_S + 10^{-12}R_S$ instead of $R_S$ in the gap definition.

The second equation in problem (*36*) is replaced by a condition on the normal reaction forces $f_{n_c}^{\mathcal{C}}(M)$ (i.e. Lagrange multipliers) evaluated for any point $M$ in $\mathcal{C}$. If the reaction force $f_{n_c}^{\mathcal{C}}(M)$ is positive, the point is removed from $\mathcal{C}$. As introduced in the gap condition, a tolerance could also be applied if necessary. However, in the present paper, a strict condition allowed to obtain results validated by Hertz theory.

5.1.3: FFT-based algorithm

In practice, the two iterative algorithms (active-set strategy and accelerated fixed-point algorithm) are gathered in Algorithm 2, an accelerated fixed-point algorithm with a conditional update of the contact zone $\mathcal{C}$. To account for the non-linear nature of the problem, the displacement of the indenter is increased incrementally.

1. Initialize $\widetilde{\boldsymbol{u}} = \widetilde{\boldsymbol{u}}$ of previous increment (nothing to do)
2. Increase increment loading (position of sphere centre) $\rightarrow O'_s$
3. Initialize active contact zone: $\mathcal{C} = \{M \in S_1^1 \ / \ g(M) < 0\}$
4. $proj_U : \boldsymbol{n_c}(M) = \ \boldsymbol{O'_s M'}/\|\boldsymbol{O'_s M'}\| \quad \forall M \in S_1^1$
5. $\quad \widetilde{\boldsymbol{u}}(M) = \boldsymbol{MO'_s} + R_s \boldsymbol{n_c}(M) \quad \forall M \in \mathcal{C}$
6. Loop $i \geq 0$
7. $\quad$ if $(ACV_U)$
8. $\quad\quad$ if $(i > 0) ACV_{storeR}(i, \widetilde{\boldsymbol{u}})$
9. $\quad\quad$ if $(i > 0 \ \& \ \mathrm{modulo}(i, mod_{ACV}) = 0) \ ACV_{accelerate} \rightarrow \widetilde{\boldsymbol{u}}$
10. $\quad\quad ACV_{storeS}(i, \widetilde{\boldsymbol{u}})$
11. $\quad \underline{\sigma} = \mathbb{C}: \underline{\nabla_{\mathrm{D}}^{\mathrm{S}}}(\widetilde{\boldsymbol{u}})$
12. $\quad \boldsymbol{p} = \mathbf{div_D}(\underline{\sigma})$
13. $\quad proj_P : \ \boldsymbol{p}(M) = \boldsymbol{p}(M) - \big((\boldsymbol{p}.\boldsymbol{n_c})\boldsymbol{n_c}\big)(M) \quad \forall M \in \mathcal{C}$
14. $\quad$ if equilibrium$(\boldsymbol{p})$: EXIT Loop *(see equation (5))*
15. $\quad g(M) = \|\boldsymbol{O'_s M'}\| - \boldsymbol{R_s}, \qquad \forall M \in S_1^1$
16. $\quad \mathcal{V}_{gap}^i = \{M \in S_1^1/\mathcal{C} \ / \ g(M) < \epsilon_g\}$
17. $\quad f_{n_c}^{\mathcal{C}}(M) = \big(-\mathbf{div_D}(\underline{\sigma}).\boldsymbol{n_c}\big)(M), \qquad \forall M \in S_1^1$
18. $\quad\quad$ ( *assuming an extra null stress layer outside the cell* )
19. $\quad \mathcal{V}_{react}^i = \{M \in \mathcal{C} \ / \ f_{n_c}^{\mathcal{C}}(M) > \epsilon_r\}$
20. $\quad$ if $\big(\{\mathcal{V}_{gap}^i, \mathcal{V}_{react}^i\} = \{\mathcal{V}_{gap}^{i-1}, \mathcal{V}_{react}^{i-1}\}\big)$
21. $\quad\quad \mathcal{C}^{old} = \mathcal{C}$
22. $\quad\quad$ update $\mathcal{C} = (\mathcal{C}/\mathcal{V}_{react}^{i-1}) \cup \mathcal{V}_{gap}^i$
23. $\quad\quad$ *if* $\big(\mathcal{C} \neq \mathcal{C}^{old}\big) \ i = 0$ (to reinitialize ACV arrays)
24. $\quad \boldsymbol{p} = \boldsymbol{p} - \mathbf{div_D}\Big(\mathbb{R}_0: \underline{\nabla_{\mathrm{D}}}(\widetilde{\boldsymbol{u}})\Big)$
25. $\quad$ if $(ACV_P)$
26. $\quad\quad$ if $(i > 0) ACV_{storeR}(i, \widetilde{\boldsymbol{p}})$
27. $\quad\quad$ if $(i > 0 \ \& \ \mathrm{modulo}(i, mod_{ACV}) = 0) ACV_{accelerate} \rightarrow \widetilde{\boldsymbol{p}}$
28. $\quad\quad ACV_{storeS}(i, \widetilde{\boldsymbol{p}})$
29. $\quad \widehat{\boldsymbol{p}} = DT(\boldsymbol{p})$
30. $\quad \widehat{\widetilde{\boldsymbol{u}}} = \widehat{\mathbb{G}}_{\mathrm{D}}(\widehat{\boldsymbol{p}})$
31. $\quad \widetilde{\boldsymbol{u}} = iDT(\widehat{\widetilde{\boldsymbol{u}}})$
32. $\quad proj_U : \boldsymbol{n_c}(M) = \ \boldsymbol{O'_s M'}/\|\boldsymbol{O'_s M'}\| \quad \forall M \in S_1^1$
33. $\quad \widetilde{\boldsymbol{u}}(M) = \boldsymbol{MO'_s} + R_s \boldsymbol{n_c}(M) \quad \forall M \in \mathcal{C}$

*Algorithm 2 : Description of the FFT-based algorithm proposed for spherical indentation (for an increment of loading) using a small strain framework. Compared to Algorithm 1, the main differences are highlighted in green for the update of the contact zone, and in red for the functions $proj_U$ and $proj_P$, specialized for the indentation application. For the present application, the tolerances for contact detection are $\epsilon_g = 10^{-12} R_s$ and $\epsilon_r = 0$. For field equations, if a point (M) is not specified (for the sake of clarity) the equation is evaluated at all the points of the unit-cell*

The goal of the algorithm is to find the solution $\widetilde{\boldsymbol{u}}$ after an incremental displacement of the indenter. The initial guess for $\widetilde{\boldsymbol{u}}$ is the field at the end of previous increment and the initial contact zone $\mathcal{C}$ is defined as the set of points violating the gap condition (i.e. $g(M) < 0$) after the incremental displacement of the sphere.

The accelerated fixed-point algorithm is very close to Algorithm 1, with a specialization of the functions $proj_U$ and $proj_P$ (red lines) to apply the Dirichlet constraints in the contact zone $\mathcal{C}$ as explained in section 5.1.1, and a conditional update of the contact zone (green lines). For that purpose, at each iteration $i$, the set of points violating the gap condition $\mathcal{V}_{gap}^{i}$ (first equation in problem (*36*)) is evaluated as well as the set of points violating the normal stress condition $\mathcal{V}_{react}^{i}$ (second equation in problem (*36*)). Instead of using $(\boldsymbol{n_c}.\underline{\sigma}.\boldsymbol{n_c}) > 0$, the condition is applied to the normal reaction forces, $f_{n_c}^{\mathcal{C}} > 0$, with $f_{n_c}^{\mathcal{C}} = -\mathbf{div_D}(\underline{\sigma}).\boldsymbol{n_c}$, evaluated on the points of $\mathcal{C}$ (i.e. a subset of face $S_1^1$) using the discrete derivation of $\underline{\sigma}$ considering an additional layer of null stress on the external side of $S_1^1$. Before updating the contact zone $\mathcal{C}$, the solution field must be precise enough, but it is useless to iterate until a very precise solution as soon as the contact zone is not correct. Hence, it is proposed that the contact zone is updated if the sets of points $\mathcal{V}_{gap}^{i}$ and $\mathcal{V}_{react}^{i}$ are both identical to the sets of points evaluated at the previous iteration. In other terms, the contact zone is updated if the sets of points candidates for the update are stable from one iteration to another. Note that an additional condition on the equilibrium condition (with a higher tolerance than the one used for the exit condition) is also required to ensure a sufficient precision before update (a value of $10^{-2}$ is used compared to the standard $10^{-4}$ tolerance). Importantly, concerning convergence acceleration, the iteration indicator $i$ is reset to 0 when the contact zone is modified (line 23 in Algorithm 2) so that convergence acceleration is restarted (otherwise the arrays storing residual and solution fields would mix fields obtained for two different contact zones, or in other words, for two different problems).

To summarize, the fixed-active-set solver uses the equilibrium criterion with a tolerance of $10^{-4}$ that will be the final condition once the active-set region is no longer updated during iterations. The criterion for updating the active-set is twofold: the equilibrium criterion must be satisfied with a sufficient precision (here $10^{-2}$) and the set of points violating the contact conditions must be stable from one iteration to another.

The active-set update considered here differs from the primal-dual active-set formulation, in which primal and dual variables are combined through a parameter $c$ [54] [64]. Instead, the primal condition on the gap and the dual condition on the contact multiplier are monitored separately, and the active-set is updated according to the respective violations of these conditions. This naïve active-set implementation was sufficient for simulating the simple case of the rigid spherical indentation, but the dual-primal active-set formulation should be introduced and tested in the future. Note that in practice, a tolerance is used to define the set of points violating the gap $\mathcal{V}_{gap}^{i}$ that could be affected by numerical precision: the contact condition is tested using $R_s + 10^{-12}\,R_s$ instead of $R_s$ for the sphere radius used in the gap definition. This was sufficient to avoid spurious oscillations probably due to a wrong affectation of borderline nodes, together with the use of the TETRA2 FD scheme. Note also that several nodes can enter or leave the active-set and this could generate cycling issues (i.e. cycling from one active-set to another without finding a stable active-set).  However, in the present simple application, the question of cycling has not been observed.

If the present algorithm is rather simple, it is efficient enough to demonstrate its ability to solve the simple problem of spherical indentation. Of course, for more complex situations, the algorithm could benefit from various improvements such as the primal-dual active-set method [54][64], the detection of cycling and the large experience of the contact mechanics community acquired with finite element codes.

## 5.2 : Finite strain implementation

The SS implementation proposed above combines equations that do not distinguish initial and deformed configurations and contact conditions (in the contact zone $\mathcal{C}$) applied in the deformed configuration. For a more rigorous treatment, a FS implementation is proposed.

To solve the contact problem with a fixed contact zone $\mathcal{C}$ in step 1 (see section 5.1.1), the set of equations gathered in problem (*37*) can be reformulated in the finite strain setting using the first Piola-Kirchhoff (or Boussinesq), non-symmetric, second-order stress tensor $\underline{\Pi}$:

$$\begin{cases} \mathbf{div_D}\left(\underline{\Pi}\right) + f_{n_c}^{\mathcal{C}} \boldsymbol{n_c} = 0 & \text{in } \Omega^{\infty} \\ \underline{\tilde{F}} = \underline{1} + \underline{\nabla_D}(\tilde{\boldsymbol{u}}) & \text{in } \Omega^{\infty} \\ \underline{\Pi} = \mathcal{F}\left(\underline{\tilde{F}}\right) & \text{in } \Omega \\ \tilde{\boldsymbol{u}}(M) = \boldsymbol{MO_s'} + R_s \boldsymbol{n_c} & \text{in } \mathcal{C} \\ \tilde{\boldsymbol{u}} \text{ appropriate symmetries} & \text{in } \Omega^{\infty} \\ \underline{\Pi} \text{ appropriate symmetries} & \text{in } \Omega^{\infty} \end{cases} \tag{39}$$

In that case, the precise form of the "behavior" law $\underline{\Pi} = \mathcal{F}\left(\underline{\tilde{F}}\right)$ is intentionally let general (a simple example can be $\underline{\mathrm{S}} = \mathbb{C}\colon \underline{\mathrm{E}}$, with $\underline{\Pi} = \underline{\tilde{F}}.\underline{\mathrm{S}}$ and $\underline{\mathrm{S}}$ the second Piola-Kirchhoff tensor, and $\underline{\mathrm{E}} = \frac{1}{2}(\underline{\tilde{F}}^T\underline{\tilde{F}} - \underline{1})$, the Green-Lagrange strain tensor for an elastic material).

To update the contact zone in step 2, the contact conditions on the signs of both the gap and the reaction are exactly the same as in section 5.1.2. In the SS implementation, the reaction $f_{n_c}^{\mathcal{C}} = -\mathbf{div_D}\left(\underline{\sigma}\right).\boldsymbol{n_c}$ is evaluated on the points of $\mathcal{C}$ taking into account an additional layer of null stress on the external side of $S_1^1$. In the FS implementation, the same method is used to evaluate $f_{n_c}^{\mathcal{C}}$, simply replacing $\underline{\sigma}$ by $\underline{\Pi}$ so that $f_{n_c}^{\mathcal{C}} = -\mathbf{div_D}\left(\underline{\Pi}\right).\boldsymbol{n_c}$.

As the two steps of the algorithm are very similar between small and finite strain frameworks, the FFT-based algorithm implemented for SS (see Algorithm 2) is only slightly modified to account for FS (see Algorithm 3).

1. Initialize $\widetilde{\boldsymbol{u}} = \widetilde{\boldsymbol{u}}$ of previous increment (nothing to do)
2. Increase increment loading (position of sphere centre) $\rightarrow O_s'$
3. Initialize active contact zone: $\mathcal{C} = \{M \in S_1^1 \;/\; g(M) < 0\}$
4. $proj_U : \boldsymbol{n_c}(M) = \boldsymbol{O_s'M'}/\|\boldsymbol{O_s'M'}\| \quad \forall M \in S_1^1$
5. $\quad \widetilde{\boldsymbol{u}}(M) = \boldsymbol{MO_s'} + R_s \boldsymbol{n_c}(M) \quad \forall M \in \mathcal{C}$
6. Loop $i \geq 0$
7. $\quad$ if $(ACV_U)$
8. $\quad\quad$ if $(i > 0) ACV_{storeR}(i, \widetilde{\boldsymbol{u}})$
9. $\quad\quad$ if $(i > 0$ & $\text{modulo}(i, mod_{ACV}) = 0)\; ACV_{accelerate} \rightarrow \widetilde{\boldsymbol{u}}$
10. $\quad\quad ACV_{storeS}(i, \widetilde{\boldsymbol{u}})$
11. $\quad \underline{\Pi} = \mathcal{F}\left(\underline{1} + \underline{\nabla_D}(\widetilde{\boldsymbol{u}})\right)$
12. $\quad \boldsymbol{p} = \mathbf{div_D}(\underline{\Pi})$
13. $\quad proj_P : \boldsymbol{p}(M) = \boldsymbol{p}(M) - \left((\boldsymbol{p}.\boldsymbol{n_c})\boldsymbol{n_c}\right)(M) \quad \forall M \in \mathcal{C}$
14. $\quad$ if equilibrium$(\boldsymbol{p})$: EXIT Loop *(see equation (5))*
15. $\quad g(M) = \|\boldsymbol{O_s'M'}\| - \boldsymbol{R_s}, \quad \forall M \in S_1^1$
16. $\quad \mathcal{V}_{gap}^i = \{M \in S_1^1/\mathcal{C} \;/\; g(M) < \epsilon_g\}$
17. $\quad f_{n_c}^{\mathcal{C}}(M) = \left(-\mathbf{div_D}(\underline{\Pi}).\boldsymbol{n_c}\right)(M), \quad \forall M \in S_1^1$
18. $\quad\quad$ *( assuming an extra null stress layer outside the cell)*
19. $\quad \mathcal{V}_{react}^i = \{M \in \mathcal{C} \;/\; f_{n_c}^{\mathcal{C}}(M) > \epsilon_r\}$
20. $\quad$ if $\left(\{\mathcal{V}_{gap}^i, \mathcal{V}_{react}^i\} = \{\mathcal{V}_{gap}^{i-1}, \mathcal{V}_{react}^{i-1}\}\right)$
21. $\quad\quad \mathcal{C}^{old} = \mathcal{C}$
22. $\quad\quad$ update $\mathcal{C} = (\mathcal{C}/\mathcal{V}_{react}^{i-1}) \cup \mathcal{V}_{gap}^i$
23. $\quad\quad$ *if* $\left(\mathcal{C} \neq \mathcal{C}^{old}\right)\; i = 0$ (to reinitialize ACV arrays)
24. $\quad \boldsymbol{p} = \boldsymbol{p} - \mathbf{div_D}\left(\mathbb{R}_0 : \underline{\nabla_D}(\widetilde{\boldsymbol{u}})\right)$
25. $\quad$ if $(ACV_P)$
26. $\quad\quad$ if $(i > 0) ACV_{storeR}(i, \widetilde{\boldsymbol{p}})$
27. $\quad\quad$ if $(i > 0$ & $\text{modulo}(i, mod_{ACV}) = 0) ACV_{accelerate} \rightarrow \widetilde{\boldsymbol{p}}$
28. $\quad\quad ACV_{storeS}(i, \widetilde{\boldsymbol{p}})$
34. $\quad \widehat{\boldsymbol{p}} = DT(\boldsymbol{p})$
35. $\quad \widehat{\widetilde{\boldsymbol{u}}} = \mathbb{G}_D(\widehat{\boldsymbol{p}})$
36. $\quad \widetilde{\boldsymbol{u}} = iDT\left(\widehat{\widetilde{\boldsymbol{u}}}\right)$
29. $\quad proj_U : \boldsymbol{n_c}(M) = \boldsymbol{O_s'M'}/\|\boldsymbol{O_s'M'}\| \quad \forall M \in S_1^1$
30. $\quad \widetilde{\boldsymbol{u}}(M) = \boldsymbol{MO_s'} + R_s \boldsymbol{n_c}(M) \quad \forall M \in \mathcal{C}$

*Algorithm 3 : Description of the FFT-based algorithm proposed for spherical indentation (for an increment of loading) using a finite strain framework. Compared to Algorithm 1, the main differences are highlighted in green for the update of the contact zone, and in red for the functions $proj_U$ and $proj_P$, specialized for the indentation application. For the present application, the tolerances for contact detection are $\epsilon_g = 10^{-12} R_s$ and $\epsilon_r = 0$. For field equations, if a point $(M)$ is not specified (for the sake of clarity) the equation is evaluated at all the points of the unit-cell*

## 5.3 : Application

The example treated in this section consists in simulating the indentation of a cube with a spherical indenter. The spherical indenter is centered with respect to the contact face. The sphere diameter is equal to the cube edge length of 1mm and the applied displacement indenter is $\delta$. The behavior law is assumed to be elastic isotropic with the Lamé coefficients of steel ($\lambda = 120$ GPa, $\mu = 80$ GPa). For SS assumption, the behavior law is given by ($\underline{\sigma} = \lambda tr(\underline{\varepsilon})\underline{1} + 2\mu\underline{\varepsilon}$) and for FS, the Saint Venant-Kirchhoff model is used replacing the Cauchy stress $\underline{\sigma}$ by the second Piola-Kirchhoff stress tensor $\underline{S}$ and the linearized strain $\underline{\varepsilon}$ by the Green-Lagrange strain $\underline{E}$.

For the sake of validation, results are compared to the results of the Hertz theory regarding the reaction force $F_{Hertz}$ as a function of the indenter displacement $\delta$ :

$$F_{Hertz} = \frac{4E}{3(1-\nu^2)}\sqrt{R\delta^3} \tag{40}$$

With $E$ and $\nu$ the elastic Young modulus and Poisson ratio and $R$ the sphere radius. The validity of Hertz theory being limited to small $\delta/R$ ratio (with $R = 0.5$mm), the simulation covers ratios below 2% ($\delta_1 = 0.01$mm) for a validation purpose, and ratios up to 20% ($\delta_{max} = 0.1$mm) to compare the difference between small and finite strain simulations. A reference setup is chosen with a spatial resolution of $65^3$ voxels and 80 loading steps. Other resolutions and loading steps are used below for sensitivity analysis. Note that for the comparison with Hertz theory, that relies on a half-space infinite domain, the numerical solution may suffer from the finite -size of the domain. However, restricting our comparison up to $\delta_1 = 0.01$mm, the ratio between the contact radius ($\sqrt{R\delta_1}$ according to Hertz theory) and the cell size (1mm) is 0.07. Looking at the good agreement between Hertz theory and the numerical results, this ratio appears sufficient for the comparison. An analysis of the sensitivity w.r.t. the cell size or the choice of boundary conditions would be of interest but goes beyond the scope of the present paper.

The finite difference scheme chosen for this application is the TETRA2 scheme. Simulations with the reference setup are run using the $\mathbb{R}_0$ tensor proposed by Paux [44] (i.e. $\mathbb{R}_0 : \underline{\nabla u} = \lambda_0 \underline{\nabla u} \odot \underline{I} + 2\mu_0 \underline{\nabla u}$, see section 3.2.3) with Lamé coefficients $(\lambda_0, \mu_0)$ proportional to the material Lamé coefficients $(\lambda, \mu)$. Before going further, the convergence is briefly studied as a function of the preconditioner $\mathbb{R}_0$ to find an approximate optimal choice, i.e. the reference parameters $(\lambda_0, \mu_0)$. For that purpose, the total number of iterations (summing over the 80 steps) is reported in Table 8. For both FS and SS simulations, an optimal value is found around a factor of 5 (i.e. the reference parameters $(\lambda_0, \mu_0)$ = 5x$(\lambda, \mu)$). This value is kept for SS simulations and, for FS simulations, due to the non-convergence observed for lower factors, a higher value of 10 is used. It can be noticed that, in Table 8, the average number of iterations per step is between 92 and 45.

| | $\times 1$ | $\times 2$ | $\times 5$ | $\times 10$ | $\times 20$ |
|---|---|---|---|---|---|
| Small strain | 5442 | 4089 | 3614 | 4184 | 5253 |
| Finite strain | No Conv. | No Conv. | 4863 | 5354 | 7413 |

*Table 8 : Total number of iterations using 80 loading steps and a resolution of $65^3$. The factor multiplies the elastic behaviour of the material to define the Lamé coefficients of the preconditioner $\mathbb{C}_0$ (i.e. reference material). No convergence is declared when the number of iterations in a single step reaches 2000.*

Taking advantage of the given BCs, to evaluate the reaction force in the FS framework, we simply use $F = A_0 \boldsymbol{e_1}.\langle\underline{\Pi}\rangle.\boldsymbol{e_1}$, where $A_0$ is the area of face $S_1^0$ (and $S_1^1$) in the initial configuration and $\langle\underline{\Pi}\rangle$ is the volume average of the Piola-Kirchhoff stress tensor over the VE. In the SS framework, we use $F = A_0 \boldsymbol{e_1}.\langle\underline{\sigma}\rangle.\boldsymbol{e_1}$, replacing the stress tensor.

Figure 8 and Figure 9 draw the evolution of the force as a function of the indentation depth for a given spatial resolution of $65^3$ and different loading steps (20, 40, 80 and 160). Figure 8 presents SS simulation results and FS results are displayed on Figure 9. First, the robustness of the method is emphasized when comparing various loading steps: the agreement between the results demonstrates that, over this range of loading steps, the simulation results are independent of this choice, for both SS and FS cases. Second, the comparison between the Hertz theory and SS simulations is made for small $\delta/R$ ratios that correspond to validity domain of the Hertz theory (see zoom part in Figure 8, with $\delta$ below 0.01mm corresponding to $\delta/R$ =2%). The very good agreement validates both the method and its implementation. The same conclusion can be drawn from the zoom part in Figure 9, validating also the FS case, at least in the range of small $\delta/R$ ratios. For larger ratios, up to 20%, it can be noticed that the agreement is still quite good for the SS case. However, the difference with the FS case increases significantly with $\delta/R$, demonstrating the importance of considering FS in that case. Regarding the cumulated number of iterations in Figure 10, it is worth noting that it increases with the number of loading steps, in a similar way for both the SS and FS cases. To understand this, it should be highlighted that, from one step to another, the loading is not proportional as the contact zone increases. Hence, the initialization of the iterative algorithm (line 1 in Algorithm 2 and Algorithm 3) do not consider the results of the previous steps (such as done classically with a linear extrapolation in standard FFT-based solvers with proportional loadings). However, the mean number of iterations per step decreases with the step size (35, 45, 57 and 65 iterations respectively for 160, 80, 40 and 20 loading steps, in the SS case, and 53, 67, 89, 128 in the FS case). Proposing a better initialization could be an interesting future prospect to reduce the computation cost. Lastly, for a given number of loading steps, the number of iterations is higher for a FS simulation by a factor of approximately 1.5.

After modifying the loading step for a given resolution in the previous paragraph, Figure 11 and Figure 12 examine the spatial resolution sensitivity for a given number of loading steps (80, in the present case). For the given range of spatial resolutions (from $33^3$ to $129^3$) the comparison exhibits a very small resolution sensitivity of the Force-Indentation depth curves, for both SS or FS cases, for large and small depth (insets in the figures). However, regarding the cumulated number of iterations in Figure 13, it is worth noting that it increases with the spatial resolution, in a similar way for both the SS and FS cases. A part of the explanation comes with the number of times the contact zone needs to be updated during the steps. For resolutions of 33, 65 and 129, these numbers are respectively 97, 178 and 270 (for SS simulations). Actually, the larger the number of elements in the contact zone, the more contact zone readjustments are expected. However, these numbers do not solely explain the gaps observed Figure 13 between curves for resolution 33, 65 and 129. A deeper analysis is proposed in appendix A2 reporting the evolution of the number of updates, of the number of nodes in the contact zone, but also additional results on simulations considering fixed active-sets and equality kinematic constraints. It must be highlighted from this last series of simulations the moderate mesh sensitivity of the convergence with a dependence in $d^{-0.2}$.

As a final result, Figure 14 displays the unit-cell in its deformed configuration at the end of the loading, for both SS and FS simulations. The surface of the indenter in its final position is also represented. The deformed shape is qualitatively consistent with expectations: the displacement is smooth without oscillations; the surface of the indenter coincides with the deformed shape in the contact zone and there is no sign of adhesion with the indenter in the neighborhood of the contact zone. The difference between the SS and FS cases, existing on the macroscopic curve (reaction force-indentation depth) when the $\delta/R$ ratio increases (see for example the comparison between Figure 8 and Figure 9), is also observed on the vertical displacement, on Figure 14 for $\delta/R$=20% (i.e. last loading step), whose displacement iso-contours are more flat for the FS case.

To summarize, the comparison with Hertz theory for small $\delta/R$ ratios partially validates the new FFT-based solvers proposed in Algorithm 2 and Algorithm 3 to account for spherical indentation and unilateral contact mechanics, respectively for SS and FS cases. Note that a deeper validation is proposed in appendix A1 reporting additional comparisons with Hertz theory on the radius of the contact area, on the maximum contact pressure and on the contact pressure profiles, but also, additional verifications on the evaluation of the total reaction force and on the sign of contact forces. For higher ratios, SS and FS results depart from each other, demonstrating the need to account for the non-approximated equations (i.e. FS framework instead of SS hypothesis). The robustness of the method is demonstrated through spatial resolution and loading step sensitivity studies: the macroscopic curve is very stable around the setup used as reference ($65^3$ resolution and 80 loading steps). Finally, it is observed that the number of iterations increases with both the spatial resolution and the number of loading steps. A part of this increase is consistent with the algorithms (especially their initializations) and a future prospect will be to reduce it.

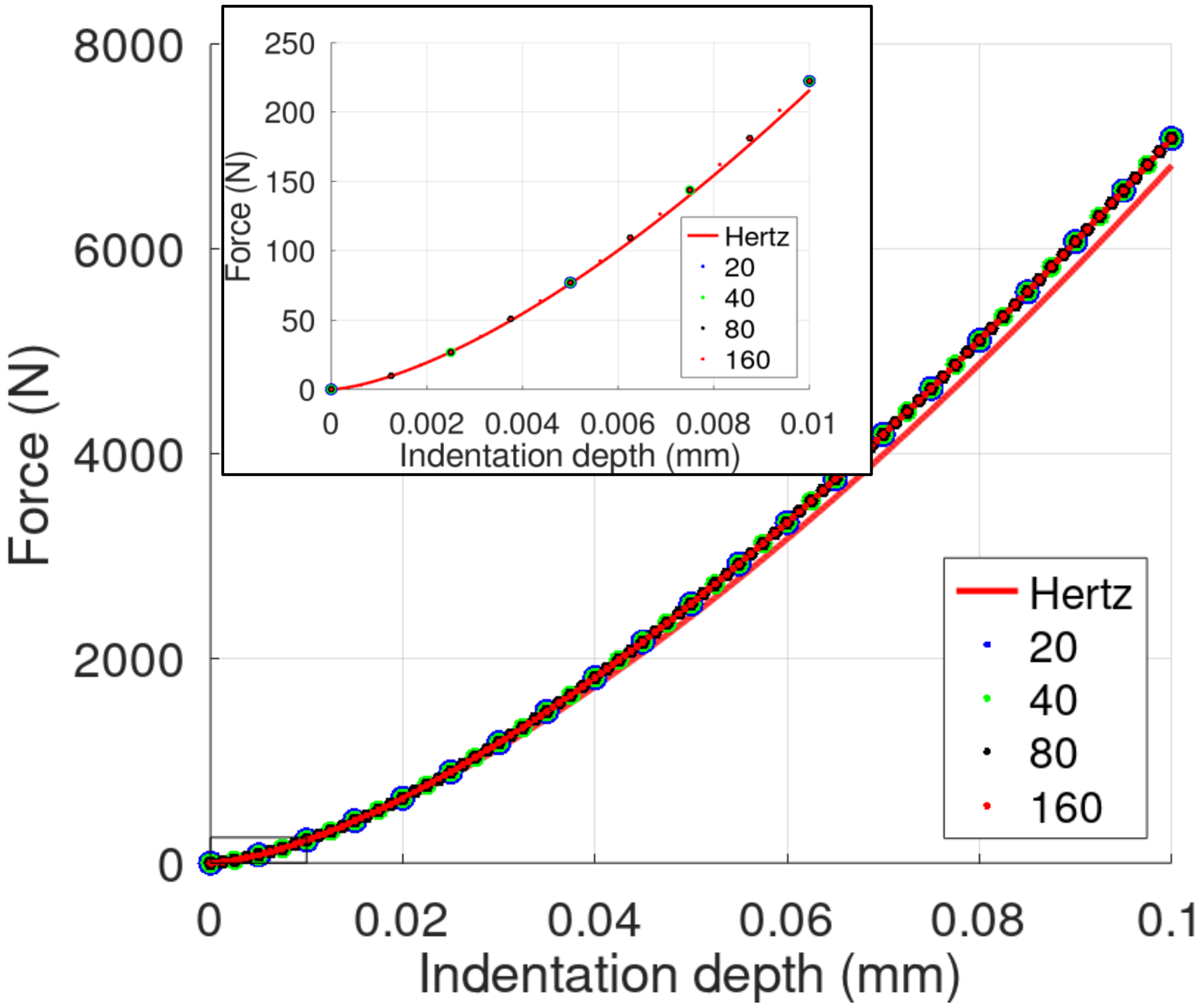

*Figure 8 : Force-indentation depth curves from small strain simulations with a spatial resolution of $65^3$ and 20, 40, 80 and 160 loading steps. The red continuous line represents the curve obtained with the Hertz theory.*

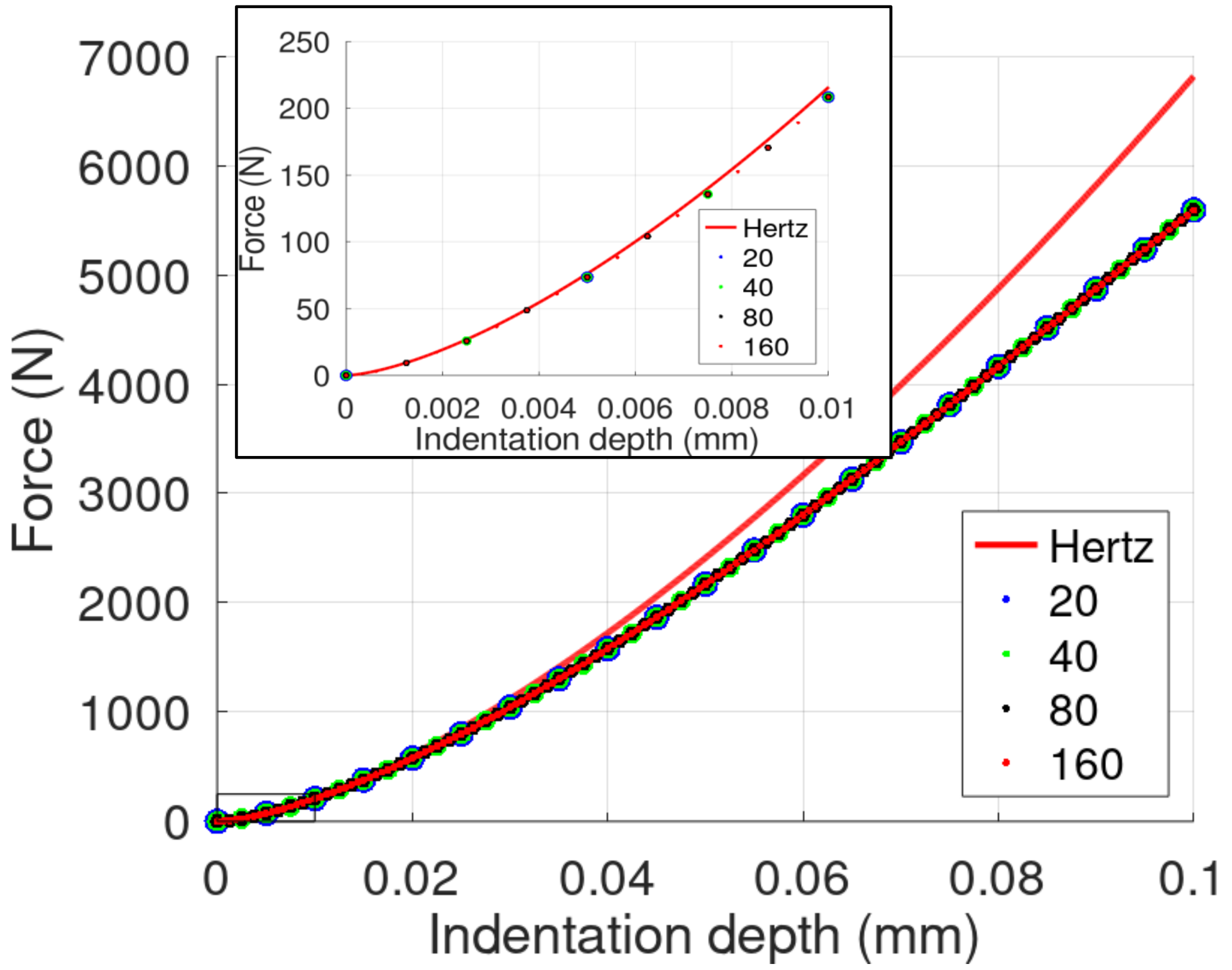

*Figure 9 : Force-indentation depth curves from finite strain simulations with a spatial resolution of* $65^3$ *and 20, 40, 80 and 160 loading steps. The red continuous line represents the curve obtained with the Hertz theory.*

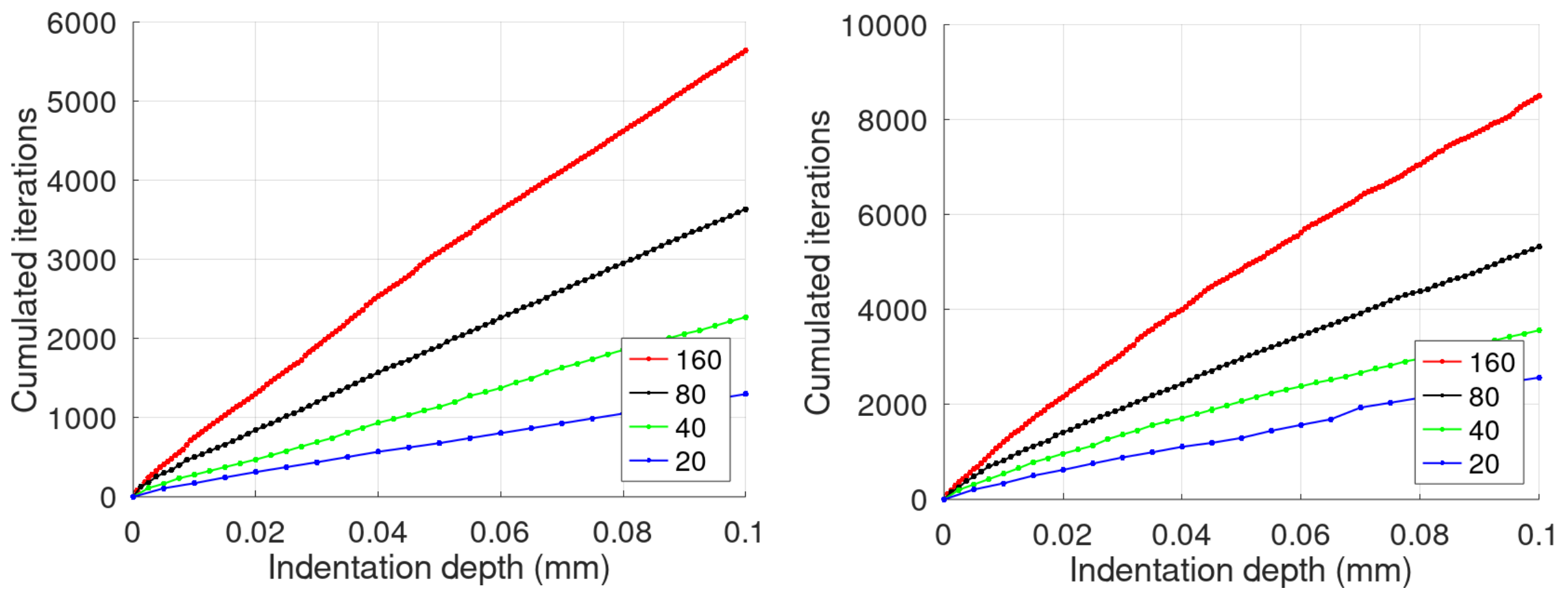

*Figure 10 : Cumulated number of iterations for simulations with 20, 40, 80 and 160 loading steps and a spatial resolution of* $65^3$ *: small strain case (left) and finite strain case (right).*

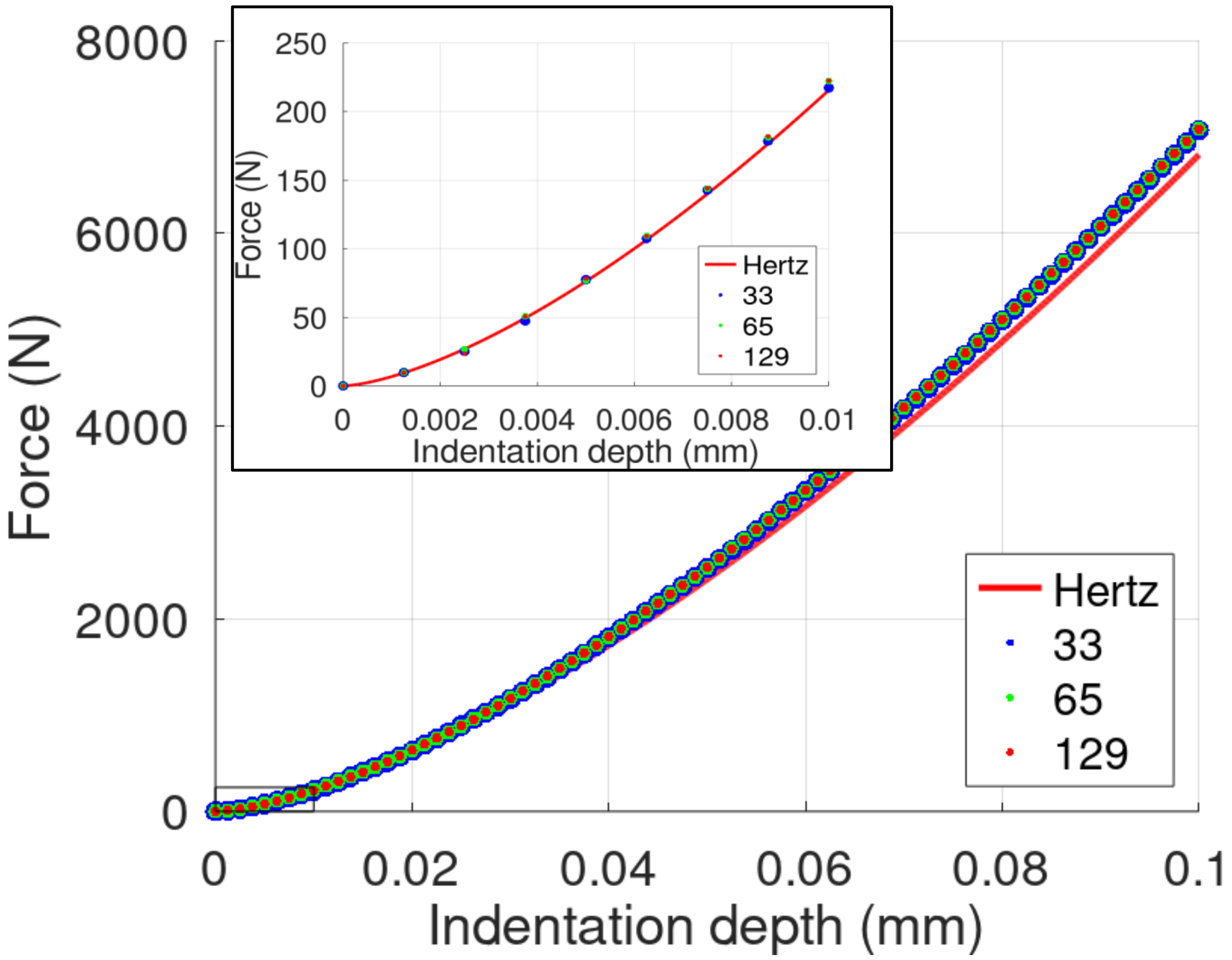

*Figure 11 : Force-indentation depth curves from small strain simulations with 80 loading steps and a spatial resolution of* $33^3$*,* $65^3$ *and* $129^3$*. The red continuous line represents the curve obtained with the Hertz theory.*

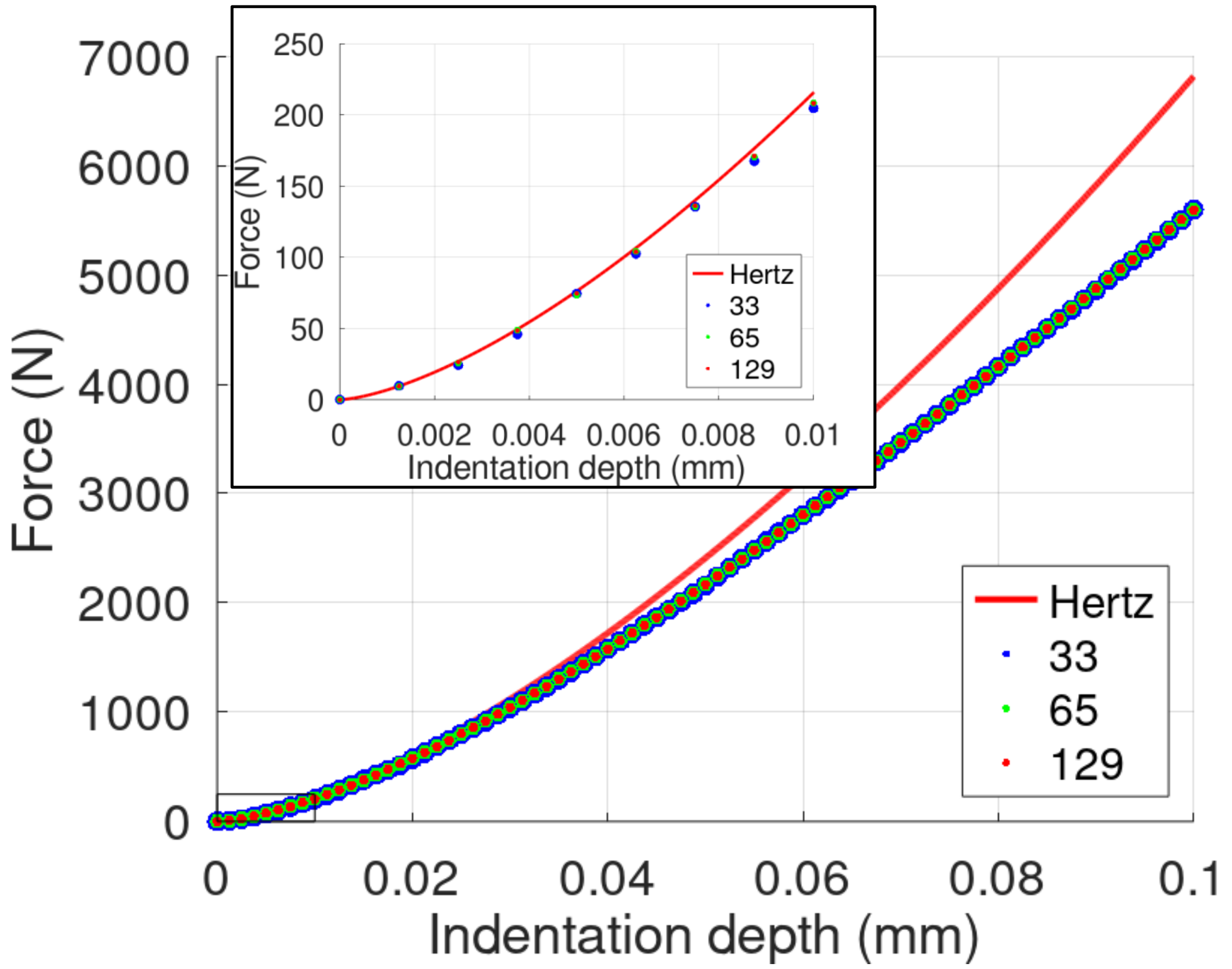

*Figure 12 : Force-indentation depth curves from finite strain simulations with 80 loading steps and a spatial resolution of* $33^3$, $65^3$ *and* $129^3$. *The red continuous line represents the curve obtained with the Hertz theory.*

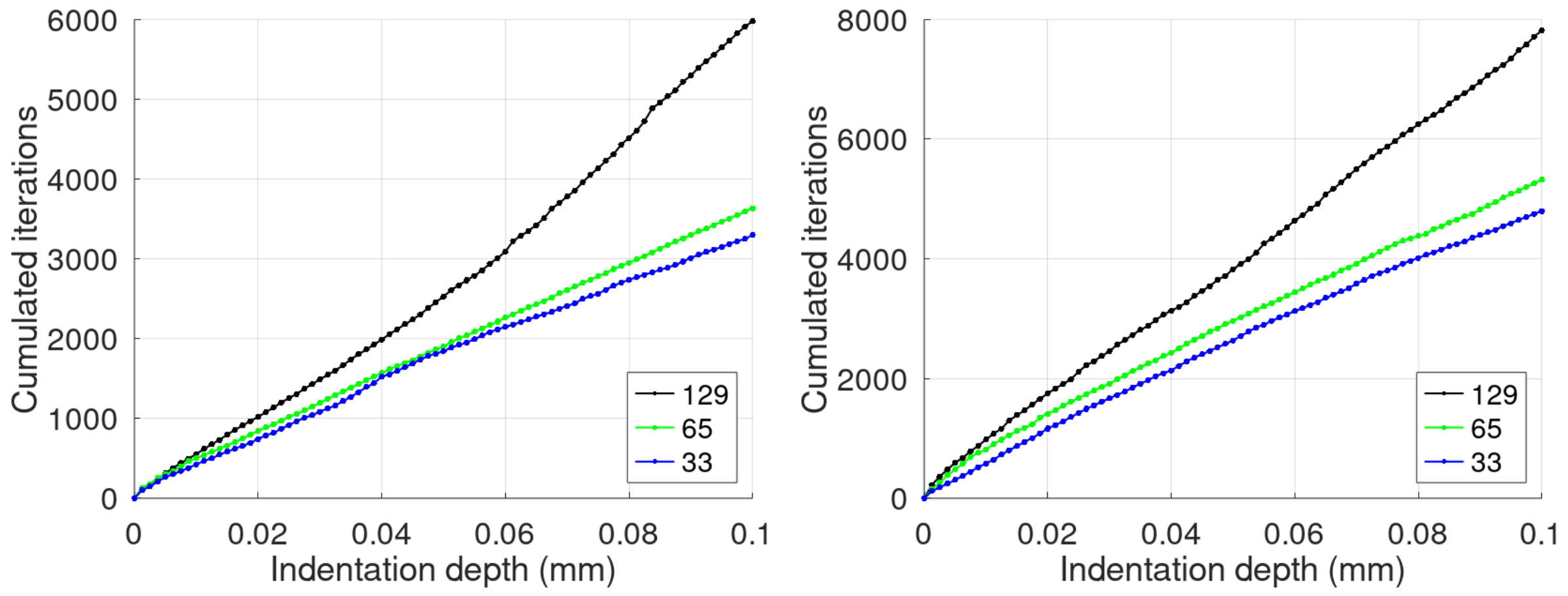

*Figure 13 : Cumulated number of iterations for simulations with 80 loading steps and a spatial resolution of* $33^3$, $65^3$ *and* $129^3$, *small strain case on the left side and finite strain case on the right side.*

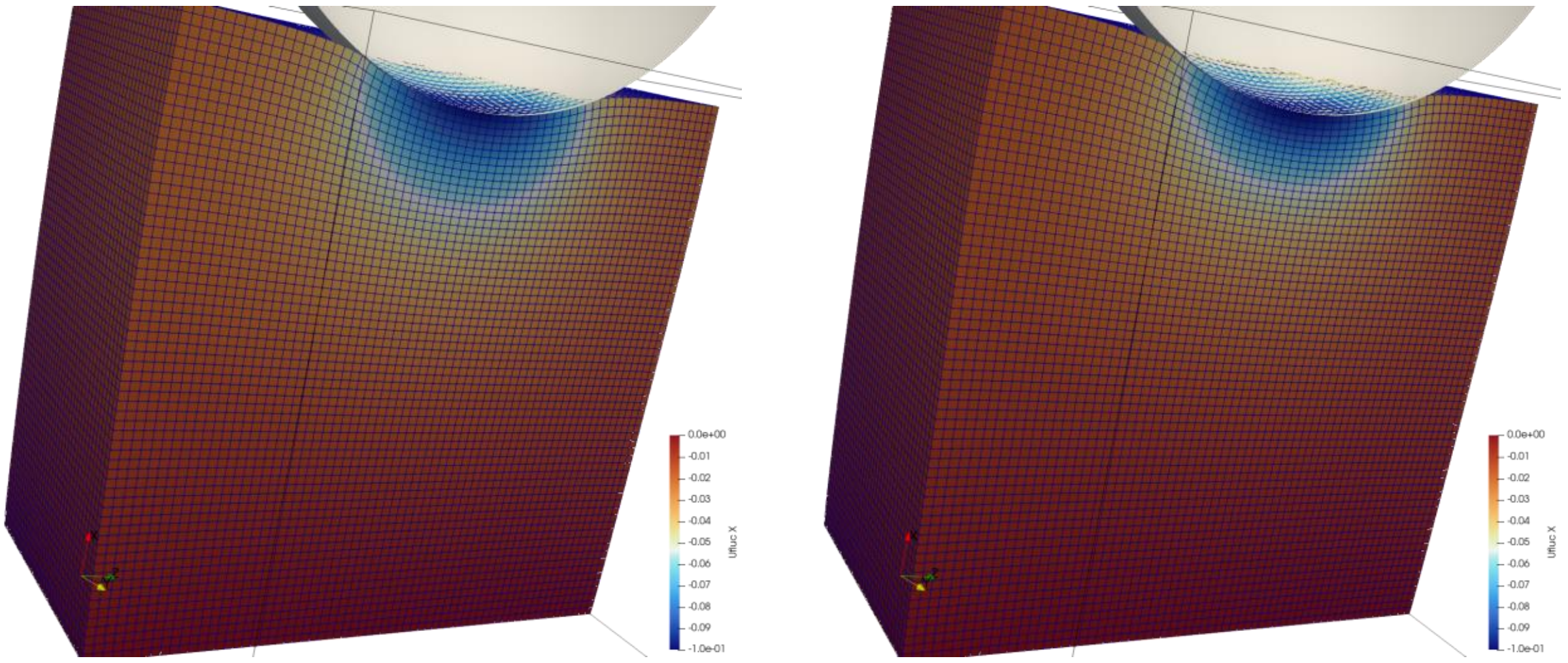

*Figure 14 : Deformed configurations at the end of the loading for a resolution of $65^3$ for small strain (left) and finite strain (right) simulations.*

5.4: Discussion: minimization under inequality constraint

Following the discussion in section 4.3 looking at problems with local Dirichlet conditions as minimization problems subject to equality constraints, the present discussion considers a minimization problem with an inequality constraint given by the definition of the gap $g$ (below $[g([\tilde{u}])]$ vanishes if $[\tilde{u}] = [0]$). Constraint is applied on the surface of potential contact $S_1^1$ so that $[g_0] \in \mathbb{R}^m$, where $m$ is equal to the number of nodes in $S_1^1$ and a scalar condition for the gap). For the SS case, the minimization problem (*18*) becomes:

$$\inf_{[\tilde{u}]\in\mathbb{R}^n}\big(J([\tilde{u}])\big) = \inf_{[\tilde{u}]\in\mathbb{R}^n}\left(\frac{1}{2}[\tilde{u}]^T[\![K]\!][\tilde{u}]\right) \quad \text{with } [\tilde{u}] \text{ subject to } [g([\tilde{u}])] - [g_0] \geq 0 \text{ on } S_1^1 \tag{41}$$

As in section 4.3, the Lagrangian of this problem is considered leading to the definition of a saddle point.

$$\sup_{[f]\in\mathbb{R}^m, [f]\leq 0} \inf_{[\tilde{u}]}\big(J([\tilde{u}]) + [f]^T([g([\tilde{u}])] - [g_0])\big) \tag{42}$$

Note that in that case, the set of Lagrange multipliers $[f] \in \mathbb{R}^m$ is defined on the surface of potential contact $S_1^1$ and their admissible space is limited to negative values. The solution widely adopted by the contact mechanics community, is given by the Karush-Kuhn-Tucker conditions reported in a simplified way (i.e. assuming $g([\tilde{u}])$ linear).

$$\begin{cases} \begin{bmatrix} [\![K]\!] & [\![\partial_{\tilde{u}} g]\!]^T \\ [\![\partial_{\tilde{u}} g]\!] & [\![0]\!] \end{bmatrix} \begin{bmatrix} [\tilde{u}] \\ [f] \end{bmatrix} = \begin{bmatrix} [0] \\ [g_0] \end{bmatrix} \\ f_i \leq 0; \quad g_i - g_{0_i} \geq 0; \quad \text{and } f_i(g_i - g_{0_i}) = 0 \end{cases} \tag{43}$$

Index $i$ runs over all the nodes of the potential contact surface $S_1^1$. In practice, such problems can be solved using FE codes with an active-set method: the first line is solved with Lagrange multipliers restricted to the points of the contact zone (active-set), then the second line is used to detect where the constraints are violated and to update the contact zone (active-set), and then return to the first line to solve the linear problem with the updated contact zone (active-set). The simulation of the spherical indentation reported above, demonstrates the ability of the proposed FFT-based solver to solve such minimization problems subject to inequality constraints. The methodology developed here for contact mechanics can be adapted to other physical equation subjected to inequality constraints.

## 6. Conclusion and future prospects

The starting point of the present work has been briefly described in section 3. The most important points are: the use of a displacement-based iterative algorithm, the efficient acceleration convergence technique used to accelerate the fixed-point algorithm, the use of DTTs to account for possibly non-periodic BCs per face and per component, and finally, the possibility to further explore the use of the TETRA2 finite difference scheme recently proposed by Finel [28] and its comparison with the HEX1 scheme.

Then, section 4 has proposed two simple modifications of the algorithm to account for local Dirichlet BCs (i.e. Dirichlet BCs that are not prescribed through DTTs on the faces of the unit-cell). The first one consists in correcting the current unknown displacement field so that the local Dirichlet BCs are satisfied (i.e. just set the prescribed values in the current field). The second one consists in introducing reaction forces, at points where local Dirichlet are prescribed, to balance the stress divergence so that the equilibrium is systematically satisfied (i.e. just set null values in the current residual field $\boldsymbol{p}$). A non-trivial application is proposed through the loading of a half-CT (Compact Tension)-like specimen. The loading is applied on a single line of nodes inside the volume element, and a null displacement for the component 3 is applied on the surface corresponding to the ligament located ahead of the crack. The rest of the surface is traction free (Neumann BC). While the HEX1 scheme (known for its possible instability associated with hourglass phenomenon) did not converge, results obtained with the TETRA2 scheme have been validated w.r.t. a FE simulation. To go beyond, the possibility of applying Dirichlet BCs to a single node has been explored, focusing on the TETRA2 scheme. It is emphasized and explained from Finel's description [28] that it is impossible to apply the displacement to a single node. Hence, the first alternative approach applies the Dirichlet BCs to the center of a voxel, so that a linear relationship is imposed between the 8 nodes of the voxel. The second approach prescribes the displacement of a single node but redistributes the reaction force on neighboring nodes. As reaction forces and Lagrange multipliers are closely related, this last approach could be referred to as "diffuse" Lagrange multipliers. The two methods have been tested and compared successfully. At the end of section 4, the local equations used to introduce the algorithm are discussed from the viewpoint of minimization with equality constraints and the introduction of Lagrange multipliers. The proposed FFT/DTT-based modified algorithm is quite original when compared to other FFT-based solver's extensions to apply Dirichlet BCs that generally follow a 'coupled' resolution solving together the displacement field and the reaction forces. Our approach has been connected to established methods for solving saddle point systems and especially to 'null-space' methods [2], that allows for a resolution on the displacement field, the reaction forces being evaluated as a post-processing. A drawback of the method is that the convergence proved to be mesh sensitive, that is a common issue when solving saddle point systems [2]. However, on the three investigated examples (section 4.1.3, section 4.2.4 and appendix A2), the sensitivity proved to be highly case dependent but was never 'catastrophic' (the worst case was a dependency in $d^{-0.6}$).

As a practical conclusion to this section, the proposed FFT-based solver has demonstrated its ability to solve minimization with equality constraints, including relations between displacements of different nodes. To the best of authors' knowledge, the proposed concept of "diffuse" Lagrange multipliers is also novel.

Finally, the question of minimization problems with inequality constraints is investigated with a focus on the simulation of contact mechanics through a simple application (a rigid spherical indenter on a cube volume element assuming frictionless contact). The modification of the previous algorithm proposed to apply local Dirichlet BCs is further extended to incorporate a so-called active-set method to deal with the evolution of the contact zone. The active-set method consists in simulating the problem with Dirichlet BCs for a given contact zone, then verifying where the contact conditions are violated, then modifying the contact zone (i.e. the active-set) and re-run the simulation with local Dirichlet BCs but with the new contact zone. This procedure is run iteratively until stability of the contact zone. This active-set method is closely introduced within the accelerated fixed-point algorithm. The algorithm is then proposed in the context of small strain assumption but also in the context of finite strain. The robustness of the method has been established through different sensitivity analysis w.r.t. spatial resolution and loading discretization. For small (indentation depth / sphere radius) ratio corresponding to the validity range of the Hertz theory, the results obtained with the SS and FS implementations are very close to the results given by the Hertz theory, validating at least partially the two implementations. When increasing the ratio, the curves obtained with SS and FS depart from each other. Note that this difference reflects both the better treatment of the kinematics in FS simulations and the choice made for the constitutive behavior.

As a future prospect, the present work widely opens the door of structural analysis to FFT-based solvers. In addition, if two different finite difference schemes (HEX1 and TETRA2) are used, it should not obscure the ability of the method to account for various types of finite element discretization (see for instance Schneider et al. [26] or Ladecký et al. [65]). Similarly, if the method is proposed with a regular grid discretization, it must be emphasized that a grid adaptation can also be taken into account (see the works of Zecevic et al. [66], Bellis and Ferrier [53] or Bignonnet et al. [67]). Note that composite voxels [68] [69] could also be thought to apply BCs in the case of immersed boundaries. At the end, gathering all these advances in a single parallel implementation, for example the AMITEX* code, a partial re-writing of the AMITEX code [18] developed by the first author, could be a very nice project.

The extension of the proposed algorithms to non-linear behaviors is rather straightforward. Actually, the evaluation of the stress (the Cauchy stress for the small strain framework or the first Piola-Kirchhoff stress) by the behavior law from the displacement gradient is completely generic. In practice in the AMITEX/AMITEX* code, it relies on the classical UMAT format. The linear relations used in the present paper correspond to very simple UMAT implementations.

The extension of the finite strain algorithm (in the context of contact mechanics, but the contact condition could also be removed) to very large deformation is still an open question. Actually, the algorithm relies on a Total Lagrangian formulation that is limited to moderate deformations. If it can be sufficient for a large number of applications in the context of mechanics of materials, it can be a limitation for polymer materials, metal forming processes and others. To overcome this limitation and deal with very large deformations, it could be interesting to investigate an Updated Lagrangian strategy together with adaptive mesh refinement [70] that could rely on the work of Bellis and Ferrier [53].

Finally, if the methodology proposed to solve minimization problems with equality or inequality constraints is introduced here in the framework of local continuum mechanics, the method is completely general and can be applied to various sets of equations related to different physics, or non-local equations in mechanics. As an example, the phase field model for fracture mechanics could benefit from this proposition to account for the irreversibility of damage (the damage variable must not decrease), instead of using a penalty method [71] or the "history variable" proposed by Miehe et al. [72] (leaving the variational framework associated with the phase field method). In that case, the active-set will gather internal nodes (not only surface nodes as for the contact mechanics problem) concerned by the irreversibility constraint.

## Acknowledgments

This research project was funded by the PTC-SN program (Programmes Transversaux de Compétences - Simulation Numérique) of the French Alternative Energies and Atomic Energy Commission (CEA), France. Authors are grateful to Cédric Flageul (Curiosity Group, Pprime Institute, CNRS - University of Poitiers - ENSMA) and Yushan Wang (Université Paris-Saclay, UVSQ, CNRS, CEA, Maison de la Simulation) for fruitfull discussions and help. We also would like to thank the reviewers for their careful reading and suggestions to improve the present manuscript.

APPENDIX A - Additional results regarding the indentation simulation in section 5.3

A1 – Comparison with Hertz theory

In section 5.3, good agreement with Hertz theory is observed on the reaction-force vs indentation depth curve. The purpose of the present appendix is to investigate this comparison in more detail.

Recall that in the simulation $d_x = d_y = d_z$ and $n_x = n_y = n_z$, so $d$ and $n$ are used below.

The Hertz theory provides an analytical expression of the radius of the contact zone $a = \sqrt{R\delta}$. On the other hand, from the numerical simulation, this radius is evaluated from the number of nodes in the contact zone ($N_c$) using the surface of the contact zone $S_c = \pi a^2$ and its approximation $S_c \sim N_c d^2$, leading to $a = \sqrt{\frac{N_c}{\pi}}d$. The comparison on Figure 15 exhibits a nice agreement up to the highest $\delta/R$ ratio (20%). Refining the mesh improves the agreement. The steps, especially observed for the lowest resolution, indicate an evolution of the depth without update of the contact zone (hence $N_c$ remains constant).

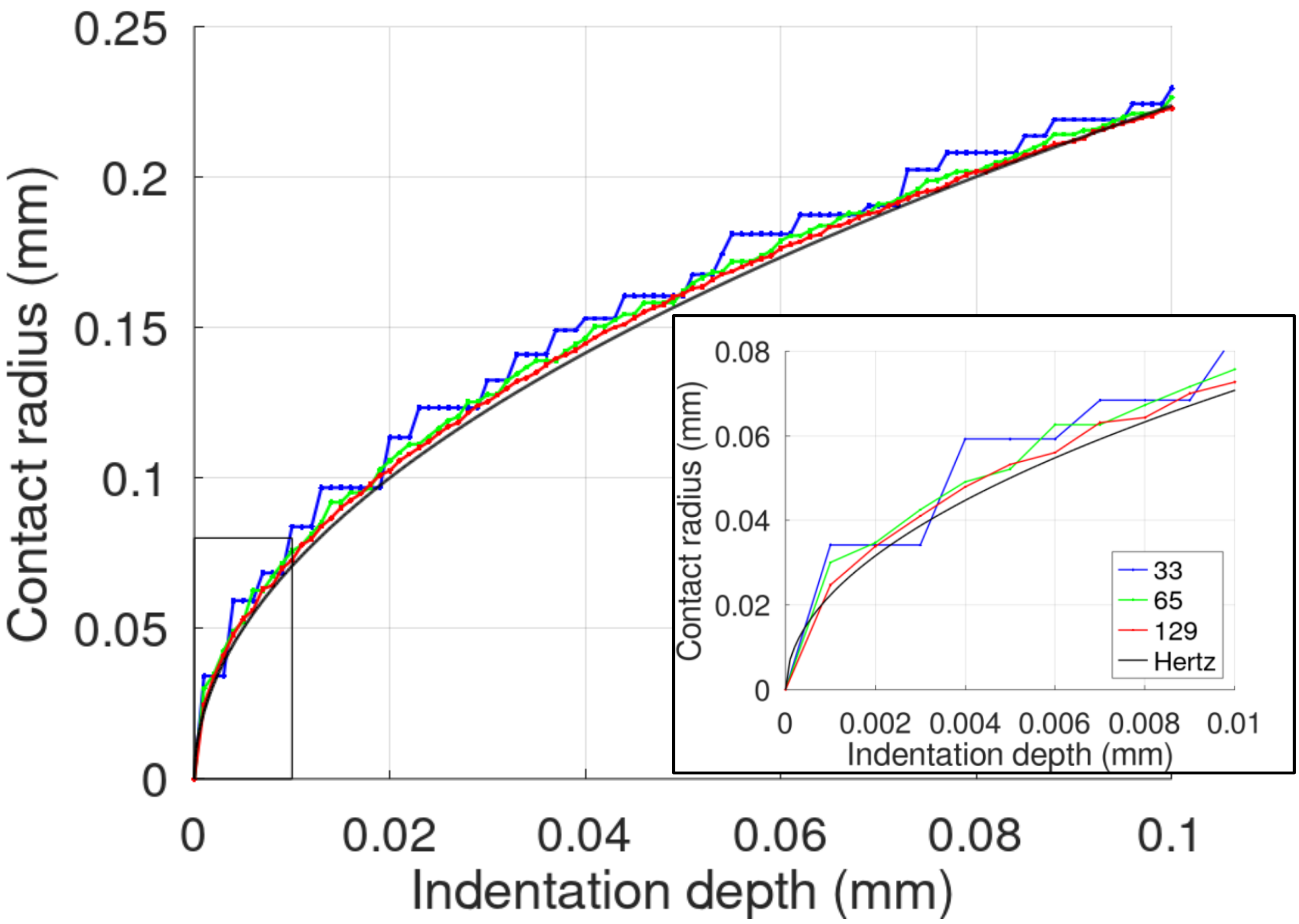

*Figure 15 : Contact radius-indentation depth curves for small strain simulations with spatial resolutions of* $33^3$, $65^3$ *and* $129^3$. *The black continuous line represents the curve obtained with the Hertz theory.*

The Hertz theory also provides an analytical expression of the contact pressure profile $P = P_0\sqrt{1 - \frac{y^2+z^2}{a^2}}$, with $a$ defined previously and $P_0 = \frac{2E^*}{\pi}\sqrt{\delta/R}$ with $E^* = E/(1-\nu^2)$. For the simulation, the contact pressure is from $f_{n_c}^C$ evaluated at each point of the contact zone using $P =$ $f_{n_c}^{C} d/2$. Note that, as the resolutions ($n$) are odd, the maximum pressure in the simulation occurs at a half voxel of the center of the contact zone, and the contact pressure profile is evaluated at a half-voxel of the symmetry axis. For a precise comparison, these exact positions are taken into account in the analytical expression (it explains the null maximum contact pressure observed for very small indentation depth on Figure 16, especially visible for the lowest resolution). The comparison of the maximum contact pressure on Figure 16 exhibits a nice agreement in the range of low $\delta/R$ ratio (the zoom in the figure), improved when increasing the resolution. For larger $\delta/R$ ratio the analytical and numerical results depart from each other. The comparison of the contact pressure profiles is made on Figure 17 for a small $\delta/R$ ratio (i.e. 0.01/0.5=2%). Once again, the agreement is very nice and improves with increasing resolution.

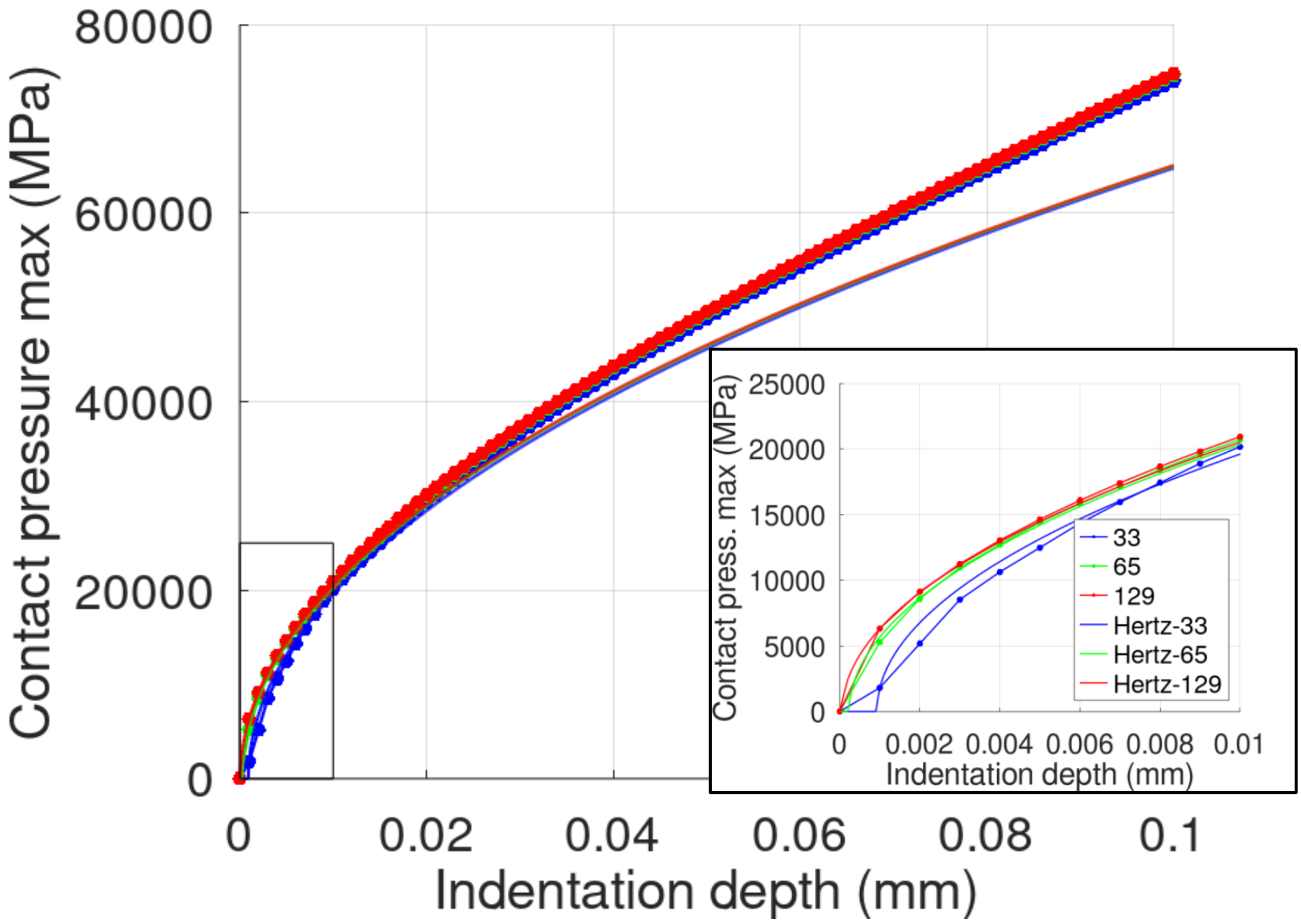

*Figure 16 : Maximum contact pressure in simulations as a function of indentation depth for small strain simulations with spatial resolutions of $33^3$, $65^3$ and $129^3$. Note that, due to odd resolutions, the maximum contact pressure is evaluated at a half voxel from the center of the contact zone.*

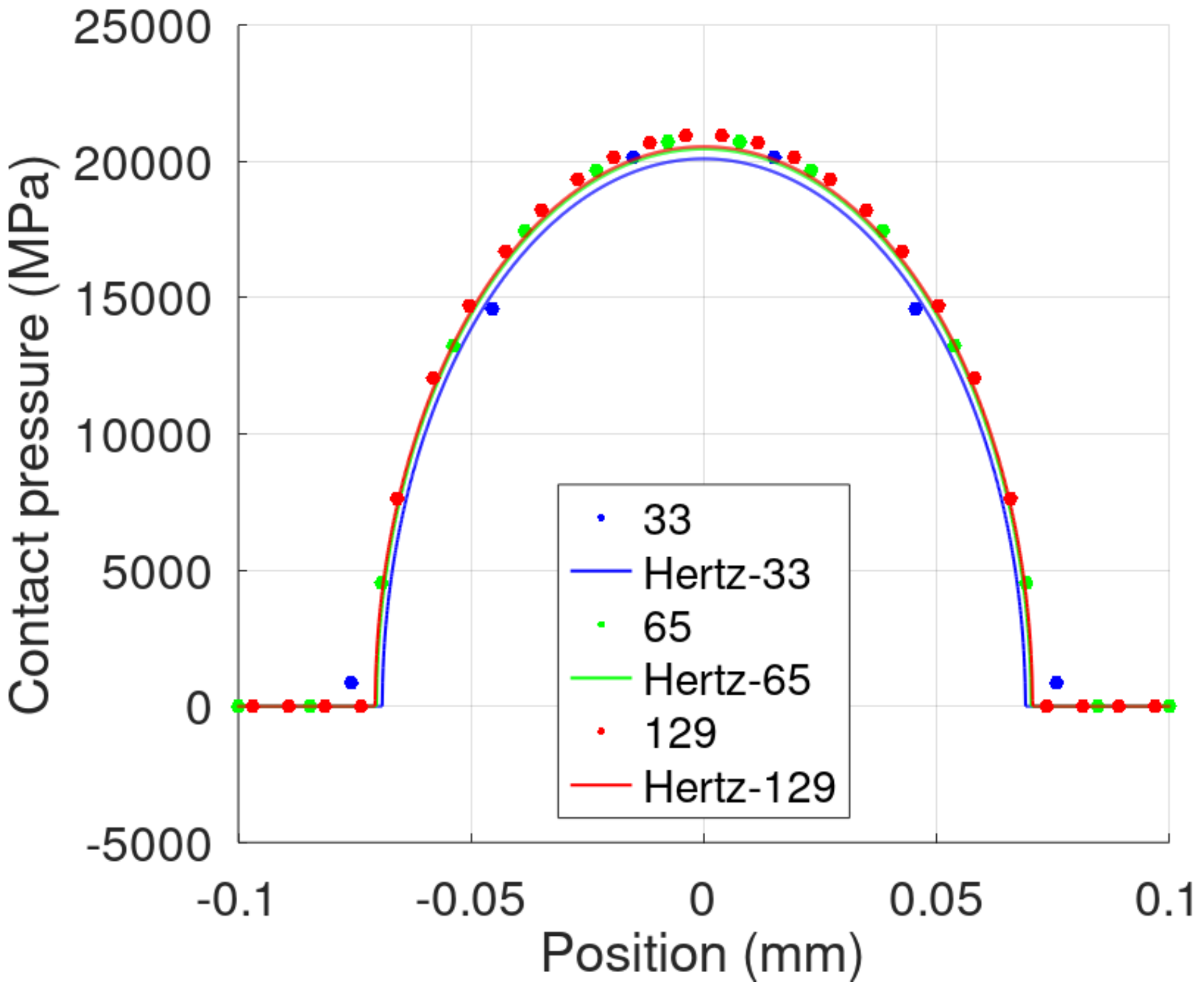

*Figure 17 : Contact pressure profiles in simulations for an indentation depth $\delta = 0.01mm$, using the small strains assumption with spatial resolutions of $33^3$, $65^3$ and $129^3$. Note that, due to odd resolutions, the contact pressure profiles are evaluated at a half voxel from the symmetry axis of the contact zone.*

To discuss the quality of the simulation, the evaluation of the reaction force evaluated in section 5.3 from the average stress on the unit-cell is compared to the reaction force evaluated from the local reaction forces in the contact zone using the sum of $f_{n_c}^{C} d^3/2$ over all the points of the contact zone. The relative difference between these two approaches, observed on Figure 18-left, is consistent with the convergence tolerance $10^{-4}$. Another interesting point is to check that the contact-zone reaction forces remain negative. For that purpose, the maximum 'signed' contact pressure is displayed on Figure 18-right. The quantity is normalized by the maximum contact pressure (unsigned) evaluated at each loading step. It is clear that this maximum value remains negative whatever the loading step and resolution. This confirms that the contact condition on the reaction forces is satisfied for all the points in the contact zone. As expected, it is also observed that this value tends to zero when increasing the resolution. Actually, when refining the mesh, the description of the contact pressure under the contact zone is refined and the probability to find a node with contact pressure close to zero increases.

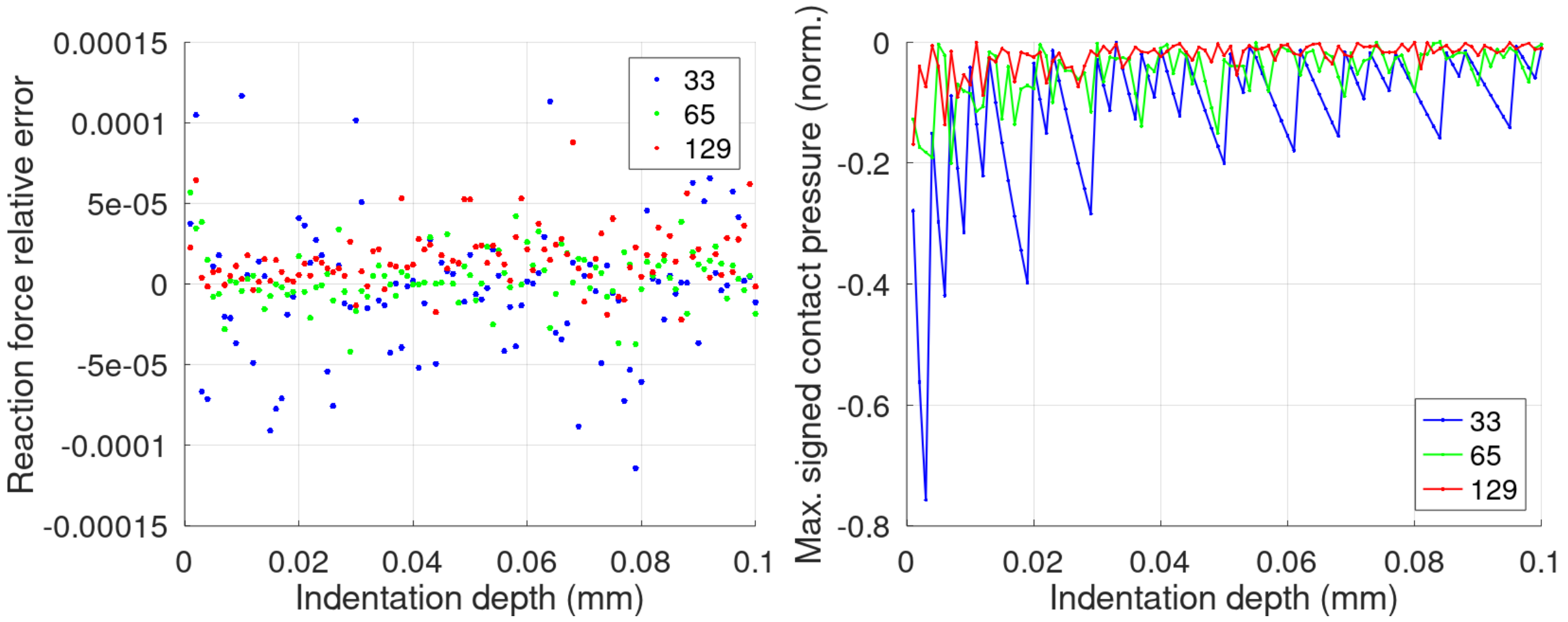

*Figure 18 : Relative difference between the total reaction force evaluated from the contact zone or from a volume average (left). Evolution of the maximum signed contact pressure over the contact zone (right). The quantity is normalized by the maximum contact pressure (unsigned). Simulations are performed using small strain assumptions with resolutions* $33^3$, $65^3$ *and* $129^3$.

## A2 – Convergence analysis w.r.t. mesh refinement

As observed in Figure 13 the cumulated number of iterations increases with spatial resolution, and the evolution is more important from resolution 65 to 129 than from 33 to 65. In order to investigate more deeply this observation, the number of nodes in the contact zone as well as the number of contact zone updates are displayed on Figure 19 as a function of the loading (i.e. indentation depth).

The evolutions of the number of updates are rather linear and, as expected, this number increases with resolution. However, the increase between resolution 33 and 65 is rather similar to the increase between resolutions 65 and 129, hence, it cannot, by itself, explain the curves displayed on Figure 13.

The evolution of the number of nodes in the contact zone evolves quite linearly but with a higher difference between resolution 65 and 129 than between 33 and 65. Note that, as expected from the equation $S_c = N_c d^2 = \pi a^2$ (see A1) and the Hertz theory ($a^2 = R\delta$), the number of contact nodes approximately follows the linear relation $N_c = \frac{\pi R}{d^2}\delta$ with a dependency of the slope in $d^{-2}$. This larger sensitivity of the number of contact nodes with respect to the resolution could explain the results of Figure 13.

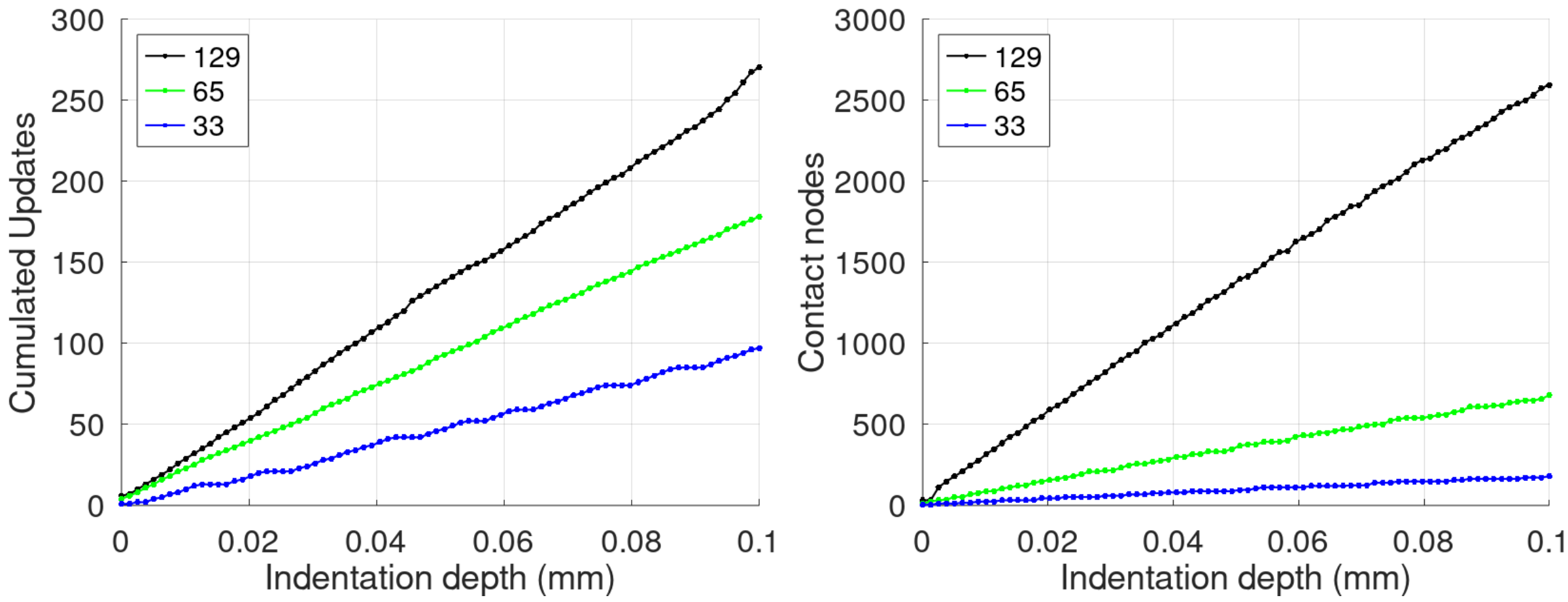

*Figure 19 : Number of cumulated contact zone updates (left) and number of contact nodes (right) as a function of the indentation depth for simulations with different spatial resolution of* $33^3$, $65^3$ *and* $129^3$ *(small strain case with 80 loading steps).*

To further explore this point, a new series of simulations has been performed using a fixed active-set and a single loading step, the active-set being defined by the nodes violating initially the gap condition. Using a fixed active-set, each simulation takes into account a kinematic equality constraint (null gap) applied to the points of the active-set. The numbers of iterations at convergence (using the same convergence tolerance as used for the contact simulations, i.e. $10^{-4}$) are reported in Table 9 for the three resolutions and different indentation depths. Disregarding a few points attributed to the non-smooth convergence of the convergence acceleration method, the general trend is that the number of iterations increases with the resolution and the loading. However, for a constant indentation depth, the increase is not very important, even if the number of points in the active-set also depends on $d^{-2}$ (from a simple geometric analysis $N_c = \frac{\pi(2R-\delta)}{d^2}\delta$ ). Hence, this result tends to contradict the idea proposed just above to explain the results of Figure 13 that remains an open question.

As a final comment of Table 9, it must be mentioned that for a given depth, the number of nodes in the active-set increases by ~15 when increasing resolution from 33 to 129, and for a given resolution, it increases by ~14 when increasing the depth from 0.005 to 0.075. Hence, for a similar increase of nodes in the active-set (~14-15), the number of iterations is multiplied by 1.3 (on average) when increasing the resolution for a given depth, and by 1.7 (on average) when increasing the depth for a given resolution. An estimate of the mesh sensitivity deduced from Table 9 gives a moderate dependence in $d^{-0.2}$.

| $\delta(mm)$ / $n_x$ | 0.005 | 0.01 | 0.025 | 0.05 | 0.075 | 0.1 |
|---|---|---|---|---|---|---|
| 33 | 54 | 56 | 78 | 86 | 86 | 130 |
| 65 | 56 | 58 | 96 | 148 | 94 | 140 |
| 129 | 62 | 78 | 106 | 126 | 121 | 158 |

*Table 9 : Number of iterations at convergence with a fixed active-set (see text for the details of the simulation), for various resolutions* $n_x$ *and indentation depth* $\delta$.